\documentclass{aa}

\usepackage{graphicx}
\usepackage{txfonts}
\usepackage{xcolor}
\usepackage{amsmath}
\DeclareMathOperator\arsinh{arsinh}
\usepackage{float}

\usepackage{hyperref}
\usepackage{multirow}
\usepackage{appendix}
\usepackage{makecell}

\usepackage{natbib}
\bibpunct{(}{)}{;}{a}{}{,} 

\begin{document}

   \title{\textsc{MaxiMask} and \textsc{MaxiTrack}: Two new tools for identifying contaminants in astronomical images using convolutional neural networks}

  \titlerunning{\textsc{MaxiMask} and \textsc{MaxiTrack}}
  
   \author{M. Paillassa\inst{1}, E. Bertin\inst{2}, H. Bouy\inst{1}}

   \institute{\inst{1} Laboratoire d'astrophysique de Bordeaux, Univ. Bordeaux, CNRS, B18N, all\'ee Geoffrey Saint-Hilaire, 33615 Pessac, France \\ \inst{2} Sorbonne Universit\'e, CNRS, UMR 7095, Institut d'Astrophysique de Paris, 98 bis bd Arago, 75014 Paris, France.}
   
   \date{Received 18 July 2019 / Accepted 9 December 2019}

 
   \abstract{
     In this work, we propose two convolutional neural network classifiers for detecting contaminants in astronomical images. Once trained, our classifiers are able to identify various contaminants, such as cosmic rays, hot and bad pixels, persistence effects, satellite or plane trails, residual fringe patterns, nebulous features, saturated pixels, diffraction spikes, and tracking errors in images. They encompass a broad range of ambient conditions, such as seeing, image sampling, detector type, optics, and stellar density.
     
     The first classifier, \textsc{MaxiMask}, performs semantic segmentation and generates bad pixel maps for each contaminant, based on the probability that each pixel belongs to a given contaminant class.
     
     The second classifier, \textsc{MaxiTrack}, classifies entire images and mosaics, by computing the probability for the focal plane to be affected by tracking errors.
     
     We gathered training and testing data from real data originating from various modern charged-coupled devices and near-infrared cameras, that are augmented with image simulations. We quantified the performance of both classifiers and show that \textsc{MaxiMask} achieves state-of-the-art performance for the identification of cosmic ray hits. Thanks to a built-in Bayesian update mechanism, both classifiers can be tuned to meet specific science goals in various observational contexts.
     }


   \maketitle

%

\section{Introduction}

Catalogs extracted from astronomical images are at the heart of modern observational astrophysics.
Minimizing the number of spurious detections in these catalogs has become increasingly important because the noise added by such contaminants can, in many cases, compromise the scientific objectives of a survey.
Properly identifying and flagging spurious detections yields substantial scientific gains, but it is complicated by the numerous types of contaminants that pollute images. Some of them stem from the detector electronics (e.g., dead or hot pixels, persistence, saturation), from the optics (diffraction along the optical path, scattered and stray light), from post-processing (e.g., residual fringes), while others are the results of external events (cosmic rays, satellites, tracking errors).
The amount of data produced by modern astronomical surveys makes visual inspection impossible in most cases.
For this reason, developing fully automated methods to separate contaminants from true astrophysical sources is a critical issue in modern astronomical survey pipelines.

Most current pipelines rely on a fine prior knowledge of their instruments to detect and mask electronic contaminants \citep[e.g.,][]{HSC_Pipeline,2018PASP..130g4501M} and to some extent optical contaminants \citep[e.g.,][]{Kawanomoto2016, Komiyama2016}. Cosmic ray hits can be identified by rejecting outliers in the timeline, provided that multiple consecutive exposures are available, by using algorithms sensitive to their peculiar shapes, such as Laplacian edge detection  \citep[e.g., LA Cosmic,][]{2001PASP..113.1420V} or wavelets \citep[e.g.,][]{2008StMet...5..373O}. The Radon transform or the Hough transform have often been used to detect streaks caused by artificial satellites or planes in images \citep[e.g,][]{2002SPIE.4847..123V, nir2018optimal}.

In this work, we want to overcome some of the drawbacks of the above mentioned methods. 
First, the typical data volume produced by modern surveys requires that the software is largely unsupervised and as efficient as possible. 
Second, we aim to develop a robust and versatile tool for the community at large and therefore want to avoid the pitfall inherent in software that is tailored to a single or a handful of instruments, without compromising on performance.
Third, we would like to have a unified tool able to detect many contaminants at once.
Finally, we want to assign to each pixel a probability of belonging to a given contaminant class rather than Boolean flags.
These constraints lead us to choose machine learning techniques and in particular supervised learning and convolutional neural networks (CNNs).

Supervised learning is a field of machine learning dealing with models that can learn regression or classification tasks based on a data set containing the inputs and the expected outputs.
During the learning process, model parameters are adjusted iteratively to improve the predictions made from the input data.
The learning procedure itself consists of minimizing a loss function that measures the discrepancy between model predictions and the expected values.
Minimization is achieved through stochastic gradient descent.
We recommend \cite{ruder2016overview} for an overview of gradient descent based optimization algorithms.

Convolutional neural networks \citep{lecun1995convolutional} are particulary well-suited to identifying patterns in images.
Unlike previous approaches that would involve hand crafted feature detectors, such as SIFT descriptors \citep{lowe1999object}, CNN models operate directly on pixel data.
This is made possible by the use of trainable convolution kernels to detect features in images.
Convolution is shift-equivariant, which allows the same features to be detectable at any image location.

CNNs are now widely used in various computer vision tasks, including image classification, that is assigning a label to a whole image \citep{krizhevsky2012imagenet, simonyan2014very, szegedy2015going}, and semantic segmentation, that is assigning a label to each pixel \citep{long2015fully, badrinarayanan2017segnet, garcia2017review}.

In this work, we propose to identify contaminants using both image classification and semantic segmentation.

In the following, we first describe the images that we used and how we built our data sets. Then, we focus on the neural network architecture that we used. Finally, we evaluate the models performance on test sets and on real data.

\section{Data}

In this section we describe the data used to train our two neural networks. 
We distinguish between two types of contaminants:
On the one hand, local contaminants, that affects only a fraction of the image at specific locations.
This includes cosmic rays, hot columns and lines, dead columns and lines, dead clustered pixels, hot pixels, dead pixels, persistence, satellite trails, residual fringe patterns, ``nebulosity'', saturated pixels, diffraction spikes, and over scanned pixels. These add up to 12 classes. 
On the other hand, global contaminants, that affects the whole image, such as tracking errors.

\subsection{Local contaminant data}

For local contaminants, we choose to build training samples by adding defects to uncontaminated images in order to have a ground truth for each contaminant.
In this section we first describe the library of astronomical images used for our analysis, then focus on the selection of uncontaminated images, and finally describe the way each contaminant is added.

\subsubsection{Library of real astronomical images}
In an effort to have the most realistic dataset, we choose to use real data as much as possible and take advantage of the private archive of wide-field images gathered for the COSMIC-DANCE survey \citep{2013A&A...554A.101B}.
The COSMIC-DANCE library offers several advantages.
First, it includes images from many past and present optical and near-infrared wide-field cameras.
Images cover a broad range of detector types and ground-based observing sites, ensuring that our dataset is representative of most modern astronomical wide-field instruments.
Table~\ref{tab:instruments} gives an overview of the properties of the cameras used to build the image database.
Second, most problematic exposures featuring tracking/guiding loss, defocusing or strong fringing were already identified by the COSMIC-DANCE pipeline, providing an invaluable sample of real problematic images.

In all cases except for Megacam, DECam, UKIRT and HSC exposures, the raw  data and associated calibration frames were downloaded and processed using standard procedures with an updated version of \emph{Alambic} \citep{2002SPIE.4847..123V}, a software suite developed and optimized for the processing of large multi-chip imagers.
In the case of Megacam, the exposures processed and calibrated with the \emph{Elixir} pipeline were retrieved from the CADC archive \citep{Elixir}.
In the case of DECam, the exposures processed with the community pipeline were retrieved from the NOAO public archive \citep{2014ASPC..485..379V}.
UKIRT exposures processed by the Cambridge Astronomical Survey Unit were retrieved from the WFCAM Science Archive.
Finally, the HSC raw images were processed using the official HSC pipeline \citep{HSC_Pipeline}.
In all cases, a bad pixel map is associated to every individual image.
In the case of DECam and HSC, a data quality mask is also associated to each individual image and provides integer-value codes for pixels which are not scientifically useful or suspect, including in particular bad pixels, saturated pixels, cosmic ray hits, satellite tracks, etc.
All the images in the following consist of individual exposures and not co-added exposures. 

\begin{table}
\small
  \caption{Instruments used in this study \label{tab:instruments}}

  \begin{tabular}{lcccc} \hline\hline
    Telescope   & Instrument        &   Type & Platescale   &  Ref. \\
     &                          &     & [pixel$^{-1}$] &      \\
    \hline
    CTIO Blanco  & DECam   & CCD & 0\farcs26 & (1) \\
    CTIO Blanco  & MOSAIC2   & CCD & 0\farcs26 & (2) \\
    KPNO Mayall  & MOSAIC1   & CCD & 0\farcs26 & (2) \\
    KPNO Mayall  & NEWFIRM   & IR & 0\farcs4 & (3) \\
    CFHT   & MegaCam  & CCD & 0\farcs18 & (4) \\
    CFHT   & CFH12K   & CCD & 0\farcs21 & (5) \\
    CFHT   & UH8K   & CCD & 0\farcs21 & (6) \\
    INT & WFC & CCD & 0\farcs33 & (7) \\
    UKIRT &  WFCAM & IR & 0\farcs4 & (8)\\
    LCO Swope & Direct CCD & CCD & 0\farcs43 & (9) \\
    VST & OmegaCam & CCD & 0\farcs21 & (10) \\
    Subaru & HSC & CCD & 0\farcs17 & (11) \\
    VISTA & VIRCAM & IR & 0\farcs34 & (12) \\
    \hline
  \end{tabular}

  \tablebib{(1)~\citet{DECAM}  ; (2)~\citet{2000SPIE.3965...80W} ; (3)~\citet{2003SPIE.4841..525A} ; (4)~\citet{2003SPIE.4841...72B} ;  (5)~\citet{2000SPIE.4008.1010C} ; (6)~\citet{1995AAS...187.7305M} ; (7)~\citet{1998IEEES..16...20I} ; (8)~\citet{2007A&A...467..777C} ; (9)~\citet{Swope_NewCCD} ;  (10)~\citet{2002Msngr.110...15K} ; (11) ~\citet{2018PASJ...70S...1M} ; (12) ~\citet{2006SPIE.6269E..0XD} }

\end{table}

\subsubsection{Non-contaminated images}
None of the exposures in our library are defect-free.
The first step to create the non-contaminated dataset to be used as ``reference'' images consists in identifying the cleanest possible subset of exposures.
CFHT-Megacam ($u,r,i,z$ bands), CTIO-DECam ($g,r,i,z,Y$ bands) and Subaru-HSC ($g,r,i,z,y$ bands) exposures are found to have the best cosmetics and are selected to create the non-contaminated dataset.
The defects inevitably present in these images are handled as follows.

First, dead pixels and columns are identified from flat-field images and inpainted using Gaussian interpolation \citep[e.g.,][]{Williams1998}. Then, the vast majority of cosmic rays are detected using the Astro-SCRAPPY Python implementation \citep{curtis_mccully_2018_1482019} of LA Cosmic \citep{2001PASP..113.1420V} and also inpainted using Gaussian interpolation. Finally, given the high performance of the DECam and HSC pipelines, the corresponding images are perfect candidates for our non-contaminated datasets.
These two pipelines not only efficiently detect but also interpolate problematic pixels (in particular saturated pixels, hot and bad pixels, cosmic ray hits).
Such interpolations being a feature of several modern pipelines (e.g., various NOAO pipelines, but also the LSST pipeline), we choose to treat these pixels as regular pixels so that the networks are able to work with images originating from such pipelines. 

Patches of size $400\times400$ pixels are randomly extracted from the cleaned images. 75\% of them are used to generate training data and the remaining 25\% for test data.

The final non-contaminated dataset includes 50,000 individual images, ensuring that we have a sufficiently diverse and large amount of training data for our experiment.

A non representative training set can severely impact the performance of a CNN and result in significant biases in the classification task.
To prevent this, we measure a number of basic properties describing prototypical aspects of ground-based astronomical images to verify that their distributions in the uncontaminated dataset are wide enough and reasonably well sampled.

The measured properties include, for example, the average full-width at half-maximum (FWHM) of point-sources is estimated in each image using \textsc{PSFEx} \citep{2013ascl.soft01001B}.
This allows us to ensure that the training set covers a broad range of ambient (seeing) conditions and point spread functions (PSFs) sampling. Also, the source density (number of sources in the image divided by the physical size of the image) is measured to make sure that our training set encompasses a broad range of source crowding, from sparse cosmological fields to dense, low-galactic latitude stellar fields.

Additionally, the background is modeled in all the images following the method used by \textsc{SExtractor} \citep{bertin1996sextractor}, i.e. using a combination of $\kappa.\sigma$-clipping and mode estimation. The background model provides important parameters such as the standard deviation of the background which is required in most of the data-processing operations that follow.

\begin{figure*}[ht]
  \begin{minipage}{0.16\linewidth}
    \includegraphics[scale=1.5]{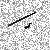}
  \end{minipage} 
  \begin{minipage}{0.16\linewidth}
    \includegraphics[scale=1.5]{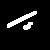}
  \end{minipage}
  \begin{minipage}{0.16\linewidth}
    \includegraphics[scale=1.5]{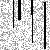}
  \end{minipage} 
  \begin{minipage}{0.16\linewidth}
    \includegraphics[scale=1.5]{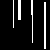}
  \end{minipage}
  \begin{minipage}{0.16\linewidth}
    \includegraphics[scale=1.5]{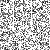}
  \end{minipage} 
  \begin{minipage}{0.16\linewidth}
    \includegraphics[scale=1.5]{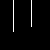}
  \end{minipage} \\
  
  \begin{minipage}{0.16\linewidth}
    \includegraphics[scale=1.5]{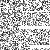}
  \end{minipage} 
  \begin{minipage}{0.16\linewidth}
    \includegraphics[scale=1.5]{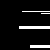}
  \end{minipage}
  \begin{minipage}{0.16\linewidth}
    \includegraphics[scale=1.5]{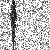}
  \end{minipage} 
  \begin{minipage}{0.16\linewidth}
    \includegraphics[scale=1.5]{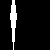}
  \end{minipage}
  \begin{minipage}{0.16\linewidth}
    \includegraphics[scale=1.5]{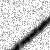}
  \end{minipage} 
  \begin{minipage}{0.16\linewidth}
    \includegraphics[scale=1.5]{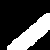}
  \end{minipage}

  \caption{Examples of contaminants and their ground truth. Top row: cosmic ray hits, hot columns, bad columns. Bottom row: bad lines, persistence, satellite trails.}
  \label{contex}
\end{figure*}

\subsubsection{Cosmic rays (CR)}

``Cosmic ray'' hits are produced by particles hitting the detector or by the photons resulting from the decay of radioactive atoms near the detector.
They appear as bright and sharp patterns with shapes ranging from dots affecting one or two pixels to long wandering tracks commonly referred to as ``worm'', depending on incidence angle and detector thickness.

We create a library of real CRs using dark frames with long exposure times from the CFH12K, HSC, MegaCam, MOSAIC, and OmegaCam cameras.
These cameras comprise both ``thick'', red-sensitive, deep depletion charged-couple devices (CCDs), more prone to long worms, and thinner, blue-sensitive devices, more prone to unresolved hits.
Dark frames are exposures taken with the shutter closed, so that the only contributors to the content of undamaged pixels are the offset, dark current, and CR hits (plus Poisson and readout noise).
A mask $\boldsymbol{M}$ of the pixels affected by CR hits in a given dark frame $\boldsymbol{D}$ can therefore easily be generated by applying a simple detection threshold.
We conservatively set this threshold to 3 $\sigma_{D}$ above the median value $m_{D}$ of $D$:
\begin{align}
  \forall p,\ M_p=\left\{\begin{matrix} 1 & \mathrm{if}\ D_p > m_{D} + 3\sigma_{D} \\ 0 & \text{otherwise}. \end{matrix}\right.
\end{align}

Among all the dark images used, a bit more than 900 million cosmic ray pixels are detected after thresholding.
Considering that the average footprint area of a cosmic ray hit is 15 pixels, this represents a richly diversified population of about 60 million cosmic ray ``objects''.

Next we dilate $\boldsymbol{M}$ with a $3\times 3$ pixel kernel to create the final $\boldsymbol{M}^{\,\rm (D)} $ mask.
This mask is used both as ground truth for the classifier, and also to generate the final ``contaminated'' image $\boldsymbol{C}$ by adding CR pixels with rescaled values to the uncontaminated image $\boldsymbol{U}$:
\begin{equation}
 \boldsymbol{C} = \boldsymbol{U} + k_{\boldsymbol C} \frac{\sigma_{\boldsymbol U}}{\sigma_{\boldsymbol D}} \boldsymbol{D} \odot \boldsymbol{M}^{(\boldsymbol{D})},
\end{equation}
where $\sigma_{\boldsymbol U}$ is the estimated standard deviation of the uncontaminated image background, $\odot$ denotes the element-wise product and $k_{\boldsymbol C}$ is a scaling factor empirically set to 1/8.
$\boldsymbol{D}$ has been background-subtracted before this operation, using a \textsc{SExtractor}-like background estimation.

A typical CR hit added to an image and its ground truth mask are shown in Fig.~\ref{contex}.

\subsubsection{Hot columns and lines, dead columns, lines, and clustered pixels, hot pixels, and dead pixels (HCL, DCL, HP, DP)}

These contaminants mainly come from electronic defects and the way the detectors are read.
They correspond to pixels having a response very different from that of neighbors, either much lower (bad pixels, traps) or much noisier (hot pixels).
These blemishes can be found as single pixels, in small clusters, or affecting a large fraction of a column or row.
We treat single pixels and clumps, columns, and lines separately, although they may often share a common origin.

All these hot or dead pixels added to the uncontaminated images are simulated. 
The number of these pixels is set as follow.

For columns and lines, a random number of columns and lines is chosen with a uniform distribution over $[1,4]$. Each column or line has a uniform length picked between 30 and the whole image height or width. It has a uniform thickness in $[1,3]$.
For punctual pixels, a random fraction of pixels is chosen with a uniform distribution between 0.0002 and 0.0005. Pixels are uniformly distributed over the image.
Clustered pixels are given a rectangular or a random convex polygonal shape. The random convex shapes are constrained to have 5 or 6 edges and to fit in $20\times20$ bounding boxes.

The values of these pixels are computed as follows.
For hot values, a uniformly distributed random base value $v$ is chosen in the interval $[15\sigma_{\boldsymbol U}, 100\sigma_{\boldsymbol U}]$.
Then hot values are generated according to the normal law $\mathcal{N}(v, (0.02 v)^2)$ so that hot values are randomly distributed over $[0.9v, 1.1v]$. 
For dead values, one of the following three equiprobable recipes is chosen at random to generate bad pixel values. Either all values are exactly 0. Either values are generated according to the normal law $\mathcal{N}\left(0, (0.02\sigma_{\boldsymbol U})^2\right)$ so that these are close to 0 values but not exactly 0. Either a random base value $v$ is chosen with a uniform distribution in the interval $[0.1 m_{\boldsymbol U}, 0.7 m_{\boldsymbol U}]$, where $m_{\boldsymbol U}$ is the median of the uncontaminated image sky background. In this case, dead pixel values are generated using the normal law $\mathcal{N}\left(v, (0.02v)^2\right)$, so that values fall in the interval $[0.9v, 1.1v]$.

Example of such column and line defaults are shown in Fig.~\ref{contex}.

\begin{figure*}[ht]
  \includegraphics[width=0.19\textwidth]{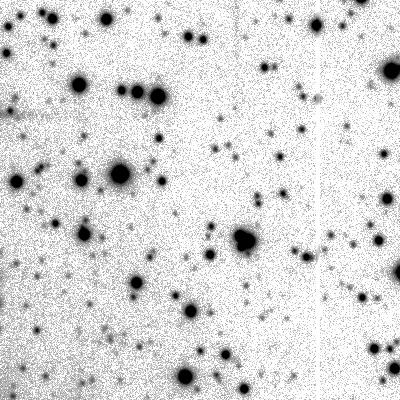}
  \includegraphics[width=0.19\textwidth]{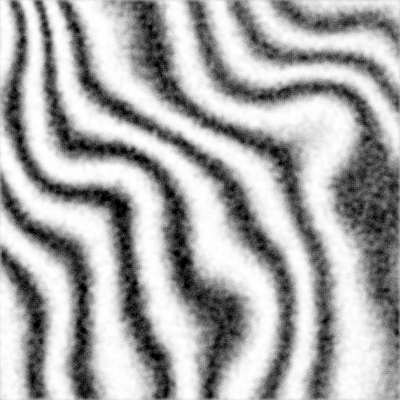}
  \includegraphics[width=0.19\textwidth]{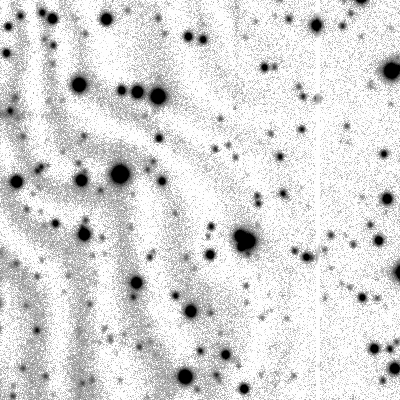}
  \includegraphics[width=0.19\textwidth]{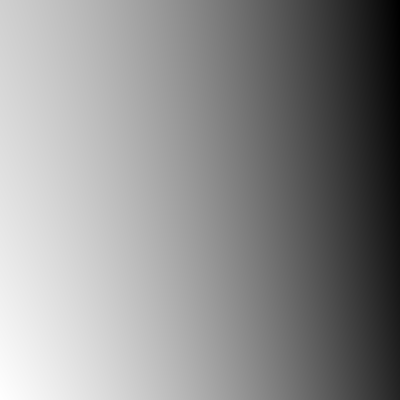}
  \includegraphics[width=0.19\textwidth]{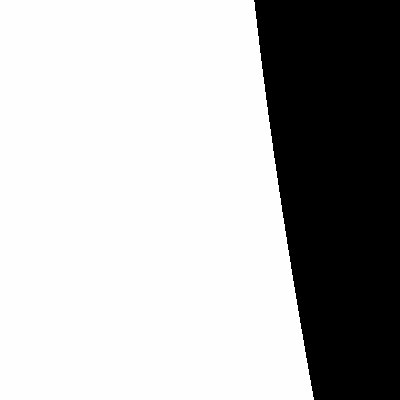}
  
  \includegraphics[width=0.19\textwidth]{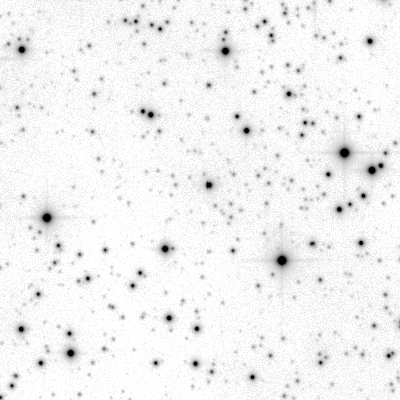}
  \includegraphics[width=0.19\textwidth]{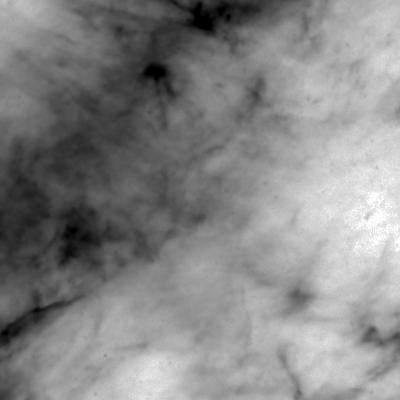}
  \includegraphics[width=0.19\textwidth]{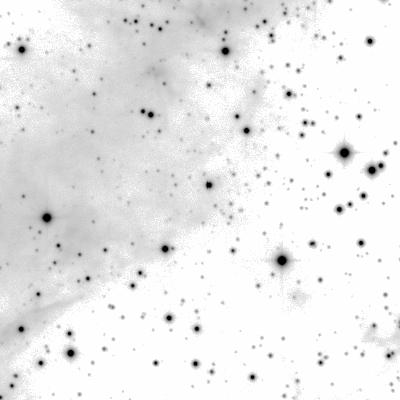}
  \includegraphics[width=0.19\textwidth]{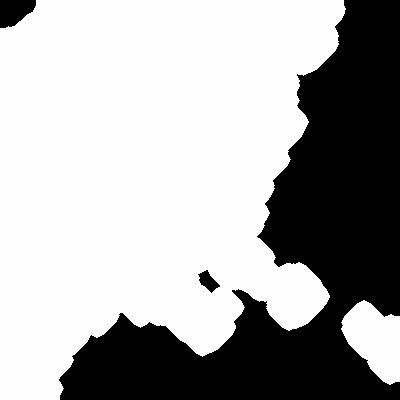}
  \caption{Examples of added fringes and nebulosities. Top: fringes; uncontaminated input exposure, smoothed fringe pattern, contaminated image, ground truth mask, polynomial envelope. Bottom: nebulosities; uncontaminated input exposure, Herschel 250~$\mu$m molecular cloud image, contaminated image, ground truth mask.} 
  \label{frneb}
\end{figure*}

\subsubsection{Persistence (P)}

Persistence occurs when overly bright pixels in a previous exposure leave a remnant image in the following exposures.

To simulate this effect in an uncontaminated image, we applied the so-called ``Fermi model'' described in \cite{2015wfc..rept...15L}.
Persistence, in units of $e^-.s^{-1}$), is modeled as a function of the initial pixel level $x_p$ and time $t$:

\begin{equation}
  f(x_p, t) = A_p\left(\frac{1}{\exp{(-\frac{x_p-x_0}{\delta x}})+1}\right) \left(\frac{x_p}{x_0}\right)^{\alpha} \left(\frac{t}{1000}\right)^{-\gamma}.
  \label{perseq}
\end{equation}

The goal of \cite{2015wfc..rept...15L} was to fit the model parameters $x_0, \delta x, \alpha, \gamma$ using observations to later predict persistence for their detector.
In our simulations, parameter values are randomized to represent various types and amounts of persistence (see Table~\ref{tab1}).
To compute the pixel value of the persistence effect, we derive the number of electrons emitted by the persistence effect during the exposure.
In the following, we note $T$ the duration of the exposure in which the persistence effect occurs, and $\Delta t$ the delay between that exposure and the previous one.
We obtain the number of ADUs collected at pixel $p$ during the interval $[\Delta t, \Delta t + T]$ by integrating Eq.\ref{perseq} and dividing by the gain $G$:

\begin{eqnarray}
  P_p & = & \frac{1}{G}\int_{\Delta t}^{\Delta t + T}f(x_p,t)\,dt\\
  & = & \frac{A_p}{G}\left(\frac{1000^\gamma}{\exp{(-\frac{x_p-x_0}{\delta x}})+1}\right) \left(\frac{x_p}{x_0}\right)^{\alpha}  \left(\frac{(\Delta t + T)^{1-\gamma} - \Delta t^{1-\gamma}}{1-\gamma}\right).
  \label{eq:persist}
\end{eqnarray}

These pixel values are then added to the uncontaminated image:

\begin{equation}
\boldsymbol{C} = \boldsymbol{U} + k_{\boldsymbol P}\,\sigma_{\boldsymbol U} \frac{\boldsymbol{P} - P_{min}}{(P_{max} - P_{min})},
\end{equation}
where $\boldsymbol{P}$ are the persistence values computed in Eq.~\ref{eq:persist}, $P_{min}$ and $P_{max}$ are the minimum and maximum of these values, and $k_{\boldsymbol P}$ is a scaling factor empirically set to 5.

\begin{table}[ht]
  \begin{tabular}{|c|c|}
    \hline
    $A_p$ & $1$ \\
    \hline
    $x_p$ ($e^-$) & \makecell{$Poisson(x_m)$ with \\ $x_m \sim \mathcal{N}(15.10^5, (0.02\times 15.10^5)^2)$} \\
    \hline
    $x_0$ ($e^-$)& $\mathcal{N}(9.10^4, (0.02\times 9.10^4)^2)$ \\
    \hline
    $\delta_x$ ($e^-$) & $\mathcal{N}(18.10^3, (0.02\times 18.10^3)^2)$ \\
    \hline
    $\alpha$ & 0.178 \\
    \hline
    $\gamma$ & 1.078 \\
    \hline
    $G$ ($e^-.s^{-1}$) & $\mathcal{N}(10, 1)$ \\
    \hline
  \end{tabular}
  \caption{Parameters used for the generation of persistence}
  \label{tab1}
\end{table}

Images of saturated stars are simulated using {\sc SkyMaker} \citep{2009MmSAI..80..422B} and binarized to generate masks of saturated pixels.
The masks define the footprints of persistence artifacts, within which the $x_p$'s are computed (Table \ref{tab1}).
An example is shown in Fig.~\ref{contex}.

\subsubsection{Trails (TRL)}
Satellites or meteors, and even planes crossing the field of view generate long trails across the frame that are quasi-rectilinear.
We simulate these motion-blurred artifacts by generating close star images with identical magnitudes along a linear path using once again {\sc SkyMaker}. We also generate a second population of trails with magnitude changes to account for satellite ``flares''.
A random, Gaussian-distributed component with a $\approx 1$~pixel standard deviation is added to every stellar coordinate to simulate jittering from atmospheric turbulence, so that the stars are not aligned along a perfect straight line.
For meteors, defocusing must be taken into account \citep{2018MNRAS.474.4837B}.
The amount of defocusing $\theta$, expressed as the apparent width of the pupil pattern in arc-seconds, is:

\begin{equation}
  \theta = \frac{180}{\pi}\times 3600 \times \frac{D}{d},
\end{equation}
where $D$ is the diameter of the primary mirror, and $d$ the meteor distance, both in meters.
$D$ and $d$ are randomly drawn from flat distributions in the intervals $[2, 8]$ and $[80,000, 120,000]$, respectively.

The ground truth mask is obtained by binarizing the satellite image at a small and arbitrary threshold above the simulated background.
This mask is then dilated using a $7\times7$ pixel structuring element.

To avoid any visible truncation, we add the whole simulated satellite image multiplied by a dilated version $\boldsymbol{M}^{(\boldsymbol{S})}$ of the ground truth mask to the uncontaminated image:
\begin{equation}
    \boldsymbol{C} =  \boldsymbol{U} + k_{\boldsymbol T}\frac{\sigma_{\boldsymbol U}}{\sigma_{\boldsymbol T}} \boldsymbol{T} \odot \boldsymbol{M}^{(\boldsymbol{T})},
\end{equation}
where $\sigma_{\boldsymbol S}$ is the standard deviation of the satellite image background, $\sigma_{\boldsymbol U}$ the standard deviation of the uncontaminated image background, and $k_{\boldsymbol T}$ is a scaling factor empirically set to 6.
An example of a satellite trail is shown in Fig.~\ref{contex}.

\subsubsection{Fringes (FR)}

Fringes are thin-film interference patterns occurring in the detectors.
The irregular shape of fringes is caused by thickness variations within the thin layers.
To add fringing to images, we use real fringe maps produced at the pre-processing level by {\sc Alambic} for all the optical CCD cameras of Table~\ref{tab:instruments}. These reconstructed fringe maps are often affected by white noise, which we mitigate by smoothed using a top-hat kernel with diameter 7 pixels.
The fringe pattern $\boldsymbol{F}$ can affect large areas in an image but not necessarily all the image.
To reproduce this effect, a random 3rd-degree 2D polynomial envelope $\boldsymbol{E}$ that covers the whole image is generated. 
The final fringe envelope ${\boldsymbol E}^{\,\rm (F)}$ is computed by normalizing $\boldsymbol{E}$ over the interval $[-5,5]$ and flattening the result using the sigmoid function:
\begin{equation}
E^{\rm (F)}_p = \left(1+\exp \left(-5\,\frac{2E_p - E_{\rm max}-E_{min}}{E_{max}-E_{min}}\right)\right)^{-1},
\end{equation}
where $E_{min}$ and $E_{max}$ are the minimum and maximum values of $E_p$, respectively.

The fringe pattern, modulated by its envelope, is then added to the uncontaminated image:
\begin{equation}
    \boldsymbol{C} =  \boldsymbol{U} + k_{\boldsymbol F}\frac{\sigma_{\boldsymbol U}}{\sigma_{\boldsymbol F}} \boldsymbol{F} \odot \boldsymbol{E}^{\,\rm (F)},
\end{equation}
where $\sigma_{\boldsymbol F}$ is the standard deviation of the fringe pattern and $k_{\boldsymbol F}$ is an empirical scaling factor set to 0.6.
The ground truth mask is computed by thresholding the 2D polynomial envelope to $-0.20$.

An example of a simulated contamination by a fringe pattern can be found in Fig.~\ref{frneb}.

\subsubsection{Nebulosity (NEB)}

Extended emission originating from dust clouds illuminated by star light or photo-dissociation regions can be present in astronomical images.
These ``nebulosities'' are not artifacts but they make the detection and measurement of overlapping stars or galaxies more difficult; they may also trigger the fringe detector.
Hence, it is useful to have them identified and properly flagged.
Because thermal distribution of dust closely matches that of reflection nebulae at shorter wavelength \citep[e.g.,][]{2013ApJ...767...80I}, we use far-infrared images of molecular clouds around star-forming regions as a source of nebulous contaminants.
We choose pipeline-processed 250~$\mu$m images obtained with the SPIRE instrument \citep{2010A&A...518L...3G} on-board the Herschel Space Observatory \citep{2010A&A...518L...1P}, which we retrieve from the Herschel Science Archive. The 250~$\mu$m channel offers the best compromise between signal-to-noise ratio and spatial resolution.
Moreover, at wavelengths of 250~$\mu$m and above, low galactic latitude fields contain mostly extended emission from the cold gas and almost no point sources (apart from a few proto-stars and proto-stellar cores). Therefore, they are perfectly suited to being added to our optical and near-infrared wide-field exposures.
We do not resize or reconvolve the SPIRE images, taking advantage of the scale-invariance of dust emission observed down to the arcsecond level in molecular clouds \citep{2016A&A...593A...4M}. 

We add the nebulous contaminant data to our uncontaminated images in the same way we do for fringes, except that there is no 2D polynomial envelope.
The whole nebulosity image is background-subtracted (using a \textsc{SExtractor}-like background estimation) to form the final nebulosity pattern $\boldsymbol{N}$ which is then added to the uncontaminated image:
\begin{equation}
    \boldsymbol{C} =  \boldsymbol{U} + k_{\boldsymbol N}\frac{\sigma_{\boldsymbol U}}{\sigma_{N}} \boldsymbol{N},
\end{equation}
where $k_{\boldsymbol N}$ is an empirical scaling factor set to 1.3.
The ground truth mask is computed by thresholding $\boldsymbol{N}$ at one sigma above 0.
This mask is then eroded with a $6$ disk diameter structuring element to remove spurious individual pixels, and dilated with a $22$ disk diameter structuring element.
An example of added nebulosity is shown in Fig.~\ref{frneb}.
The light from line-emission nebulae may not necessarily exhibit the same statistical properties as the reflection nebulae targeted for training.
However line-emission nebulae are generally brighter and in practice the classifier has no problem detecting them.

\subsubsection{Saturation and bleeding (SAT)}

Each detector pixel can accumulate only a limited number of electrons.
Once the full well limit is reached, the pixel becomes saturated.
In CCDs, charges may even \textit{overflow}, leaving saturation trails (a.k.a bleeding trails) along the transfer direction.
Such pixels are easily be identified in clean images knowing for each instrument the saturation level.

\subsubsection{Diffraction spikes}

Diffraction spikes are patterns appearing around bright stars and caused by light diffracting around the spider supporting the secondary mirror.
Given the typical cross-shape of spiders, the pattern is usually relatively easy to identify.
In some cases, the pattern can deviate significantly from a simple cross because it is affected by various effects, such as distortions, telescope attitude, the truss structure of spider arms, rough edges, or cables around the secondary mirror support, reflections on other telescope structures,...
A specific strategy was put in place to build a spikes library to be used to train the CNN.

On the one hand, MegaCam and DECam are mounted on equatorial telescopes and the orientation of spikes is usually (under standard northeast orientation) a '+' for Megacam and an 'x' for DECam\footnote{DECam images sometimes also exhibit a horizontal spike of unknown origin \citep{2016A&C....16...99M}}.
On the other hand, HSC is mounted on the alt-az Subaru telescope, and spikes do not display any preferred orientation, making their automated identification more complicated.
For this reason, we define a two-step strategy, in which, first, samples of '+'- and 'x'-shaped spikes are extracted from DECam and Megacam images, and randomly rotated to generate a library of diffraction spikes with various orientations.
The library is then used to train a new CNN that for identifying spikes in HSC images. 

\begin{figure*}[ht]
  \begin{minipage}{0.19\linewidth}
    \includegraphics[scale=0.3]{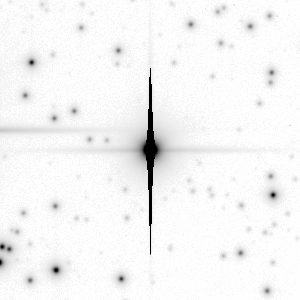}
  \end{minipage}
  \begin{minipage}{0.19\linewidth}
      \includegraphics[scale=0.3]{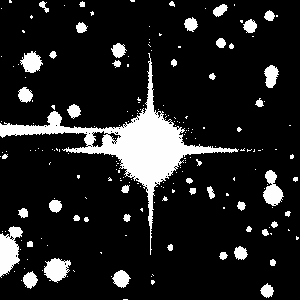}
  \end{minipage}
  \begin{minipage}{0.19\linewidth}
    \includegraphics[scale=0.3]{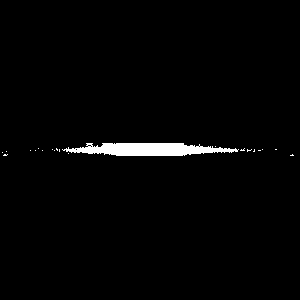} \\
    \includegraphics[scale=0.3]{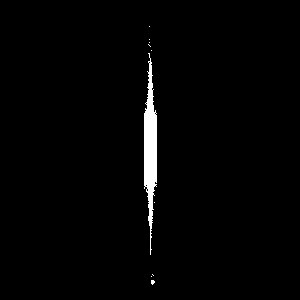}
  \end{minipage}
  \begin{minipage}{0.19\linewidth}
    \includegraphics[scale=0.3]{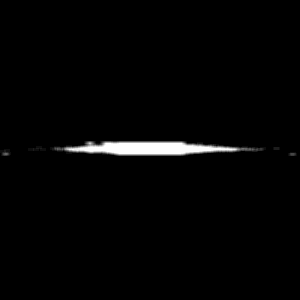} \\
    \includegraphics[scale=0.3]{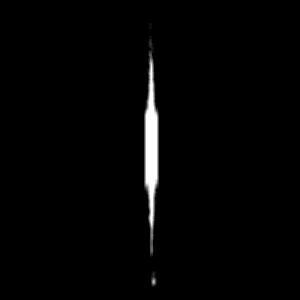}
  \end{minipage}
  \begin{minipage}{0.19\linewidth}
    \includegraphics[scale=0.3]{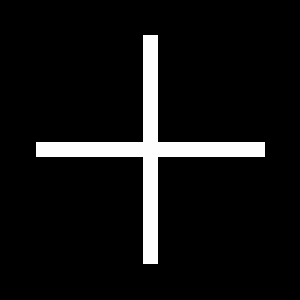}
  \end{minipage}
  
  \caption{Empirical spike flagging process. From left to right: the source image centered on a bright star candidate, the same image thresholded, the two pointwise products, the matched filtered pointwise products, the final mask drawed from the empirical size computed with the two previous masks.}
  \label{sp}
\end{figure*}

\paragraph{Megacam and DECam analysis}
We first identify the brightest stars using \textsc{SExtractor} and extract $300\times 300$ pixel image cutouts around them.
The cutouts are thresholded at three sigma above the background and binarized.
Element-wise products are computed between these binary images and large '+'-shaped (Megacam) or 'x'-shaped (DECam) synthetic masks to isolate the central stars.
Each point-wise product is then matched-filtered with a thinner version of the same pattern and binarized using an arbitrary threshold set to 15 ADUs.  
The empirical size of the spike components is estimated in these masks by measuring the maximum extent of the resulting footprint along any of the two relevant spike directions (horizontal and vertical or diagonals).
Finally, the maximum size of the two directions is kept and empirically rescaled to obtain the final spike length and width.
If the resulting size is too small, we consider that there is no spike in order to avoid false positives (e.g., a star bright enough to be detected by \textsc{SExtractor} but without obvious spikes).
Fig.~\ref{sp} gives an overview of the whole process.

\begin{figure}[ht]
  \includegraphics[scale=0.5]{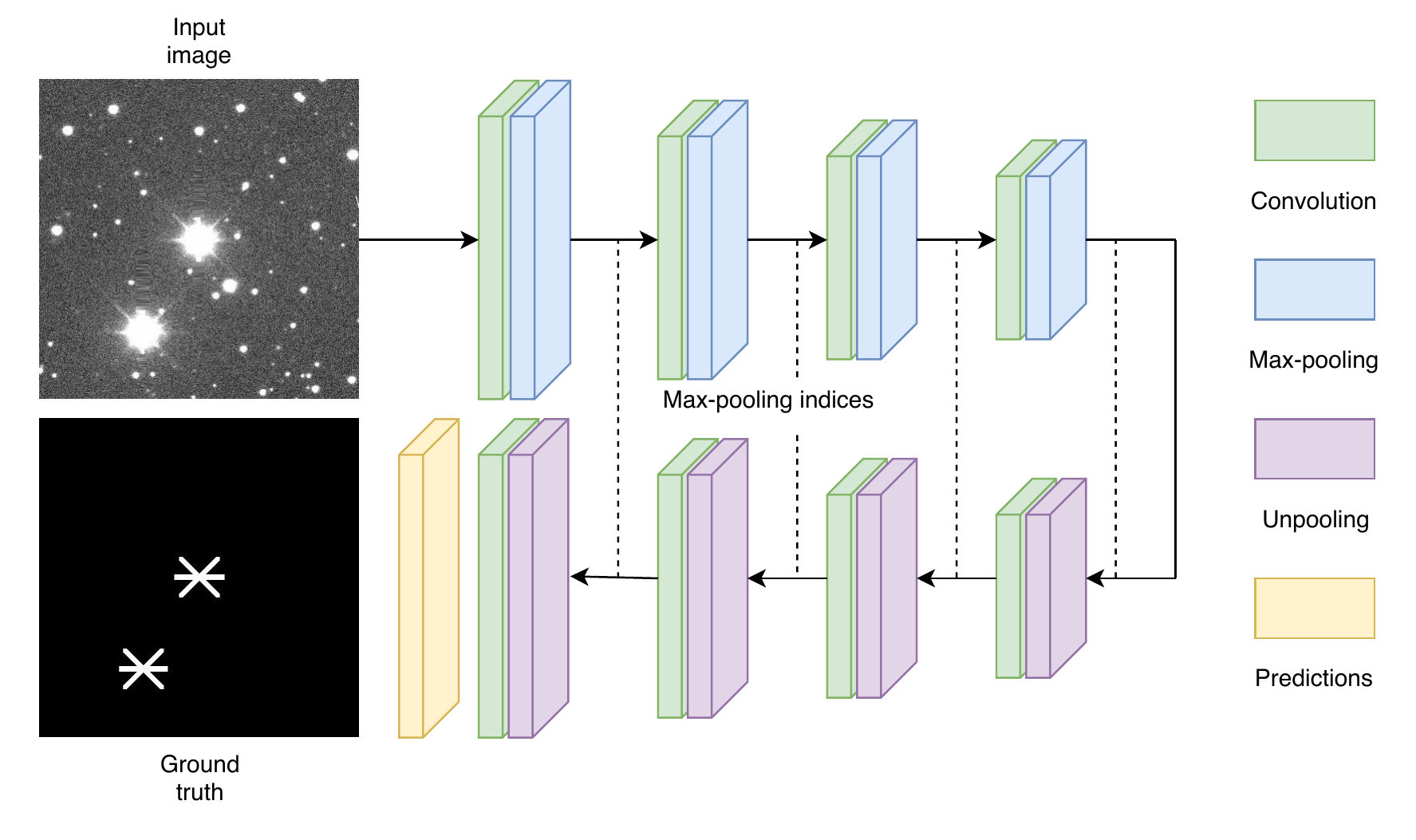}
  \caption{Neural network used specifically for spike detection.}
  \label{spike_nn}
\end{figure}

\begin{figure}[ht]
  \includegraphics[scale=0.31]{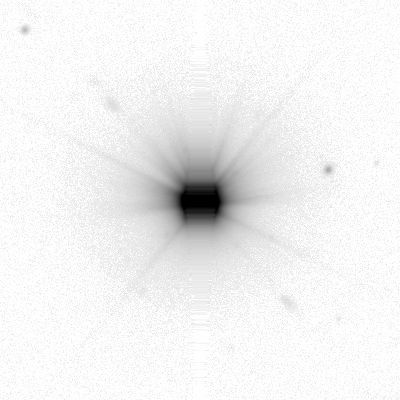}
  \includegraphics[scale=0.31]{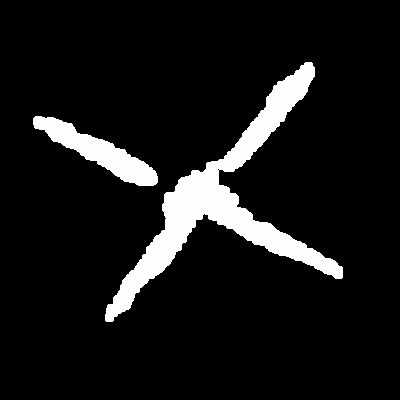}
  \caption{Example of a spike mask obtained by inference of the separate neural network.}
  \label{HSC_inf}
\end{figure}

\paragraph{HSC analysis}
We train a new neural network to identify spikes in all directions.
For that purpose, we build a new training set using the spikes identified in MegaCam and DECam images as described above and apply a random rotation between 0\degr\, and 360\degr to ensure rotational invariance.
The neural network has a simple SegNet-like convolutional-deconvolutional architecture \citep{badrinarayanan2015segnet}, but it is not based on VGG hyper-parameters \citep{simonyan2014very}.
It uses 21$\times$21, 11$\times$11, 7$\times$7 and 5$\times$5 convolutional kernels in 8, 16, 32 and 32 feature maps, respectively.
The model architecture is shown in Fig.~\ref{spike_nn}.
Activation functions are all ELU except on the last layer where it is softmax.
It is trained to minimize the softmax cross entropy loss with the Adam optimizer \citep{kingma2014adam}.
Each pixel cost is weighted to balance the disproportion between spike and background pixels. If $p_{s}$ is the spike pixel proportion in the training set, then spike pixels are weighted with $1-p_{s}$, while background pixel are weighted with $p_{s}$ (this is the two-class equivalent of the basic weighting scheme described in section \ref{localcontsec}).
Once trained we run inferences on all the brightest stars detected with  \textsc{SExtractor} in the HSC images.
Output probabilities are binarized based on the MCC (see \ref{smcc}) and the resulting mask is empirically eroded and dilated to obtain a clean mask.
An example is given in Fig.~\ref{HSC_inf}

\subsubsection{Overscan (OV)}

Overscan regions are common in CCD exposures, showing up as strips of pixels with very low values at the borders of the frame.
To avoid triggering false predictions on real data, overscans must be included in our training set.
Doing so, and although these are not truly contaminants, we find it useful to include an ``overscan'' class in the list of identified features.
Overscan regions are simulated by including random strips on the sides of images.
Pixel values in the strips are generated in the same way as bad pixel values. 

\subsubsection{Bright background (BBG) and background (BG)}

The objects of interest in this study are the contaminants. Hence, following standard computer vision terminology, all the other types of pixels, including both astronomical objects and empty sky areas, belong to the ``background''. 

We find that defining a distinct class for each of these types of background pixels helps with the training procedure.
We thus define the ``bright background'' (BBG) pixels as pixels belonging to astronomical objects\footnote{Including astrophysical sources in the ``background'' class can seem  somewhat counter-intuitive in a purely astronomical context, but for consistency we choose to follow the  computer vision terminology and meaning.} (except nebulosity) present in the uncontaminated images, and background pixels (BG) as pixels covering an empty sky area. 

Ground truth masks for bright background pixels are obtained by binarizing the image before adding the contaminants to $10\sigma_{\boldsymbol U}$.
The remaining pixels are sky background pixels, which are not affected by any labeled feature.

\subsection{Global contaminants}

We now describe the data used to identify global contaminants.

\begin{figure}[ht]
  \begin{minipage}{0.24\linewidth}
    \includegraphics[scale=0.31]{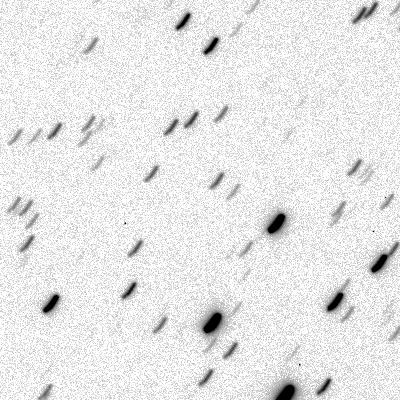}
  \end{minipage} \hspace{2cm}
  \begin{minipage}{0.24\linewidth}
    \includegraphics[scale=0.31]{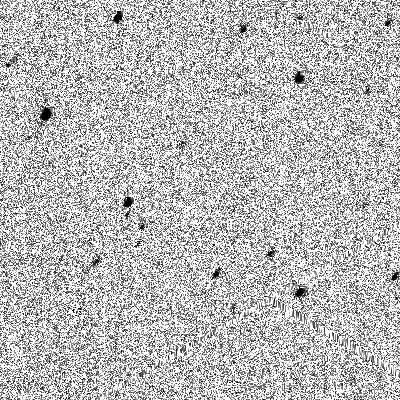}
  \end{minipage} \\
  \begin{minipage}{0.24\linewidth}
    \includegraphics[scale=0.31]{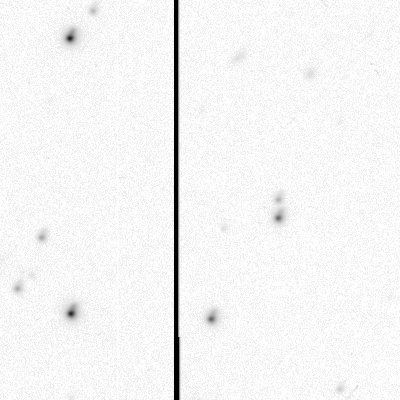}
  \end{minipage} \hspace{2cm}
  \begin{minipage}{0.24\linewidth}
    \includegraphics[scale=0.31]{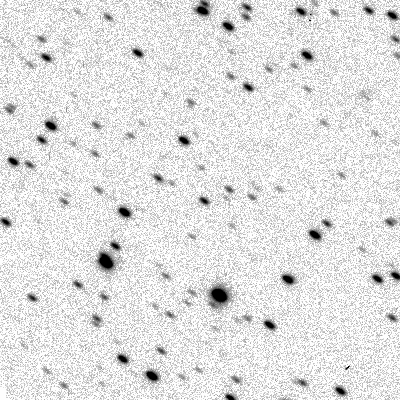}
  \end{minipage}
  \caption{Examples of images affected by tracking errors.}
  \label{trackex}
\end{figure}

Tracking errors happen when the telescope moves during an exposure due to, for instance, telescope guiding or tracking failures, wind gusts, or earthquakes.
As illustrated in Fig.~\ref{trackex}, this causes all the sources to be blurred along a path on the celestial sphere generated by the motion of the telescope.
Because tracking errors affect the entire focal plane, the analysis is performed globally on the whole image.
The library of real images affected by TR events is a compilation of exposures identified in the COSMIC-DANCE survey for the cameras of Table~\ref{tab:instruments}, and images that were gathered over the years at the UKIRT telescope, kindly provided to us by Mike~Read .

\subsection{Generating training samples}

Both types of contaminants -- global and local contaminants -- must be handled separately: they require different neural network architectures, and different training data sets as well.

\begin{figure}[ht]
  \includegraphics[scale=0.56]{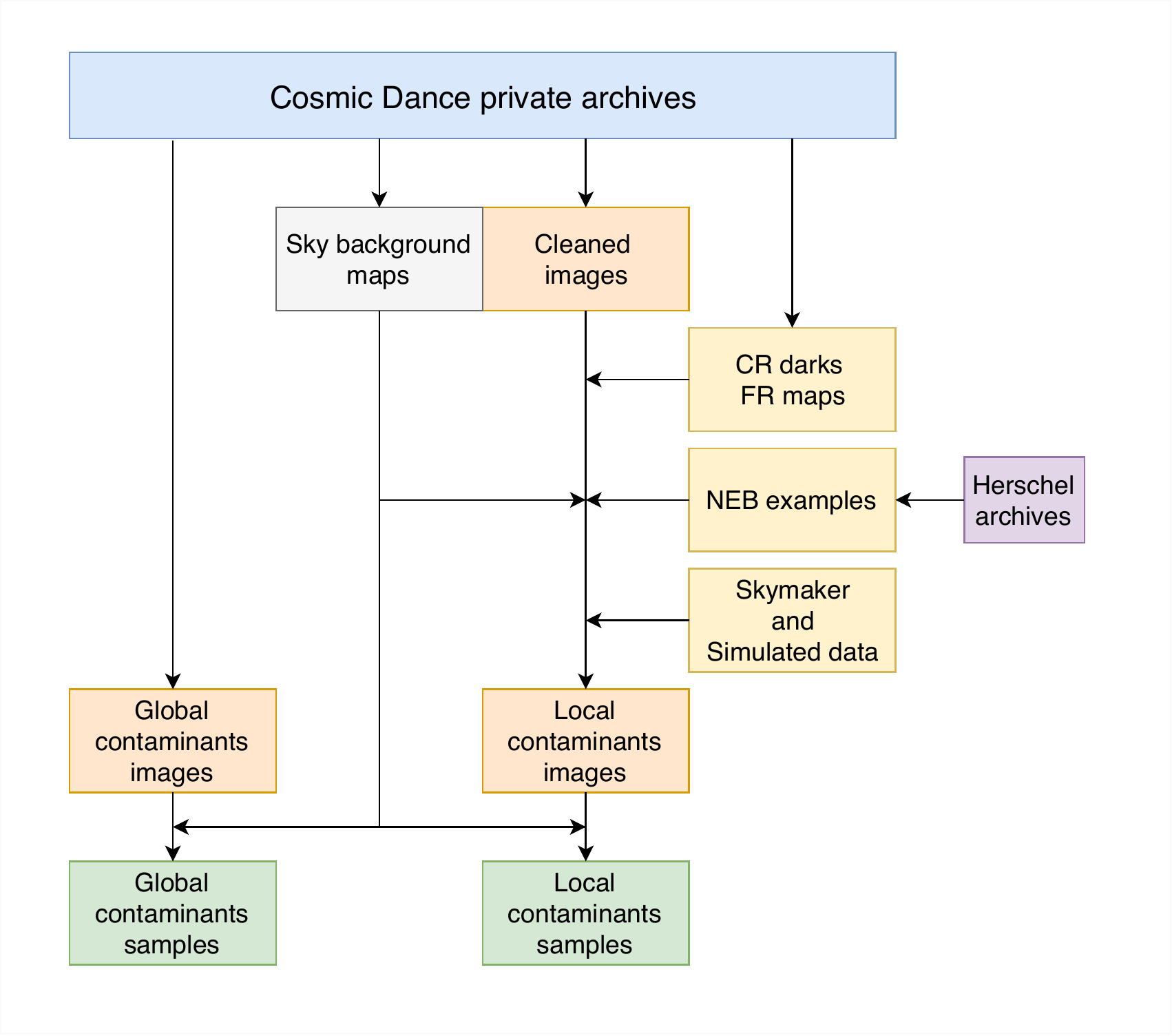}
  \caption{Schematic view of the sample production pipeline.
    All COSMIC-DANCe archive images have their background map computed.
    Clean images are built from the COSMIC-DANCe archives. Contaminants from diverse sources (COSMIC-DANCe archives, Herschel archives or simulations) are added to clean images; this step uses the background maps.
    The resulting local contaminant images are dynamically compressed (see section \ref{dcsec}) and ready to be fetched into the neural network.
    Global contaminant samples are directly obtained from the COSMIC DANCe archives and dynamically compressed.}
  \label{pip}
\end{figure}

Figure ~\ref{pip} gives a synthetic view of the sample production pipeline and the various data sources.
  
The breakdown per imaging instrument of the COSMIC DANCe dataset is listed Table~\ref{datause}.

\begin{table}[ht]
    \centering
    \begin{tabular}{lccccc}
        Instrument & Clean & CR & No TR & TR \\
        \hline
        DECam & \checkmark &  & \checkmark &  \\
        MOSAIC2 &  & \checkmark &  &  \\
        MOSAIC1 &  & \checkmark &  &  \\
        NEWFIRM &  &  & \checkmark & \checkmark \\
        Megacam & \checkmark & \checkmark & \checkmark & \checkmark \\
        CFH12K &  & \checkmark & \checkmark & \checkmark \\
        CFH8K &  &  &  & \checkmark \\
        WFC &  &  & \checkmark & \checkmark \\
        WFCAM &  &  & \checkmark & \checkmark \\
        Direct CCD (LCO Swope) &  &  & \checkmark & \checkmark \\
        VST &  & \checkmark & \checkmark & \checkmark \\
        HSC & \checkmark & \checkmark & \checkmark &  \\
        VIRCAM &  &  & \checkmark & \checkmark \\
    \end{tabular}
    
    \caption{COSMIC-DANCE archive usage per imaging instrument: \textit{Clean} is for uncontaminated images, \textit{CR} for dark images used for cosmic ray identification, \textit{No TR} is for images \textit{not} affected by tracking errors, and \textit{TR} for images affected by tracking errors.}
    \label{datause}
\end{table}

The following subsections treat about some special features of the sample generation.

\begin{figure}[ht]
  \begin{minipage}{0.48\linewidth}
    \includegraphics[scale=0.31]{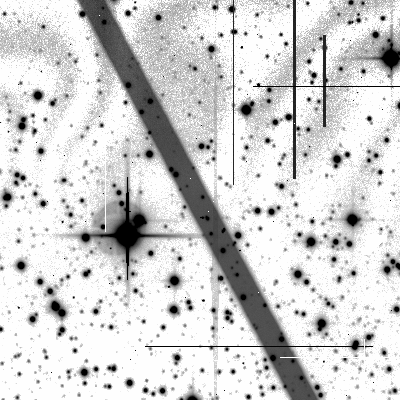}
  \end{minipage} 
  \begin{minipage}{0.48\linewidth}
    \includegraphics[scale=0.31]{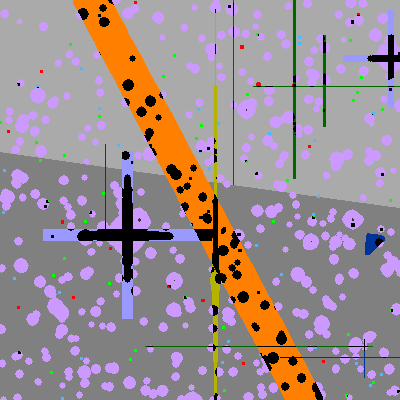}
  \end{minipage} \\
  \begin{minipage}{0.48\linewidth}
    \includegraphics[scale=0.31]{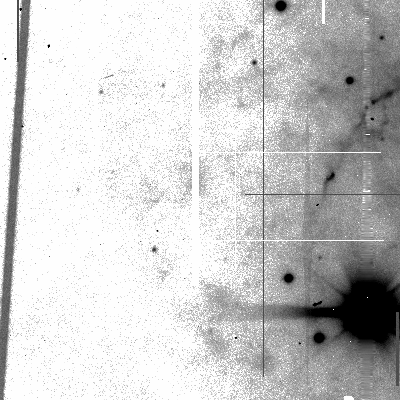}
  \end{minipage} 
  \begin{minipage}{0.48\linewidth}
    \includegraphics[scale=0.31]{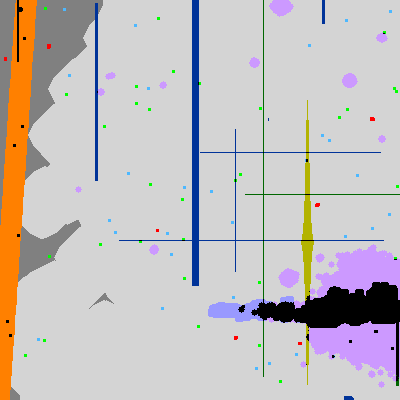}
  \end{minipage}
  \caption{Examples of input (left) and their ground truth (right). Each class is assigned a color so that the ground truth can be represented as a single image (red: CR, dark green: HCL, dark blue: BCL, green: HP, blue: BP, yellow: P, orange: TRL, gray: FR, light gray: NEB, purple: SAT, light purple: SP, brown: OV, pink: BBG, dark gray: BG).
  Pixels that belong to several classes are represented in black.
  In the interest of visualization, hot and dead pixel masks have been morphologically dilated so that they appear as $3\times 3$ pixel areas in this representation.}
  \label{sampex}
\end{figure}

\subsubsection{Local contaminants}

The order in which local contaminants are added is important. Bad columns, lines, and pixels are added last because they are static defaults defining the final value of a pixel, no matter how many photons hit them.

In our neural network architecture contaminant classes do not need to be mutually exclusive. Each pixel can be assigned several classes as several defaults can affect a given pixel (e.g., fringes and cosmic ray hit).
On the other hand, the faint background class that defines pixels not affected by any default excludes all other classes.
A list of all the contaminants included in this study are presented in Tab.~\ref{contlist}.

\begin{table}[ht]
    \centering
    \begin{tabular}{lc}
        Contaminant & Abbreviation \\ 
        \hline
        Cosmic rays & CR \\
        Hot columns/lines &  HCL \\
        Dead columns/lines/clusters & DCL \\
        Hot pixels & HP \\
        Dead pixels & DP \\
        Persistence & P \\
        Trails & TRL \\
        Fringes &  FR \\
        Nebulosities & NEB \\
        Saturated pixels & SAT \\
        Diffraction spikes & SP \\
        Overscanned pixels & OV \\
        Bright background & BBG \\
        Background & BG 
    \end{tabular}
    
    \caption{List of all the contaminants and their abbreviated names.}
    \label{contlist}
\end{table}

Fig.~\ref{sampex} shows examples of local contaminant sample input images, each with its color-coded ground truth.

\subsubsection{Global contaminants}

The global contaminant dataset contains images that have been hand labeled as affected by tracking errors or not. The images, taken from the COSMIC DANCe archives, are not cleaned, hence they are potentially affected by preexisting local contaminants. This is because the global contaminant detector is intended to be operated before the local one.

\subsubsection{Dynamic compression}
\label{dcsec}

All images are dynamically compressed before being fed to the neural networks using the following procedure:
\begin{equation*}
  \tilde{\boldsymbol{C}} = \arsinh{\left( \frac{\boldsymbol{C} - \boldsymbol{B} + \mathcal{N}(0, \sigma_{\boldsymbol U}^2)}{\sigma_{\boldsymbol U}}\right)}.
\end{equation*}

The aim of dynamic compression is to reduce the dynamic range of pixel values, which is found to help neural network convergence.
The image is first background subtracted. Then, a small random offset is added to increase robustness regarding background subtraction residuals.
The resulting image is normalized by the standard deviation of the background noise and finally compressed through the $\arsinh$ function, which has the property to behave linearly around zero and logarithmically for large (positive or negative) values. 

\subsubsection{Data augmentation}

We deploy data augmentation techniques to use our data to the maximum of its information potential. The two following data augmentation procedures are applied to the set of local contaminant training samples. First, random rotations, using as angles multiples of $90^\circ$, are applied to cosmic ray, fringe patterns, and nebulosity patterns. Secondly, some images are rebinned. When picking up a clean image, we check if the image can be $2\times 2$ rebinned with the constraint that the FWHM remains greater than 2 pixels --- the FWHM of the image was previously estimated using  \textsc{SExtractor} \citep{bertin1996sextractor}.
This value is chosen on the basis of the plate sampling offered by current ground-based imagers. If the image can be $2\times 2$ rebinned while meeting the condition above, it has a 50\% probability to be rebinned.

\section{Convolutional neural networks}

In this section, we describe the convolutional neural networks used for our analysis. The first one, \textsc{MaxiMask}, classifies pixels (``local contaminants'') while the second one, \textsc{MaxiTrack}, classifies images (``global contaminants'').

\subsection{Local contaminant neural network}
\label{localcontsec}

\subsubsection{Architecture}

The model used for the semantic segmentation of the local contaminants, \textsc{MaxiMask}, is based on \cite{badrinarayanan2015segnet} and \cite{yang2018semantic}, which both rely on a VGG-like architecture \citep{simonyan2014very}. It consists of three parts.

The first part contains single and double convolutional layers followed by max-pooling downsampling.
This enables the network to compute relevant feature maps at different scales.
During this step, max-pooling pixel indices are kept up for later reuse.

\begin{figure}[ht]
  \includegraphics[scale=0.58]{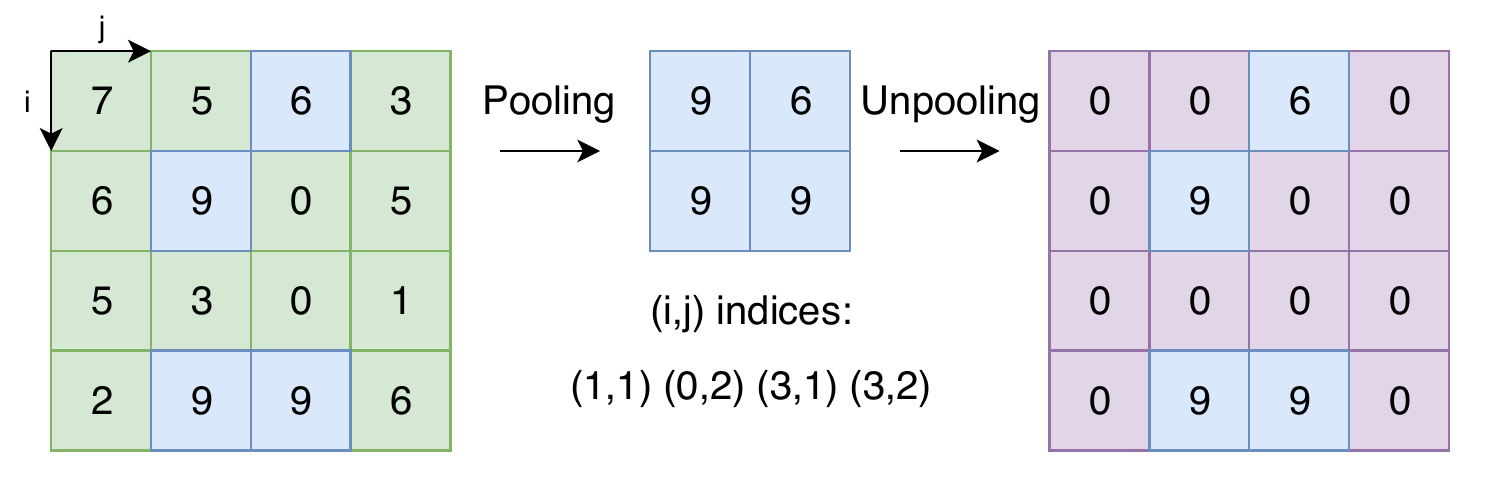}
  \caption{Example of an unpooling process. Indices of max-pooling are kept up and reused to upsample the feature maps.}
  \label{unpool}
\end{figure}

The second part also incorporates convolutional layers and recovers spatial resolution by upsampling feature maps using the max-pooling indices.
An example of unpooling is given in Fig.~\ref{unpool}.
At each resolution level, the feature maps of the first part are summed with the corresponding upsampled feature maps to make use of the maximum of information.

The third part is made of extra unpool-convolution paths (UCPs) that recover the highest image resolution from each feature map resolution so that the network can exploit the maximum of information of each resolution.
Thus, it results 5 pre-predictions, one for each resolution.

\begin{figure*}[ht]
  \includegraphics[scale=0.52]{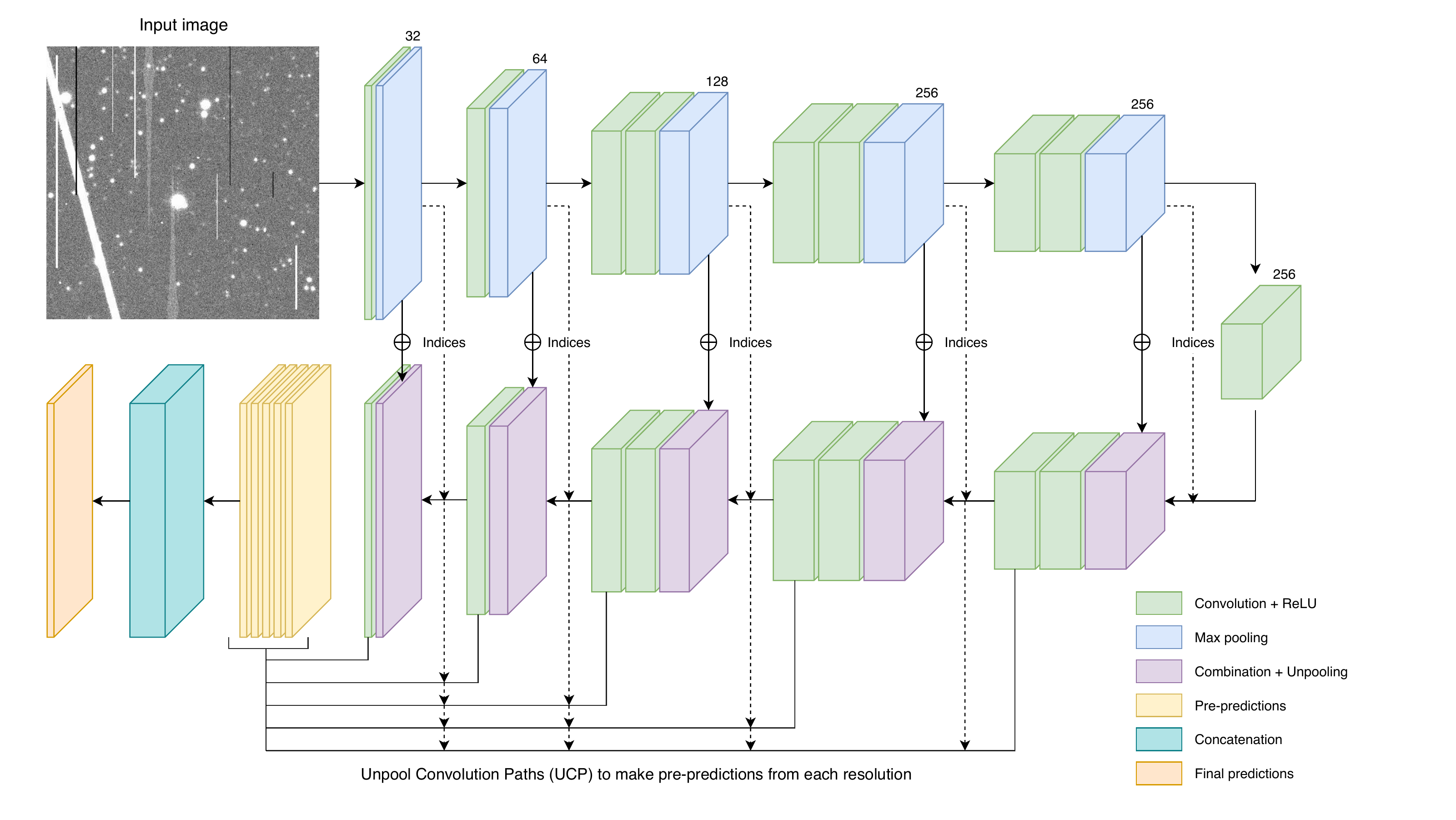}
  \caption{Scheme representation of the local contaminants neural network architecture.}
  \label{nnscheme}
\end{figure*}

The 5 pre-predictions are finally concatenated and a last convolution layer builds the final predictions.
The sigmoid activation functions in this last layer are not \textit{softmax}-normalized, to allow non-mutually exclusive classes to be assigned jointly to pixels. 
All convolutional layers use $3\times3$ kernels and apply ReLU activations.
The architecture is represented in Fig.~\ref{nnscheme} and hyperparameters are described more precisely in Table~\ref{nntab}.
The neural network is implemented using the TensorFlow library \citep{abadi2016tensorflow} on a TITAN X Nvidia GPU.

\subsubsection{Training and loss function}

Training is done for 30 epochs on 50,000 images, with mini-batches shuffled at every epoch.
The batch size is kept small (10) to maintain a reasonable memory footprint.
The model is trained end-to-end using the Adam optimizer \citep{kingma2014adam}.
The loss function $L$ is the sigmoid cross-entropy \citep{rubinstein1999cross} summed over all classes and pixels, and averaged across batch images:

\begin{equation}
\begin{aligned}
  L = - \frac{1}{\mathrm{card}({\mathcal B})} \sum _{b\in \mathcal{B}} \sum_{p\in {\cal P}} w^{\prime}_{p,b} \sum_{\omega_c\in {\cal C}}\biggl( y_{b,p,c} & \log \hat{y}_{b,p,c} \\ &
  + (1-y_{b,p,c}) \log(1-\hat{y}_{b,p,c})\biggr),
\end{aligned}
\label{loss}
\end{equation}

where $\mathcal{B}$ is the set of batch images, $\mathcal{P}$ is the set of all image pixels, $\mathcal{C}$ is the set of all contaminant classes, $w^{\prime}_{p,b}$ is a weight applied to pixel $p$ of image $b$ in the batch (see below), $\hat{y}_{b,p,c}$ is the sigmoid prediction for class $\omega_c$ of pixel $p$ of image $b$ in the batch, and $y_{b,p,c}$ is the ground truth label for class $\omega_c$ of pixel $p$ of image $b$ defined as:
  \begin{align}
    y_{b,p,c} = \left\{\begin{matrix} 1 & \mathrm{if}\ \omega_c \in \ {\mathcal C}_{p,b}\,\\ 0 & \text{otherwise} \end{matrix}\right.,
  \end{align}
where $\mathcal{C}_{p,b}\subset \mathcal{C}$ is the set of contaminant classes labeling pixel $p$ of image $b$ in the batch.
In order to improve the back-propagation of error gradients down to the deepest layers, several losses are combined.
In addition to the main sigmoid cross-entropy loss $L$ computed on the final predictions, we can compute a sigmoid cross-entropy for each of the 5 pre-predictions.
There are several ways to associate all of these losses.
Like \cite{yang2018semantic}, we find that adding respectively 33\% or 50\% of each of the 3 or 2 smallest resolution losses to the main loss works best.
The two main rules here are that the additional loss weights should sum to 1 and that higher resolution pre-predictions become less informative as they get closer to the one at full resolution.

Basic training procedures are vulnerable to strong class imbalance, which makes it more likely for the neural network to converge to a state where rare contaminants are not properly detected.
Contaminant classes are so statistically insignificant (down to one part in $10^6$ with real data, typically) that the classifier may be tricked into assigning all pixels to the background class.
To prevent this, we start by applying a basic weighting scheme to each pixel according to its class representation in the training set, that is each pixel $p$ of batch image $b$ belonging to classes in $\mathcal{C}_{p,b}$ is weighted by $w_{p,b}$ defined as
\begin{equation}
w_{p,b} = \sum \limits_{\omega_c \in \mathcal{C}_{p,b}} w_c,
\end{equation}
with
\begin{equation}
    w_c = \left(P(\omega_c|T) \sum \limits_{i} \frac{1}{P(\omega_i|T)}\right)^{-1},
    \label{weight}
\end{equation}
where $P(\omega_c|T)$ is the fraction of pixels labeled with class $\omega_c$ in the training dataset $T$.
The $P(\omega_c|T)$'s do not sum to one as many pixels belong to several classes and are thus counted several times.
We find that the weighting scheme brings slightly better results and less variability in the training if weights are computed at once from the class proportions of the whole set, instead of being recomputed for each image. From Eq.~\ref{weight} we have:

\begin{equation}
  \forall i \in \mathcal{C}, \forall j \in \mathcal{C},\ \frac{w_i}{w_j} = \frac{P(\omega_j|T)}{P(\omega_i|T)}{\rm \ \ and\ }\sum \limits_{\omega_c \in \mathcal{C}} w_c = 1.
\end{equation}

However, with this simple weighting scheme, background class pixels that are close to rare features are given very low weights, although they are decisive for classification.
To circumvent this, weight maps are smoothed with a $3 \times 3$ Gaussian kernel with unit standard deviation so that highly weighted regions spread over larger areas.
Other kernel sizes and standard deviations were tested but we find 3 and 1 to give the best results.
The resulting weights of this smoothing are the $w^{\prime}_{p,b}$ presented in the loss function of Eq.~\ref{loss}.

Finally, the solution is regularized by the l2 norm of all the $N$ network weights, by adding the following term to the total loss:
\begin{equation}
    L2_{reg} = \lambda \sum \limits_i^{N} \|\boldsymbol{k}_i\|_2,
\end{equation}
where the $\boldsymbol{k}_i$'s are the convolution kernel vectors.
$\lambda$ sets the regularization strength.
We find $\lambda=1$ to provide the best results.

\begin{table}[ht]
  \centering
  \begin{tabular}{|c|c|c|c|c|c|c|}
    \hline
    Layer & Size & \multicolumn{4}{|c|}{UCP from each resolution} \\
    \hline
    Input & 400x400x1 & & & & \\
    \cline{1-2}
    Conv & 400x400x32 & & & & \\
    Maxpool & 200x200x32 & & & & \\
    \cline{1-2}
    Conv & 200x200x64 & & & & \\
    Maxpool & 100x100x64 & & & & \\
    \cline{1-2}
    Conv & 100x100x128 & & & & \\
    Conv & 100x100x128 & & & & \\
    Maxpool & 50x50x128 & & & & \\
    \cline{1-2}
    Conv & 50x50x256 & & & & \\
    Conv & 50x50x256 & & & & \\
    Maxpool & 25x25x256 & & & & \\
    \cline{1-2}
    Conv & 25x25x256 & & & & \\
    Conv & 25x25x256 & & & & \\
    Maxpool & 13x13x256 & & & & \\
    \cline{1-2}
    Conv & 13x13x256 & & & & \\
    \cline{1-2}
    Unpooling & 25x25x256 & & & & \\
    Conv & 25x25x256 & & & & \\
    Conv & 25x25x256 & UCP & & & \\
    \cline{1-3}
    Unpooling & 50x50x256 & Idem & & & \\
    Conv & 50x50x256 & None & & & \\
    Conv & 50x50x128 & Idem & UCP & & \\
    \cline{1-4}
    Unpooling & 100x100x128 & Idem & Idem & & \\
    Conv & 100x100x128 & None & None & & \\
    Conv & 100x100x64 & Idem & Idem & UCP & \\
    \cline{1-5}
    Unpooling & 200x200x64 & Idem & Idem & Idem & \\
    Conv & 200x200x32 & Idem & Idem & Idem & UCP \\
    \cline{1-6}
    Unpooling & 400x400x32 & Idem & Idem & Idem & Idem \\
    Conv & 400x400x14 & Idem & Idem & Idem & Idem \\
    \hline
    Concat & \multicolumn{5}{|c|}{400x400x70} \\
    \hline
    Conv & \multicolumn{5}{|c|}{400x400x14} \\
    \hline
  \end{tabular}
  \caption{Description of the local contaminants neural network architecture, including map dimensions. All convolution kernels are $3\times 3$ and max-pooling kernels are $2\times 2$. All activation functions (not shown for brevity) are ReLU, except in the output layer where the sigmoid is used.}
  \label{nntab}
\end{table}

\subsection{Global contaminant neural network architecture}

\begin{figure*}[ht]
  \includegraphics[scale=0.53]{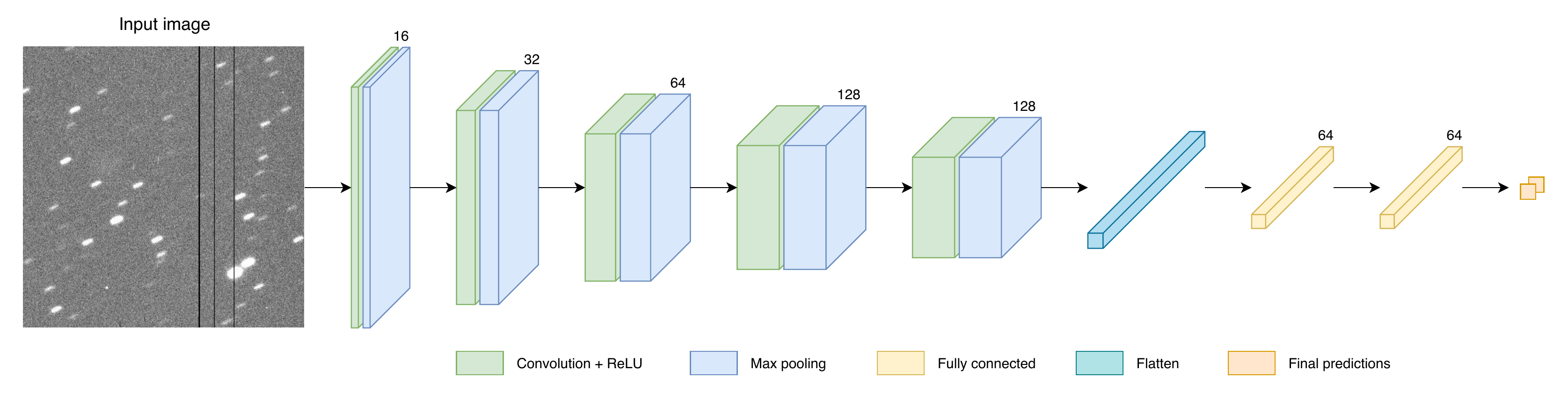}
  \caption{Scheme representation of the global contaminants neural network architecture.}
  \label{tnnscheme}
\end{figure*}

The convolutional neural network that detects global contaminants (tracking errors), \textsc{MaxiTrack}, is a simple network made of convolutional layers followed by max-pooling and fully connected layers. The architecture of the network is schematized in Fig.~\ref{tnnscheme} and detailed in Table~\ref{tnntab}. Because the two classes are mutually exclusive (affected by tracking errors or not), we adopt for the output layer a softmax activation function and a softmax cross-entropy loss function \citep{rubinstein1999cross}.
Training is done for 48 epochs on 50,000 images with a mini-batch size of 64 samples, using the Adam optimizer.

\begin{table}[ht]
  \centering
  \begin{tabular}{|c|c|}
    \cline{1-2}
    Layer & Size \\
    \cline{1-2}
    Input & 400x400x1 \\
    \hline
    Conv & 400x400x16 \\ 
    Maxpool & 200x200x16 \\ 
    \hline
    Conv & 200x200x32 \\ 
    Maxpool & 100x100x32 \\ 
    \hline
    Conv & 100x100x64 \\ 
    Maxpool & 50x50x64 \\ 
    \hline
    Conv & 50x50x128 \\ 
    Maxpool & 25x25x128 \\ 
    \hline
    Conv & 25x25x128 \\ 
    Maxpool & 13x13x128 \\ 
    \hline
    Flatten & 21632 \\
    \cline{1-2}
    Fully connected & 64 \\
    \cline{1-2}
    Fully connected & 64 \\
    \cline{1-2}
    Fully connected & 2 \\
    \cline{1-2}
  \end{tabular}
  \caption{Description of the global contaminant neural network architecture, including map dimensions.
  All convolution kernels are $9\times 9$ and max-pooling kernels are $2\times 2$.
  All activation functions (not shown for brevity) are ReLU, except in the output layer where predictions are done using softmax.}
  \label{tnntab}
\end{table}

\section{Results with test data and quality assessment}

\subsection{Local contaminants neural network}

We evaluate the quality of the results in several ways.
First, we estimate the performance of the network on test data, both quantitatively through various metrics, and qualitatively.
We verify that there is no over-fitting by checking that performance on the test set is comparable to that on the training set. 
Next, we show that performance is immune to the presence or absence of other contaminants in a given image.
We finally compare the performance of the cosmic ray detector to that of a classical algorithm.

\subsubsection{Performance metrics}

We first estimate classification performance on a benchmark test set comprising 5,000 images.
Because the network is a binary classifier for every class, we can compute a Receiver Operating Characteristic (ROC) curve for each of them.
ROC curves represent the True Positive Rate ($TPR$) versus the False Positive Rate ($FPR$):

\begin{equation}
  TPR = \frac{TP}{P} = \frac{TP}{TP+FN},
\end{equation}

\begin{equation}
  FPR = \frac{FP}{N} = \frac{FP}{TN+FP},
\end{equation}
where $P$ is the number of contaminated pixels, $TP$ is the number of true positives (contaminated pixels successfully recovered as contaminated), $FN$ is the number of false negatives  (contaminated pixels wrongly classified as non-contaminated), $N$ is the number of non-contaminated pixels, $FP$ is the number of false positives (non-contaminated pixels wrongly classified as contaminated), and $TN$ is the number of true negatives (non-contaminated pixels successfully recovered as non-contaminated).

The accuracy ($ACC$) is subsequently defined as
\begin{equation}
  ACC = \frac{TP+TN}{P+N}.
\end{equation}

The more the ROC curve bends toward the upper left part of the graph, the better the classifier. 
However with strongly imbalanced datasets, such as our pixel data, one must be very cautious with the $TPR$, $FPR$ and $ACC$ values for assessing the quality of the results.
For example, if one assumes that there are 1,000 pixels of the contaminant class ($P$) and 159,000 pixels of the background class ($N$) in a 400$\times$400 pixel sub-image, a $TPR$ of 99\% and a $FPR$ of 1\%, corresponding to an accuracy of 99\%, would actually represent a poor performance, as it would imply 990 true positives, 10 false negatives, 157,410 true negatives, and 1590 false positives. In the end, there would be more false positives $FP$ (pixels wrongly classified as contaminated) than true positives $TP$. 

For this reason the ROC curves in Fig. A~\ref{roc} are displayed with a logarithmic scale on the $FPR$ axis.
We require the $FPR$ to be very low (e.g smaller than $10^{-3}$) to consider that the network performs properly.

On the other hand, recovering the exact footprint of large, fuzzy defects is almost impossible at the level of individual pixels, which makes the classification performance for persistence, satellite trails, fringes, nebulosities, spikes and background classes look worse in Fig. A~\ref{roc} than it really is in practice.

Also, two ROC curves are drawn for cosmic rays and trails. The second one (in green) is computed using only the instances of the class that are above a specific level of the sky background. These instances were defined by retaining those which had more than a half of their pixels above $3\sigma$. These second curves shows that the network performs better on more obvious cases.

In addition to the $FPR$, $TPR$, $ACC$ and $AUC$, we use two other metrics helpful for assessing the network performance: the purity (or precision), representing the fraction of correct predictions among the positively classified samples, and the Matthews correlation coefficient \citep[MCC, ][]{matthews1975comparison}, which is an accuracy measure that takes into account the strong imbalance between classes.

\begin{equation}
  PUR = \frac{TP}{TP+FP} = \text{Purity or Precision},
\end{equation}

\begin{equation}
  MCC = \frac{ TP \times TN - FP \times FN } {\sqrt{ (TP + FP) ( TP + FN ) ( TN + FP ) ( TN + FN ) } }.
 \label{smcc}
\end{equation}

In the above example, the purity would reach only 38\% and the $MCC$ only 61\%, highlighting the classifier poor positive class discrimination.

Fig. A~\ref{pur} shows the true positive rate against the purity. 
Again, the purple curve represents how a random classifier would perform.
In these curves the best classifier would sit in the top right ($TPR=1$ and $PUR=1$).
The darkest points also represent lowest thresholds while the lighter are the highest ones.

\begin{figure*}[ht]
  \begin{minipage}{0.32\linewidth}
    \includegraphics[scale=0.41]{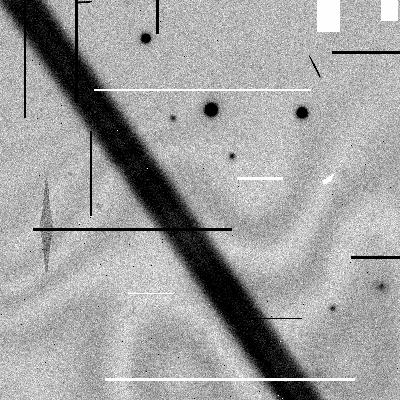}
  \end{minipage} 
  \begin{minipage}{0.32\linewidth}
    \includegraphics[scale=0.41]{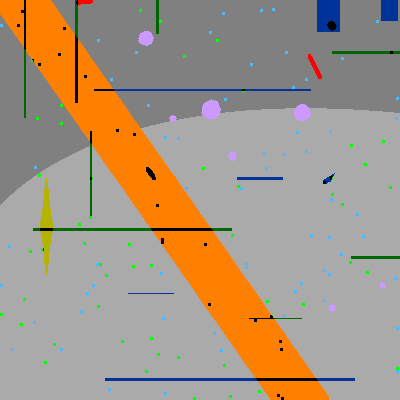}
  \end{minipage}
  \begin{minipage}{0.32\linewidth}
    \includegraphics[scale=0.41]{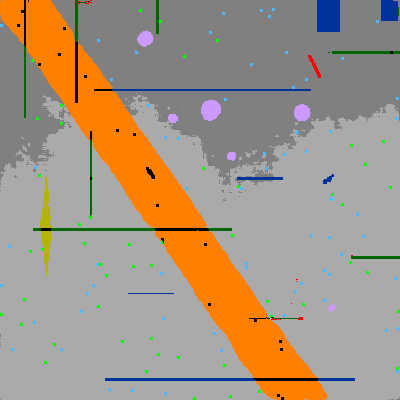}
  \end{minipage} \\
  \begin{minipage}{0.32\linewidth}
    \includegraphics[scale=0.41]{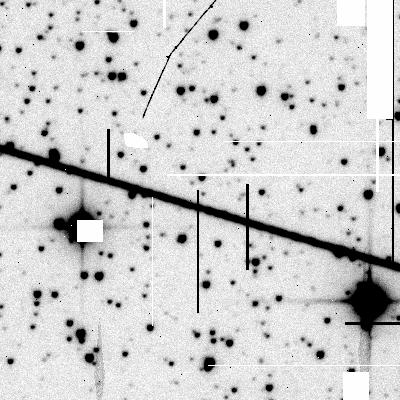}
  \end{minipage} 
  \begin{minipage}{0.32\linewidth}
    \includegraphics[scale=0.41]{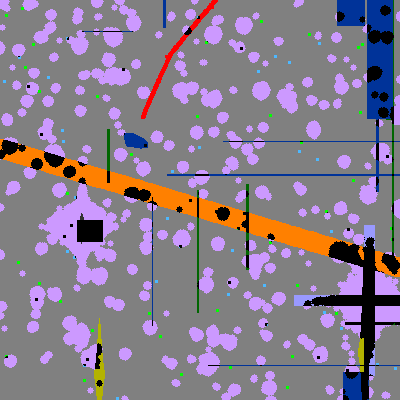}
  \end{minipage}
  \begin{minipage}{0.32\linewidth}
    \includegraphics[scale=0.41]{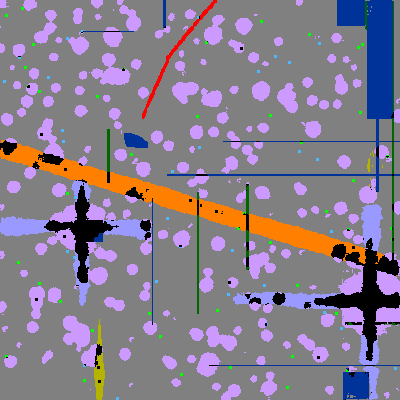}
  \end{minipage}
  \caption{Examples of qualitative results on test data. \textit{Left}: input; \textit{middle}: ground truth; \textit{right}: predictions.
  Each class is assigned a color so that the ground truth can be represented in one single image. Class predictions are done according to the threshold giving the highest MC coefficient. The color coding is identical to that of Fig. \ref{sampex}.}
  \label{qual_res}
\end{figure*}

Some qualitative results are presented in Fig.~\ref{qual_res}.
A given pixel is assigned a given class if its probability to belong to this class is higher than the best threshold in the sense of the $MCC$. 

Finally, $MCC$s are represented in Fig. A~\ref{mcc}, as a function of the output threshold.
In each curve, the threshold giving the best $MCC$ is annotated around the best $MCC$ point.
It is important to note that the best threshold depends on the modification of the prior that has been applied to the raw output probabilities.
This update of the prior is explained in section \ref{prior_sec}.

\subsubsection{Robustness regarding the context}
The MaxiMask neural network is trained using mostly images that include all contaminant classes.
Hence, we must check if the network performs equally well independently of the context, that is if it delivers equally good results for images containing, for example, a single class of contaminant.

To this aim, for every contaminant class, we generate a dataset of 1,000 images affected only by this type of contaminant (except saturated and background pixels), and another dataset of 1,000 images containing only saturated and background pixels.
We then compare the performance of MaxiMask for each class with the that obtain on the corresponding dataset.
We find that performance ($AUC$) is similar or even slightly higher for the majority of the classes.
This shows that the network is not conditioned to work only in the exact context of the training.
The results are presented in Table~\ref{indep_context}.

\begin{table}[ht]
  \centering
  \begin{tabular}{|c|c|c|}
    \hline
    Class & All contaminant & Single contaminant \\
     & set AUC & set AUC \\
    \hline
    CR & 0.96927 & 0.98314 \\
    \hline
    HCL & 0.99763 & 0.99957 \\
    \hline
    DCL & 0.99872 & 0.99976 \\
    \hline
    HP & 0.99741 & 0.99965 \\
    \hline
    DP & 0.99739 & 0.99975 \\
    \hline
    P & 0.99352 & 0.99951 \\
    \hline
    TRL & 0.99511 & 0.99813 \\
    \hline
    FR & 0.98057 & 0.93326 \\
    \hline
    NEB & 0.97895 & 0.84575 \\
    \hline
    SAT & 0.99965 & 0.99974 \\
    \hline
    SP & 0.96125 & 0.98061 \\
    \hline
    OV & 0.99997 & 1.00000 \\
    \hline
    BBG & 0.98484 & 0.99165 \\
    \hline
    BG & 0.96895 & 0.98371 \\
    \hline
  \end{tabular}
  \caption{AUC of each class depending on the test set context.}
  \label{indep_context}
\end{table}

As it can be seen, for all classes but fringes and nebulosity, performance improves when a single type of contaminant is present.
The slight improvement may come from the fact that ambiguous cases (when pixels are affected by more than one contaminant class, e.g., a cosmic ray or a hot pixel over a satellite trail) are not present in the single contaminant test set.

\subsubsection{Cosmic rays: effect of PSF undersampling and comparison with LA Cosmic \label{sec:lacosmic}}
Undersampling makes cosmic ray hits harder to distinguish from point-sources.
To solve this issue, \citet{2001PASP..113.1420V} has developed LA Cosmic, a method based on a variation of Laplacian edge detection. 
It is largely insensitive to cosmic ray morphology and PSF sampling.
LA Cosmic thus offers an excellent opportunity to test the performance of MaxiMask on undersampled exposures.

To do so, we generate two datasets containing only the cosmic ray contaminant class (plus object and background). A well sampled set of images with FWHMs larger than 2.5 pixels, and an undersampled image set with FWHMs smaller than 2.5 pixels.
We run MaxiMask and the Astro-SCRAPPY Python implementation LA Cosmic.
To make a fair comparison, LA Cosmic masks are dilated in the same way as the ground truth cosmic ray masks of MaxiMask.
However, while MaxiMask generates probability maps that can be thresholded at different levels, LA Cosmic only outputs a binary mask.
To compare the results we therefore build ROC curves for the neural network and over-plot a single point representing the result obtained with LA Cosmic.

\begin{figure}[ht]
  \includegraphics[scale=0.55]{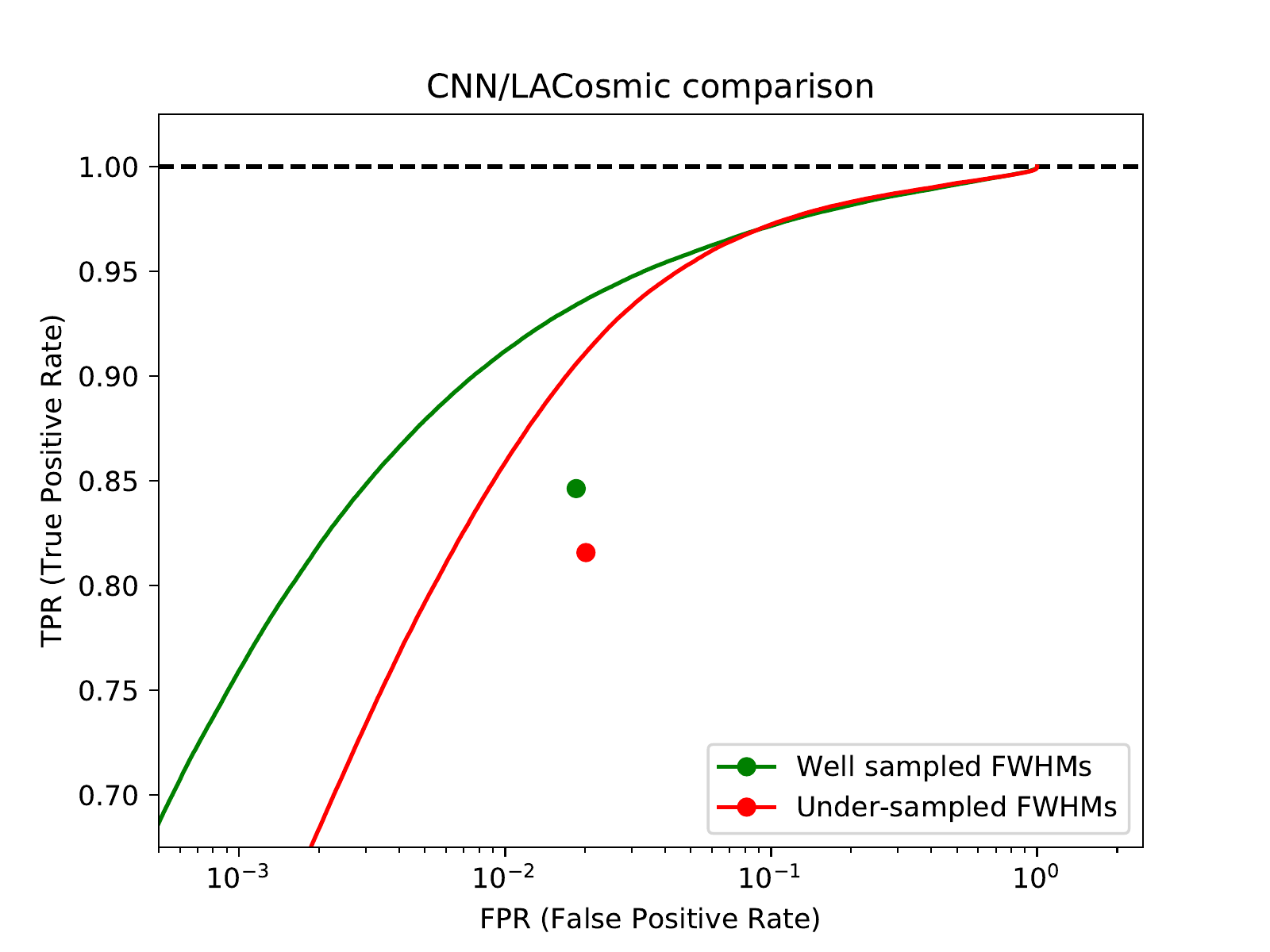}
  \caption{CR detection performance comparison with LA Cosmic.}
  \label{lacosmic}
\end{figure}

Figure \ref{lacosmic} shows that the neural network performs better than LA Cosmic in both regimes with our data.

\subsection{Global contaminants neural network}

\begin{figure}[ht]
  \includegraphics[scale=0.6]{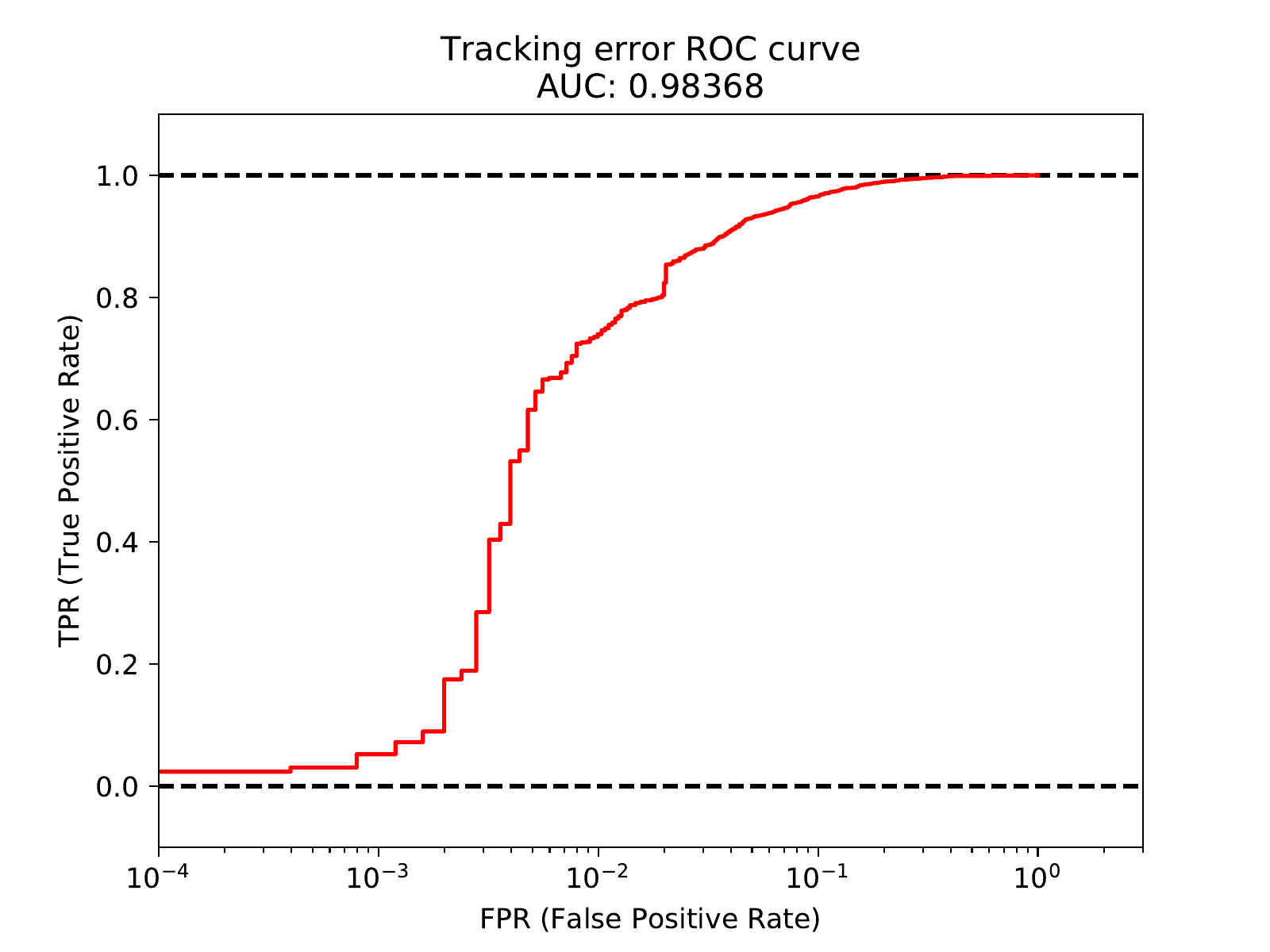}
  \caption{Global contaminant neural network ROC curve; the steps are a consequence of limited statistics. }
  \label{roctr}
\end{figure}

The ROC curve for the global contaminant neural network is shown in Fig.~\ref{roctr}. It is computed from a test set of 5,000 images.

\section{Modifying priors}
\label{prior_sec}

If one knows what class proportions are expected in the observation data, output probabilities can be updated to better match these priors \citep[e.g.,][]{saerens2002neural,bailer2008finding}.

The outputs of a \textit{perfectly trained} neural network classifier with a cross-entropy loss function can be interpreted as Bayesian posterior probabilities \citep[e.g.,][]{richard1991neural,hampshire1991,rojas1996neural}. Under this assumption and using Bayes' rule, the output for the class $\omega_c$ of the trained neural network model defined by a training set $T$ writes:
\begin{equation}
  P(\omega_c|\,\boldsymbol{x}, T) = \frac{p(\boldsymbol{x}|\omega_c, T) P(\omega_c|T)}{\sum\limits_{\omega \in \{\omega_c,\bar{\omega_c}\}} p(\boldsymbol{x}|\omega, T) P(\omega|T)},
  \label{eq1}
\end{equation}
where $\boldsymbol{x}$ is the input image data around the pixel of interest, $p(\boldsymbol{x}|\omega_c, T)$ is the distribution of $\boldsymbol{x}$ conditional to class $\omega_c$ in the training set $T$, and $P(\omega_c|T)$ is the prior probability of a pixel to belong to the class $\omega_c$ in the trained model.

As each output acts as a binary classifier, the sum is done on the class $\omega_c$ (contaminant) and its complementary $\bar{\omega_c}$ (``not the contaminant'').

With the observation data set $O$ we may similarly write:
\begin{equation}
\label{obsposterior}
  P(\omega_c|\boldsymbol{x}, O) = \frac{p(\boldsymbol{x}|\omega_c, O) P(\omega_c|O)}{\sum\limits_{\omega \in \{\omega_c,\bar{\omega_c}\}} p(\boldsymbol{x}|\omega, O) P(\omega|O)},
\end{equation}
where $P(\omega_c|O)$ is the expected fraction of pixels with class $\omega_c$ in $O$.

Now, if the appearance of defects in $O$ matches that in the training set $T$, we have $p(\boldsymbol{x}|\omega_c, T)=p(\boldsymbol{x}|\omega_c, O)$, and we can rewrite (\ref{obsposterior}) as:

\begin{eqnarray}
  P(\omega_c|\boldsymbol{x},O) & = & \frac{P(\omega_c|\boldsymbol{x},T) \frac{P(\omega_c|O)}{P(\omega_c|T)}}{\sum \limits_{\omega \in \{\omega_c,\bar{\omega_c}\}} P(\omega|\boldsymbol{x},T) \frac{P(\omega|O)}{P(\omega|T)}}\\
  & = &
  \frac{1}{1 + \left( \frac{1}{P(\omega_c|\boldsymbol{x},T)}-1\right)\frac{P(\omega_c|T)}{P(\omega_c|O)}\frac{1 - P(\omega_c|O)}{1 - P(\omega_c|T)}}.
\end{eqnarray}

If pixels were all weighted equally, the training priors $P(\omega_c|T)$ would simply be the class proportions in the training set. However, this is not the case here, and pixel weights have to be taken into account. To do so, we follow
\cite{bailer2008finding}'s approach, by using as an estimator of $P(\omega_c|T)$ the posterior mean on the test set $T'$ (which by construction is distributed identically to the training set):

\begin{equation}
\label{trainposterior}
  \hat{P}(\omega_c|T) =  \frac{1}{\mathrm{card}(T')} \sum \limits_{\boldsymbol{x} \in T'} P(\omega_c|\boldsymbol{x},T').
\end{equation}

These corrected probabilities are used to compute the MC coefficient curves in Fig.~\ref{mcc} (whereas the prior correction does not affect the ROC and purity curves).

\textsc{MaxiMask} comes with the $P(\omega_c|T)$ values already set, therefore one only needs to specify the expected class proportions in the data, that is the $P(\omega_c|O)$'s.

\section{Application to other data}

As a sanity check, we apply \textsc{MaxiMask} to data obtained from different instruments not part of the training set.
Examples of the resulting contaminant maps are shown in appendix.

Our first external check is with ZTF \citep{2019PASP..131a8002B} data.
The \textsc{MaxiMask} output for a science image featuring a prominent trail with variable amplitude is shown in Fig.~\ref{ztfex}.
We can note the ability of \textsc{MaxiMask} to properly flag both the trail and overlapping sources.

Our second external check is with the ACS instrument onboard the Hubble Space Telescope (Fig.~\ref{hstex} and ~\ref{hstex2}).
This test illustrates \textsc{MaxiMask}'s ability to distinguish cosmic rays from poorly sampled, diffraction-limited point source images.

Given the seemingly good performance of \textsc{MaxiMask} on images from instruments not part of the training set, one question that may arise is whether 
\textsc{MaxiMask} can readily be used on production for such instruments, without any retraining or transfer learning.
Our limited experience with \textsc{MaxiMask} seems to indicate that this is indeed the case, although retraining may be beneficial for specific instrumental features.
As shown here, excellent performance can be reached by training with 50,000 $400 \times 400$ images taken from three different instruments.
We think that a minimum of 10,000 $400 \times 400$ would be a good start to train on a single instrument. Assuming CCDs of approximately $2,000 \times 2,000 pixels$, thus containing 25 400x400 images, it would just need 400 CCDs, equivalent to 10 fields for a 40 CCD camera.

Our last series of tests is conducted on digital images of natural scenes (landscape, cat, human face), to check for possible inconsistencies on data that are totally unlike those from the training set.
Reassuringly, the maps produced by \textsc{MaxiMask} are consistent with the expected patterns. For instance, the cat's whiskers are identified as cosmic ray impacts, and pixels with the lowest values as bad pixels.

\section{Using \textsc{MaxiMask} and \textsc{MaxiTrack}}

\textsc{MaxiMask} and \textsc{MaxiTrack} are available at \url{https://www.github.com/mpaillassa/MaxiMask}.
\textsc{MaxiMask} is a Python module that infers probability maps from FITS images.
It can process a whole mosaic, a specific FITS image extension, or all the FITS files from a directory or a file list.
For every FITS file being processed a new FITS image is generated with the same HDU (Header Data Unit) structure as the input.
Every input image HDU has a matching contaminant map HDU in output, with one image plane per requested contaminant.
The header contains metadata related to the contaminant, including the prior and threshold used.
An option can be set to generate a single image plane for all contaminants, using a binary code for each contaminant.
Such composite contaminant maps can easily be used as flag maps, for example, in \textsc{SExtractor}.
Based on command line arguments and configuration parameters, one can select specific classes, apply updates to the priors and thresholds to the probability maps.
The code relies on the TensorFlow library and can work on both CPUs or GPUs, although the CPU version is expected to be much slower: \textsc{MaxiMask} processes about 1.2 megapixel per second with an NVidia Titan X GPU, and about 60 times less on a 2.7GHz Intel i7 dual-core CPU.
Yet, there is probabily room for improvement in processing efficiency for both the CPU and GPU versions.

\textsc{MaxiTrack} is used the same way as \textsc{MaxiMask}, except that the output is a text file indicating the probability for the input image(s) to be affected by tracking errors (one probability per extension if the image contains several HDUs).
It can also apply an update to the prior. It runs at 60 megapixels/s with an NVidia Titan X GPU and is 9 times slower on a 2.7GHz Intel i7 dual-core CPU.

\section{Summary and perspectives}

We have built a data set and trained convolutional neural network classifiers named {\sc MaxiMask} and {\sc MaxiTrack} to identify contaminants in astronomical images.
We have shown that they achieve good performance on test data, both real and simulated.
By delivering posterior probabilities, \textsc{MaxiMask} and \textsc{MaxiTrack} give the user the flexibility to set appropriate threshold levels and achieve the desired $TPR/FPR$ trade-offs depending on the scientific objectives and requirements.
Both classifiers require no input parameters or knowledge of the camera properties.

Even though the mix of contaminants in the training set is unrealistic, being dictated by training requirements, we have checked that this does not impact performance.
Output probabilities can be corrected to adapt the behavior of \textsc{MaxiMask} to any mix of contaminants in the data.

We are aware that several types of contaminants and images are missing from the current version and may be added in the future.

Local contaminants include two particularly prominent classes of contaminants: optical and electronics ghosts.
Unwanted reflections within the optics result in stray light in exposures.
These reflections can produce spurious images from bright sources commonly referred to as ``optical ghosts''.
Sometimes, reflections from very bright stars outside of the field may also be seen.
Detectors read through multiple ports also suffer from a form of electronic ghost known as cross-talk.
Electronic cross-talk causes bright sources in one of the CCD quadrants to generate a ghost pattern in other quadrants.
The ghosts may be negative or positive and are typically at the level of 1:10$^4$.
Both effects are a significant source of nuisance in wide field exposures, especially in crowded fields and deep images, where they generate false, transient sources, and can affect high precision astrometric and photometric measurements.

Another category of common issues is defocused or excessively aberrated exposures, as well as trails caused by charge transfer inefficiency, all of which  which could easily be implemented in \textsc{ MaxiTrack}.

Also, the training set used in the current version of \textsc{MaxiMask} and \textsc{MaxiTrack} does not include images from space-born telescopes nor, more generally, diffraction-limited imagers. Therefore, they are unlikely to perform optimally with such data, although limited testing indicates that they may remain usable for most features, an example of prediction on HST data is shown in Figs.~\ref{hstex} and ~\ref{hstex2}.

Finally, \textsc{MaxiMask} could be extended to not only detect contaminants, but also to generate an inpainted (i.e., ``corrected'') version of the damaged image areas wherever possible.

\begin{acknowledgements}
M.P. acknowledges financial support from the Centre National d’Etudes Spatiales (CNES) fellowship program.
We are grateful to Mike Read, of the  Royal Observatory, Edinburgh, for providing us with data from the UKIRT telescope, and to Vincent Lepetit for providing comments and suggestions that helped improve the paper.
This research has received funding from the European Research Council (ERC) under the European Union’s Horizon 2020 research and innovation programme (grant agreement No 682903, P.I. H. Bouy), and from the French State in the framework of the "Investments for the future" Program, IdEx Bordeaux, reference ANR-10-IDEX-03-02. We gratefully acknowledge the support of NVIDIA Corporation with the donation of one of the Titan Xp GPUs used for this research.
This research draws upon data distributed by the NOAO Science Archive. NOAO is operated by the Association of Universities for Research in Astronomy (AURA) under cooperative agreement with the National Science Foundation. Based on observations made with the Isaac Newton Telescope operated on the island of La Palma by the Isaac Newton Group in the Spanish Observatorio del Roque de los Muchachos of the Instituto de Astrofísica de Canarias. The Isaac Newton Telescope is operated on the island of La Palma by the Isaac Newton Group in the Spanish Observatorio del Roque de los Muchachos of the Instituto de Astrofísica de Canarias. This paper makes use of data obtained from the Isaac Newton Group Archive which is maintained as part of the CASU Astronomical Data Centre at the Institute of Astronomy, Cambridge. Based on data obtained from the ESO Science Archive Facility. This research used the facilities of the Canadian Astronomy Data Centre operated by the National Research Council of Canada with the support of the Canadian Space Agency. Based in part on data collected at Subaru Telescope which is operated by the National Astronomical Observatory of Japan and obtained from the SMOKA, which is operated by the Astronomy Data Center, National Astronomical Observatory of Japan. The Hyper Suprime-Cam (HSC) collaboration includes the astronomical communities of Japan and Taiwan, and Princeton University. The HSC instrumentation and software were developed by the National Astronomical Observatory of Japan (NAOJ), the Kavli Institute for the Physics and Mathematics of the Universe (Kavli IPMU), the University of Tokyo, the High Energy Accelerator Research Organization (KEK), the Academia Sinica Institute for Astronomy and Astrophysics in Taiwan (ASIAA), and Princeton University. Funding was contributed by the FIRST program from Japanese Cabinet Office, the Ministry of Education, Culture, Sports, Science and Technology (MEXT), the Japan Society for the Promotion of Science (JSPS), Japan Science and Technology Agency (JST), the Toray Science Foundation, NAOJ, Kavli IPMU, KEK, ASIAA, and Princeton University. This paper makes use of software developed for the Large Synoptic Survey Telescope. We thank the LSST Project for making their code available as free software at \href{http://dm.lsst.org}{http://dm.lsst.org}. The Pan-STARRS1 Surveys (PS1) have been made possible through contributions of the Institute for Astronomy, the University of Hawaii, the Pan-STARRS Project Office, the Max-Planck Society and its participating institutes, the Max Planck Institute for Astronomy, Heidelberg and the Max Planck Institute for Extraterrestrial Physics, Garching, The Johns Hopkins University, Durham University, the University of Edinburgh, Queen’s University Belfast, the Harvard-Smithsonian Center for Astrophysics, the Las Cumbres Observatory Global Telescope Network Incorporated, the National Central University of Taiwan, the Space Telescope Science Institute, the National Aeronautics and Space Administration under Grant No. NNX08AR22G issued through the Planetary Science Division of the NASA Science Mission Directorate, the National Science Foundation under Grant No. AST-1238877, the University of Maryland, and Eotvos Lorand University (ELTE) and the Los Alamos National Laboratory. Based on data collected at the Subaru Telescope and retrieved from the HSC data archive system, which is operated by Subaru Telescope and Astronomy Data Center at National Astronomical Observatory of Japan. This paper includes data gathered with the Swope telescope located at Las Campanas Observatory, Chile. Based on observations obtained with MegaPrime/MegaCam, a joint project of CFHT and CEA/DAPNIA, at the Canada-France-Hawaii Telescope (CFHT) which is operated by the National Research Council (NRC) of Canada, the Institut National des Science de l'Univers of the Centre National de la Recherche Scientifique (CNRS) of France, and the University of Hawaii. This research has made use of NASA's Astrophysics Data System Bibliographic Services. This research made use of Astropy, a community-developed core Python package for Astronomy (Astropy Collaboration, 2013, http://dx.doi.org/10.1051/0004-6361/201322068). The Herschel spacecraft was designed, built, tested, and launched under a contract to ESA managed by the Herschel/Planck Project team by an industrial consortium under the overall responsibility of the prime contractor Thales Alenia Space (Cannes), and including Astrium (Friedrichshafen) responsible for the payload module and for system testing at spacecraft level, Thales Alenia Space (Turin) responsible for the service module, and Astrium (Toulouse) responsible for the telescope, with in excess of a hundred subcontractors
\end{acknowledgements}

\bibliographystyle{aa}
\bibliography{main}

\begin{thebibliography}{56}
\expandafter\ifx\csname natexlab\endcsname\relax\def\natexlab#1{#1}\fi

\bibitem[{Abadi {et~al.}(2016)Abadi, Barham, Chen, Chen, Davis, Dean, Devin,
  Ghemawat, Irving, Isard, {et~al.}}]{abadi2016tensorflow}
Abadi, M., Barham, P., Chen, J., {et~al.} 2016, in OSDI, Vol.~16, 265--283

\bibitem[{{Autry} {et~al.}(2003){Autry}, {Probst}, {Starr}, {Abdel-Gawad},
  {Blakley}, {Daly}, {Dominguez}, {Hileman}, {Liang}, {Pearson}, {Shaw}, \&
  {Tody}}]{2003SPIE.4841..525A}
{Autry}, R.~G., {Probst}, R.~G., {Starr}, B.~M., {et~al.} 2003, in Society of
  Photo-Optical Instrumentation Engineers (SPIE) Conference Series, Vol. 4841,
  Society of Photo-Optical Instrumentation Engineers (SPIE) Conference Series,
  ed. {M.~Iye \& A.~F.~M.~Moorwood}, 525--539

\bibitem[{Badrinarayanan {et~al.}(2015)Badrinarayanan, Kendall, \&
  Cipolla}]{badrinarayanan2015segnet}
Badrinarayanan, V., Kendall, A., \& Cipolla, R. 2015, arXiv preprint
  arXiv:1511.00561

\bibitem[{Badrinarayanan {et~al.}(2017)Badrinarayanan, Kendall, \&
  Cipolla}]{badrinarayanan2017segnet}
Badrinarayanan, V., Kendall, A., \& Cipolla, R. 2017, IEEE transactions on
  pattern analysis and machine intelligence, 39, 2481

\bibitem[{Bailer-Jones {et~al.}(2008)Bailer-Jones, Smith, Tiede, Sordo, \&
  Vallenari}]{bailer2008finding}
Bailer-Jones, C.~A., Smith, K., Tiede, C., Sordo, R., \& Vallenari, A. 2008,
  Monthly Notices of the Royal Astronomical Society, 391, 1838

\bibitem[{{Bekte{\v s}evi{\'c}} {et~al.}(2018){Bekte{\v s}evi{\'c}},
  {Vinkovi{\'c}}, {Rasmussen}, \& {Ivezi{\'c}}}]{2018MNRAS.474.4837B}
{Bekte{\v s}evi{\'c}}, D., {Vinkovi{\'c}}, D., {Rasmussen}, A., \&
  {Ivezi{\'c}}, {\v Z}. 2018, \mnras, 474, 4837

\bibitem[{{Bellm} {et~al.}(2019){Bellm}, {Kulkarni}, {Graham}, {Dekany},
  {Smith}, {Riddle}, {Masci}, {Helou}, {Prince}, {Adams}, {Barbarino},
  {Barlow}, {Bauer}, {Beck}, {Belicki}, {Biswas}, {Blagorodnova}, {Bodewits},
  {Bolin}, {Brinnel}, {Brooke}, {Bue}, {Bulla}, {Burruss}, {Cenko}, {Chang},
  {Connolly}, {Coughlin}, {Cromer}, {Cunningham}, {De}, {Delacroix}, {Desai},
  {Duev}, {Eadie}, {Farnham}, {Feeney}, {Feindt}, {Flynn}, {Franckowiak},
  {Frederick}, {Fremling}, {Gal-Yam}, {Gezari}, {Giomi}, {Goldstein},
  {Golkhou}, {Goobar}, {Groom}, {Hacopians}, {Hale}, {Henning}, {Ho}, {Hover},
  {Howell}, {Hung}, {Huppenkothen}, {Imel}, {Ip}, {Ivezi{\'c}}, {Jackson},
  {Jones}, {Juric}, {Kasliwal}, {Kaspi}, {Kaye}, {Kelley}, {Kowalski},
  {Kramer}, {Kupfer}, {Landry}, {Laher}, {Lee}, {Lin}, {Lin}, {Lunnan},
  {Giomi}, {Mahabal}, {Mao}, {Miller}, {Monkewitz}, {Murphy}, {Ngeow},
  {Nordin}, {Nugent}, {Ofek}, {Patterson}, {Penprase}, {Porter}, {Rauch},
  {Rebbapragada}, {Reiley}, {Rigault}, {Rodriguez}, {van Roestel}, {Rusholme},
  {van Santen}, {Schulze}, {Shupe}, {Singer}, {Soumagnac}, {Stein}, {Surace},
  {Sollerman}, {Szkody}, {Taddia}, {Terek}, {Van Sistine}, {van Velzen},
  {Vestrand}, {Walters}, {Ward}, {Ye}, {Yu}, {Yan}, \&
  {Zolkower}}]{2019PASP..131a8002B}
{Bellm}, E.~C., {Kulkarni}, S.~R., {Graham}, M.~J., {et~al.} 2019, \pasp, 131,
  018002

\bibitem[{{Bertin}(2009)}]{2009MmSAI..80..422B}
{Bertin}, E. 2009, \memsai, 80, 422

\bibitem[{{Bertin}(2013)}]{2013ascl.soft01001B}
{Bertin}, E. 2013, {PSFEx: Point Spread Function Extractor}, Astrophysics
  Source Code Library

\bibitem[{Bertin \& Arnouts(1996)}]{bertin1996sextractor}
Bertin, E. \& Arnouts, S. 1996, Astronomy and Astrophysics Supplement Series,
  117, 393

\bibitem[{{Bosch} {et~al.}(2018){Bosch}, {Armstrong}, {Bickerton}, {Furusawa},
  {Ikeda}, {Koike}, {Lupton}, {Mineo}, {Price}, {Takata}, {Tanaka}, {Yasuda},
  {AlSayyad}, {Becker}, {Coulton}, {Coupon}, {Garmilla}, {Huang}, {Krughoff},
  {Lang}, {Leauthaud}, {Lim}, {Lust}, {MacArthur}, {Mandelbaum}, {Miyatake},
  {Miyazaki}, {Murata}, {More}, {Okura}, {Owen}, {Swinbank}, {Strauss},
  {Yamada}, \& {Yamanoi}}]{HSC_Pipeline}
{Bosch}, J., {Armstrong}, R., {Bickerton}, S., {et~al.} 2018, \pasj, 70, S5

\bibitem[{{Boulade} {et~al.}(2003){Boulade}, {Charlot}, {Abbon}, {Aune},
  {Borgeaud}, {Carton}, {Carty}, {Da Costa}, {Deschamps}, {Desforge},
  {Eppell{\'e}}, {Gallais}, {Gosset}, {Granelli}, {Gros}, {de Kat}, {Loiseau},
  {Ritou}, {Rouss{\'e}}, {Starzynski}, {Vignal}, \&
  {Vigroux}}]{2003SPIE.4841...72B}
{Boulade}, O., {Charlot}, X., {Abbon}, P., {et~al.} 2003, in Society of
  Photo-Optical Instrumentation Engineers (SPIE) Conference Series, Vol. 4841,
  Society of Photo-Optical Instrumentation Engineers (SPIE) Conference Series,
  ed. {M.~Iye \& A.~F.~M.~Moorwood}, 72--81

\bibitem[{{Bouy} {et~al.}(2013){Bouy}, {Bertin}, {Moraux}, {Cuillandre},
  {Bouvier}, {Barrado}, {Solano}, \& {Bayo}}]{2013A&A...554A.101B}
{Bouy}, H., {Bertin}, E., {Moraux}, E., {et~al.} 2013, \aap, 554, A101

\bibitem[{{Casali} {et~al.}(2007){Casali}, {Adamson}, {Alves de Oliveira},
  {Almaini}, {Burch}, {Chuter}, {Elliot}, {Folger}, {Foucaud}, {Hambly},
  {Hastie}, {Henry}, {Hirst}, {Irwin}, {Ives}, {Lawrence}, {Laidlaw}, {Lee},
  {Lewis}, {Lunney}, {McLay}, {Montgomery}, {Pickup}, {Read}, {Rees}, {Robson},
  {Sekiguchi}, {Vick}, {Warren}, \& {Woodward}}]{2007A&A...467..777C}
{Casali}, M., {Adamson}, A., {Alves de Oliveira}, C., {et~al.} 2007, \aap, 467,
  777

\bibitem[{{Cuillandre} {et~al.}(2000){Cuillandre}, {Luppino}, {Starr}, \&
  {Isani}}]{2000SPIE.4008.1010C}
{Cuillandre}, J.-C., {Luppino}, G.~A., {Starr}, B.~M., \& {Isani}, S. 2000, in
  Presented at the Society of Photo-Optical Instrumentation Engineers (SPIE)
  Conference, Vol. 4008, Society of Photo-Optical Instrumentation Engineers
  (SPIE) Conference Series, ed. M.~{Iye} \& A.~F. {Moorwood}, 1010--1021

\bibitem[{{Dalton} {et~al.}(2006){Dalton}, {Caldwell}, {Ward}, {Whalley},
  {Woodhouse}, {Edeson}, {Clark}, {Beard}, {Gallie}, {Todd}, {Strachan},
  {Bezawada}, {Sutherland}, \& {Emerson}}]{2006SPIE.6269E..0XD}
{Dalton}, G.~B., {Caldwell}, M., {Ward}, A.~K., {et~al.} 2006, in \procspie,
  Vol. 6269, Society of Photo-Optical Instrumentation Engineers (SPIE)
  Conference Series, 62690X

\bibitem[{{Flaugher} {et~al.}(2010){Flaugher}, {Abbott}, {Annis}, {Antonik},
  {Bailey}, {Ballester}, {Bernstein}, {Bernstein}, {Bonati}, {Bremer},
  {Briones}, {Brooks}, {Buckley-Geer}, {Campa}, {Cardiel-Sas}, {Castander},
  {Castilla}, {Cease}, {Chappa}, {Chi}, {da Costa}, {DePoy}, {Derylo}, {De
  Vicente}, {Diehl}, {Doel}, {Estrada}, {Eiting}, {Elliott}, {Finley},
  {Frieman}, {Gaztanaga}, {Gerdes}, {Gladders}, {Guarino}, {Gutierrez},
  {Grudzinski}, {Hanlon}, {Hao}, {Holland}, {Honscheid}, {Huffman}, {Jackson},
  {Karliner}, {Kau}, {Kent}, {Krempetz}, {Krider}, {Kozlovsky}, {Kubik},
  {Kuehn}, {Kuhlmann}, {Kuk}, {Lahav}, {Lewis}, {Lin}, {Lorenzon}, {Marshall},
  {Mart{\'{\i}}nez}, {McKay}, {Merritt}, {Meyer}, {Miquel}, {Morgan}, {Moore},
  {Moore}, {Nord}, {Ogando}, {Olsen}, {Peoples}, {Plazas}, {Roe}, {Roodman},
  {Rossetto}, {Sanchez}, {Scarpine}, {Schalk}, {Schindler}, {Schmidt},
  {Schmitt}, {Schubnell}, {Schultz}, {Selen}, {Serrano}, {Shaw}, {Simaitis},
  {Slaughter}, {Smith}, {Spinka}, {Stefanik}, {Stuermer}, {Sypniewski},
  {Talaga}, {Tarle}, {Thaler}, {Tucker}, {Walker}, {Weaverdyck}, {Wester},
  {Woods}, {Worswick}, \& {Zhao}}]{DECAM}
{Flaugher}, B.~L., {Abbott}, T.~M.~C., {Annis}, J., {et~al.} 2010, in
  \procspie, Vol. 7735, Ground-based and Airborne Instrumentation for Astronomy
  III, 77350D

\bibitem[{Garcia-Garcia {et~al.}(2017)Garcia-Garcia, Orts-Escolano, Oprea,
  Villena-Martinez, \& Garcia-Rodriguez}]{garcia2017review}
Garcia-Garcia, A., Orts-Escolano, S., Oprea, S., Villena-Martinez, V., \&
  Garcia-Rodriguez, J. 2017, arXiv preprint arXiv:1704.06857

\bibitem[{{Griffin} {et~al.}(2010){Griffin}, {Abergel}, {Abreu}, {Ade},
  {Andr{\'e}}, {Augueres}, {Babbedge}, {Bae}, {Baillie}, {Baluteau}, {Barlow},
  {Bendo}, {Benielli}, {Bock}, {Bonhomme}, {Brisbin}, {Brockley-Blatt},
  {Caldwell}, {Cara}, {Castro-Rodriguez}, {Cerulli}, {Chanial}, {Chen},
  {Clark}, {Clements}, {Clerc}, {Coker}, {Communal}, {Conversi}, {Cox},
  {Crumb}, {Cunningham}, {Daly}, {Davis}, {de Antoni}, {Delderfield}, {Devin},
  {di Giorgio}, {Didschuns}, {Dohlen}, {Donati}, {Dowell}, {Dowell}, {Duband},
  {Dumaye}, {Emery}, {Ferlet}, {Ferrand}, {Fontignie}, {Fox}, {Franceschini},
  {Frerking}, {Fulton}, {Garcia}, {Gastaud}, {Gear}, {Glenn}, {Goizel},
  {Griffin}, {Grundy}, {Guest}, {Guillemet}, {Hargrave}, {Harwit}, {Hastings},
  {Hatziminaoglou}, {Herman}, {Hinde}, {Hristov}, {Huang}, {Imhof}, {Isaak},
  {Israelsson}, {Ivison}, {Jennings}, {Kiernan}, {King}, {Lange}, {Latter},
  {Laurent}, {Laurent}, {Leeks}, {Lellouch}, {Levenson}, {Li}, {Li},
  {Lilienthal}, {Lim}, {Liu}, {Lu}, {Madden}, {Mainetti}, {Marliani}, {McKay},
  {Mercier}, {Molinari}, {Morris}, {Moseley}, {Mulder}, {Mur}, {Naylor},
  {Nguyen}, {O'Halloran}, {Oliver}, {Olofsson}, {Olofsson}, {Orfei}, {Page},
  {Pain}, {Panuzzo}, {Papageorgiou}, {Parks}, {Parr-Burman}, {Pearce},
  {Pearson}, {P{\'e}rez-Fournon}, {Pinsard}, {Pisano}, {Podosek}, {Pohlen},
  {Polehampton}, {Pouliquen}, {Rigopoulou}, {Rizzo}, {Roseboom}, {Roussel},
  {Rowan-Robinson}, {Rownd}, {Saraceno}, {Sauvage}, {Savage}, {Savini},
  {Sawyer}, {Scharmberg}, {Schmitt}, {Schneider}, {Schulz}, {Schwartz},
  {Shafer}, {Shupe}, {Sibthorpe}, {Sidher}, {Smith}, {Smith}, {Smith},
  {Spencer}, {Stobie}, {Sudiwala}, {Sukhatme}, {Surace}, {Stevens}, {Swinyard},
  {Trichas}, {Tourette}, {Triou}, {Tseng}, {Tucker}, {Turner}, {Vaccari},
  {Valtchanov}, {Vigroux}, {Virique}, {Voellmer}, {Walker}, {Ward}, {Waskett},
  {Weilert}, {Wesson}, {White}, {Whitehouse}, {Wilson}, {Winter}, {Woodcraft},
  {Wright}, {Xu}, {Zavagno}, {Zemcov}, {Zhang}, \&
  {Zonca}}]{2010A&A...518L...3G}
{Griffin}, M.~J., {Abergel}, A., {Abreu}, A., {et~al.} 2010, \aap, 518, L3

\bibitem[{Hampshire~II \& Pearlmutter(1991)}]{hampshire1991}
Hampshire~II, J.~B. \& Pearlmutter, B. 1991, in Connectionist Models
  (Elsevier), 159--172

\bibitem[{{Ienaka} {et~al.}(2013){Ienaka}, {Kawara}, {Matsuoka}, {Sameshima},
  {Oyabu}, {Tsujimoto}, \& {Peterson}}]{2013ApJ...767...80I}
{Ienaka}, N., {Kawara}, K., {Matsuoka}, Y., {et~al.} 2013, \apj, 767, 80

\bibitem[{{Ives}(1998)}]{1998IEEES..16...20I}
{Ives}, D. 1998, IEEE Spectrum, 16, 20

\bibitem[{Kawanomoto {et~al.}(2016{\natexlab{a}})Kawanomoto, Komiyama, \&
  Yagi}]{Kawanomoto2016}
Kawanomoto, S., Komiyama, Y., \& Yagi, M. 2016{\natexlab{a}}, in Subaru Users'
  Meeting FY2016

\bibitem[{Kawanomoto {et~al.}(2016{\natexlab{b}})Kawanomoto, Yagi, \&
  Kawanomoto}]{Komiyama2016}
Kawanomoto, Y., Yagi, M., \& Kawanomoto, S. 2016{\natexlab{b}}, in Subaru
  Users' Meeting FY2016

\bibitem[{Kingma \& Ba(2014)}]{kingma2014adam}
Kingma, D.~P. \& Ba, J. 2014, arXiv preprint arXiv:1412.6980

\bibitem[{Krizhevsky {et~al.}(2012)Krizhevsky, Sutskever, \&
  Hinton}]{krizhevsky2012imagenet}
Krizhevsky, A., Sutskever, I., \& Hinton, G.~E. 2012, in Advances in neural
  information processing systems, 1097--1105

\bibitem[{{Kuijken} {et~al.}(2002){Kuijken}, {Bender}, {Cappellaro},
  {Muschielok}, {Baruffolo}, {Cascone}, {Iwert}, {Mitsch}, {Nicklas},
  {Valentijn}, {Baade}, {Begeman}, {Bortolussi}, {Boxhoorn}, {Christen},
  {Deul}, {Geimer}, {Greggio}, {Harke}, {H{\"a}fner}, {Hess}, {Hess}, {Hopp},
  {Ilijevski}, {Klink}, {Kravcar}, {Lizon}, {Magagna}, {M{\"u}ller},
  {Niemeczek}, {de Pizzol}, {Poschmann}, {Reif}, {Rengelink}, {Reyes},
  {Silber}, \& {Wellem}}]{2002Msngr.110...15K}
{Kuijken}, K., {Bender}, R., {Cappellaro}, E., {et~al.} 2002, The Messenger,
  110, 15

\bibitem[{LeCun {et~al.}(1995)LeCun, Bengio, {et~al.}}]{lecun1995convolutional}
LeCun, Y., Bengio, Y., {et~al.} 1995, The handbook of brain theory and neural
  networks, 3361, 1995

\bibitem[{Long {et~al.}(2015)Long, Shelhamer, \& Darrell}]{long2015fully}
Long, J., Shelhamer, E., \& Darrell, T. 2015, in Proceedings of the IEEE
  conference on computer vision and pattern recognition, 3431--3440

\bibitem[{{Long} {et~al.}(2015){Long}, {Baggett}, \&
  {MacKenty}}]{2015wfc..rept...15L}
{Long}, K.~S., {Baggett}, S.~M., \& {MacKenty}, J.~W. 2015, {Persistence in the
  WFC3 IR Detector: an Improved Model Incorporating the Effects of Exposure
  Time}, Tech. rep.

\bibitem[{Lowe {et~al.}(1999)}]{lowe1999object}
Lowe, D.~G. {et~al.} 1999in , 1150--1157

\bibitem[{{Magnier} \& {Cuillandre}(2004)}]{Elixir}
{Magnier}, E.~A. \& {Cuillandre}, J.-C. 2004, \pasp, 116, 449

\bibitem[{Matthews(1975)}]{matthews1975comparison}
Matthews, B.~W. 1975, Biochimica et Biophysica Acta (BBA)-Protein Structure,
  405, 442

\bibitem[{McCully {et~al.}(2018)McCully, Crawford, Kovacs, Tollerud, Betts,
  Bradley, Craig, Turner, Streicher, Sipocz, Robitaille, \&
  Deil}]{curtis_mccully_2018_1482019}
McCully, C., Crawford, S., Kovacs, G., {et~al.} 2018, astropy/astroscrappy:
  v1.0.5 Zenodo Release

\bibitem[{{Melchior} {et~al.}(2016){Melchior}, {Sheldon}, {Drlica-Wagner},
  {Rykoff}, {Abbott}, {Abdalla}, {Allam}, {Benoit-L{\'e}vy}, {Brooks},
  {Buckley-Geer}, {Carnero Rosell}, {Carrasco Kind}, {Carretero}, {Crocce},
  {D'Andrea}, {da Costa}, {Desai}, {Doel}, {Evrard}, {Finley}, {Flaugher},
  {Frieman}, {Gaztanaga}, {Gerdes}, {Gruen}, {Gruendl}, {Honscheid}, {James},
  {Jarvis}, {Kuehn}, {Li}, {Maia}, {March}, {Marshall}, {Nord}, {Ogando},
  {Plazas}, {Romer}, {Sanchez}, {Scarpine}, {Sevilla-Noarbe}, {Smith},
  {Soares-Santos}, {Suchyta}, {Swanson}, {Tarle}, {Vikram}, {Walker}, {Wester},
  \& {Zhang}}]{2016A&C....16...99M}
{Melchior}, P., {Sheldon}, E., {Drlica-Wagner}, A., {et~al.} 2016, Astronomy
  and Computing, 16, 99

\bibitem[{{Metzger} {et~al.}(1995){Metzger}, {Luppino}, \&
  {Miyazaki}}]{1995AAS...187.7305M}
{Metzger}, M.~R., {Luppino}, G.~A., \& {Miyazaki}, S. 1995, in Bulletin of the
  American Astronomical Society, Vol.~27, Bulletin of the American Astronomical
  Society, 1389--+

\bibitem[{{Miville-Desch{\^e}nes} {et~al.}(2016){Miville-Desch{\^e}nes}, {Duc},
  {Marleau}, {Cuillandre}, {Didelon}, {Gwyn}, \&
  {Karabal}}]{2016A&A...593A...4M}
{Miville-Desch{\^e}nes}, M.-A., {Duc}, P.-A., {Marleau}, F., {et~al.} 2016,
  \aap, 593, A4

\bibitem[{{Miyazaki} {et~al.}(2018){Miyazaki}, {Komiyama}, {Kawanomoto}, {Doi},
  {Furusawa}, {Hamana}, {Hayashi}, {Ikeda}, {Kamata}, {Karoji}, {Koike},
  {Kurakami}, {Miyama}, {Morokuma}, {Nakata}, {Namikawa}, {Nakaya}, {Nariai},
  {Obuchi}, {Oishi}, {Okada}, {Okura}, {Tait}, {Takata}, {Tanaka}, {Tanaka},
  {Terai}, {Tomono}, {Uraguchi}, {Usuda}, {Utsumi}, {Yamada}, {Yamanoi},
  {Aihara}, {Fujimori}, {Mineo}, {Miyatake}, {Oguri}, {Uchida}, {Tanaka},
  {Yasuda}, {Takada}, {Murayama}, {Nishizawa}, {Sugiyama}, {Chiba}, {Futamase},
  {Wang}, {Chen}, {Ho}, {Liaw}, {Chiu}, {Ho}, {Lai}, {Lee}, {Jeng}, {Iwamura},
  {Armstrong}, {Bickerton}, {Bosch}, {Gunn}, {Lupton}, {Loomis}, {Price},
  {Smith}, {Strauss}, {Turner}, {Suzuki}, {Miyazaki}, {Muramatsu}, {Yamamoto},
  {Endo}, {Ezaki}, {Ito}, {Kawaguchi}, {Sofuku}, {Taniike}, {Akutsu}, {Dojo},
  {Kasumi}, {Matsuda}, {Imoto}, {Miwa}, {Suzuki}, {Takeshi}, \&
  {Yokota}}]{2018PASJ...70S...1M}
{Miyazaki}, S., {Komiyama}, Y., {Kawanomoto}, S., {et~al.} 2018, \pasj, 70, S1

\bibitem[{{Morganson} {et~al.}(2018){Morganson}, {Gruendl}, {Menanteau},
  {Carrasco Kind}, {Chen}, {Daues}, {Drlica-Wagner}, {Friedel}, {Gower},
  {Johnson}, {Johnson}, {Kessler}, {Paz-Chinch{\'o}n}, {Petravick}, {Pond},
  {Yanny}, {Allam}, {Armstrong}, {Barkhouse}, {Bechtol}, {Benoit-L{\'e}vy},
  {Bernstein}, {Bertin}, {Buckley-Geer}, {Covarrubias}, {Desai}, {Diehl},
  {Goldstein}, {Gruen}, {Li}, {Lin}, {Marriner}, {Mohr}, {Neilsen}, {Ngeow},
  {Paech}, {Rykoff}, {Sako}, {Sevilla-Noarbe}, {Sheldon}, {Sobreira}, {Tucker},
  {Wester}, \& {DES Collaboration}}]{2018PASP..130g4501M}
{Morganson}, E., {Gruendl}, R.~A., {Menanteau}, F., {et~al.} 2018, \pasp, 130,
  074501

\bibitem[{Nir {et~al.}(2018)Nir, Zackay, \& Ofek}]{nir2018optimal}
Nir, G., Zackay, B., \& Ofek, E.~O. 2018, arXiv preprint arXiv:1806.04204

\bibitem[{{Ord{\'e}novic} {et~al.}(2008){Ord{\'e}novic}, {Surace},
  {Torr{\'e}sani}, \& {Ll{\'e}baria}}]{2008StMet...5..373O}
{Ord{\'e}novic}, C., {Surace}, C., {Torr{\'e}sani}, B., \& {Ll{\'e}baria}, A.
  2008, Statistical Methodology, 5, 373

\bibitem[{{Pilbratt} {et~al.}(2010){Pilbratt}, {Riedinger}, {Passvogel},
  {Crone}, {Doyle}, {Gageur}, {Heras}, {Jewell}, {Metcalfe}, {Ott}, \&
  {Schmidt}}]{2010A&A...518L...1P}
{Pilbratt}, G.~L., {Riedinger}, J.~R., {Passvogel}, T., {et~al.} 2010, \aap,
  518, L1

\bibitem[{Rheault {et~al.}(2014)Rheault, Mondrik, DePoy, Marshall, \&
  Suntzeff}]{Swope_NewCCD}
Rheault, J.-P., Mondrik, N.~P., DePoy, D.~L., Marshall, J.~L., \& Suntzeff,
  N.~B. 2014, Spectrophotometric calibration of the Swope and duPont telescopes
  for the Carnegie supernova project 2

\bibitem[{Richard \& Lippmann(1991)}]{richard1991neural}
Richard, M.~D. \& Lippmann, R.~P. 1991, Neural computation, 3, 461

\bibitem[{Rojas(1996)}]{rojas1996neural}
Rojas, R. 1996, Neural Computation, 8, 41

\bibitem[{Rubinstein(1999)}]{rubinstein1999cross}
Rubinstein, R. 1999, Methodology and computing in applied probability, 1, 127

\bibitem[{Ruder(2016)}]{ruder2016overview}
Ruder, S. 2016, arXiv preprint arXiv:1609.04747

\bibitem[{Saerens {et~al.}(2002)Saerens, Latinne, \&
  Decaestecker}]{saerens2002neural}
Saerens, M., Latinne, P., \& Decaestecker, C. 2002, Neural computation, 14, 21

\bibitem[{Simonyan \& Zisserman(2014)}]{simonyan2014very}
Simonyan, K. \& Zisserman, A. 2014, arXiv preprint arXiv:1409.1556

\bibitem[{Szegedy {et~al.}(2015)Szegedy, Liu, Jia, Sermanet, Reed, Anguelov,
  Erhan, Vanhoucke, Rabinovich, {et~al.}}]{szegedy2015going}
Szegedy, C., Liu, W., Jia, Y., {et~al.} 2015, Cvpr

\bibitem[{{Valdes} {et~al.}(2014){Valdes}, {Gruendl}, \& {DES
  Project}}]{2014ASPC..485..379V}
{Valdes}, F., {Gruendl}, R., \& {DES Project}. 2014, in Astronomical Society of
  the Pacific Conference Series, Vol. 485, Astronomical Data Analysis Software
  and Systems XXIII, ed. N.~{Manset} \& P.~{Forshay}, 379

\bibitem[{{van Dokkum}(2001)}]{2001PASP..113.1420V}
{van Dokkum}, P.~G. 2001, \pasp, 113, 1420

\bibitem[{{Vandame}(2002)}]{2002SPIE.4847..123V}
{Vandame}, B. 2002, in Society of Photo-Optical Instrumentation Engineers
  (SPIE) Conference Series, Vol. 4847, Astronomical Data Analysis II, ed. J.-L.
  {Starck} \& F.~D. {Murtagh}, 123--134

\bibitem[{Williams(1998)}]{Williams1998}
Williams, C. K.~I. 1998, Prediction with Gaussian Processes: From Linear
  Regression to Linear Prediction and Beyond, ed. M.~I. Jordan (Dordrecht:
  Springer Netherlands), 599--621

\bibitem[{{Wolfe} {et~al.}(2000){Wolfe}, {Armandroff}, {Blouke}, {Rector},
  {Reed}, {Saha}, {Schommer}, {Smith}, {Smith}, \&
  {Walker}}]{2000SPIE.3965...80W}
{Wolfe}, T., {Armandroff}, T., {Blouke}, M.~M., {et~al.} 2000, in Society of
  Photo-Optical Instrumentation Engineers (SPIE) Conference Series, Vol. 3965,
  Society of Photo-Optical Instrumentation Engineers (SPIE) Conference Series,
  ed. {M.~M.~Blouke, N.~Sampat, G.~M.~Williams, \& T.~Yeh}, 80--91

\bibitem[{Yang {et~al.}(2018)Yang, Wu, Zhao, \& Guan}]{yang2018semantic}
Yang, T., Wu, Y., Zhao, J., \& Guan, L. 2018, Cognitive Systems Research

\end{thebibliography}
 

\begin{appendix}

\newpage

\section{Performance metric curves and qualitative tests}

  \begin{figure*}[ht]
    \begin{minipage}{0.48\linewidth}
      \includegraphics[scale=0.50]{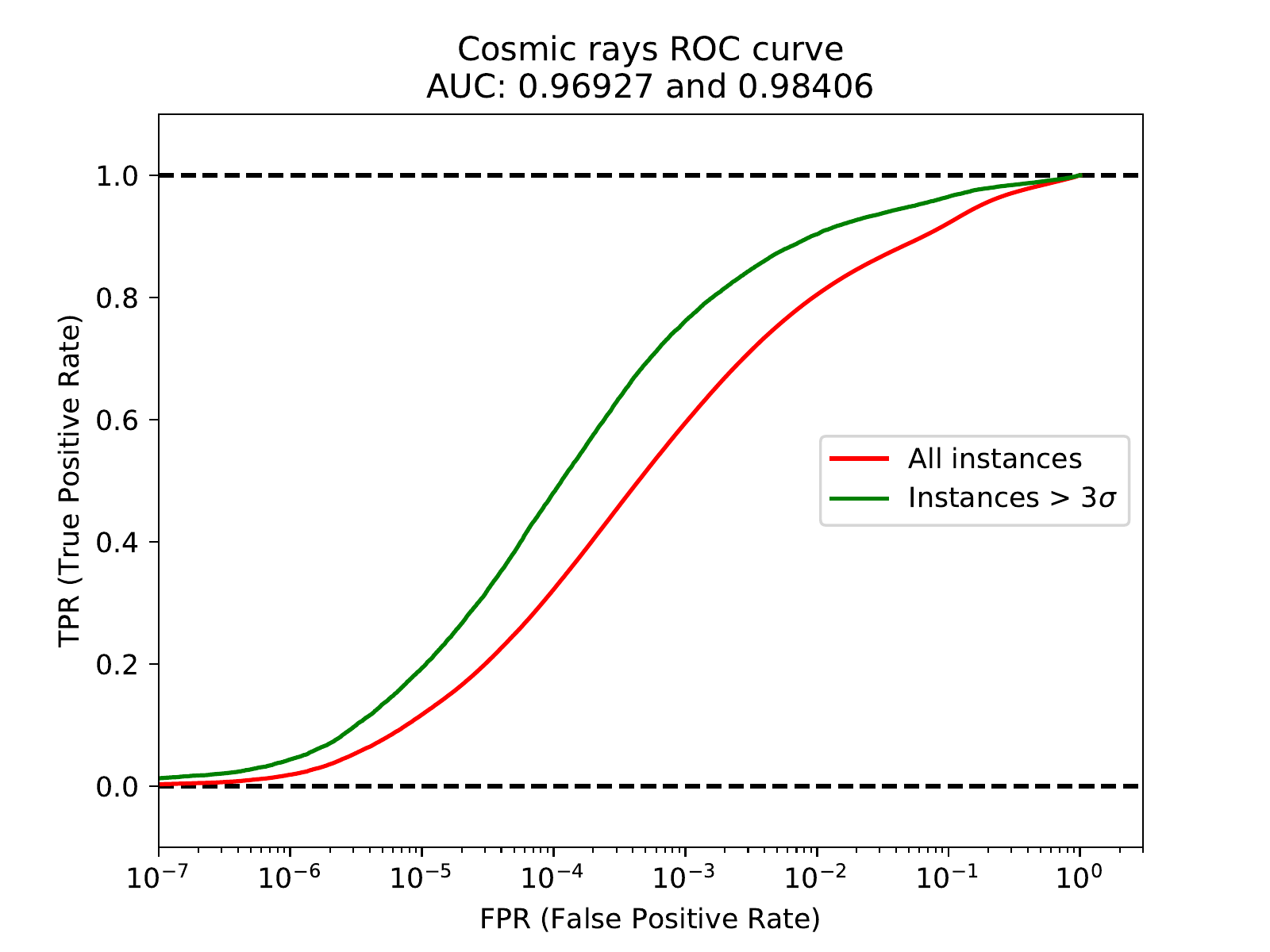}
    \end{minipage} 
    \begin{minipage}{0.48\linewidth}
      \includegraphics[scale=0.50]{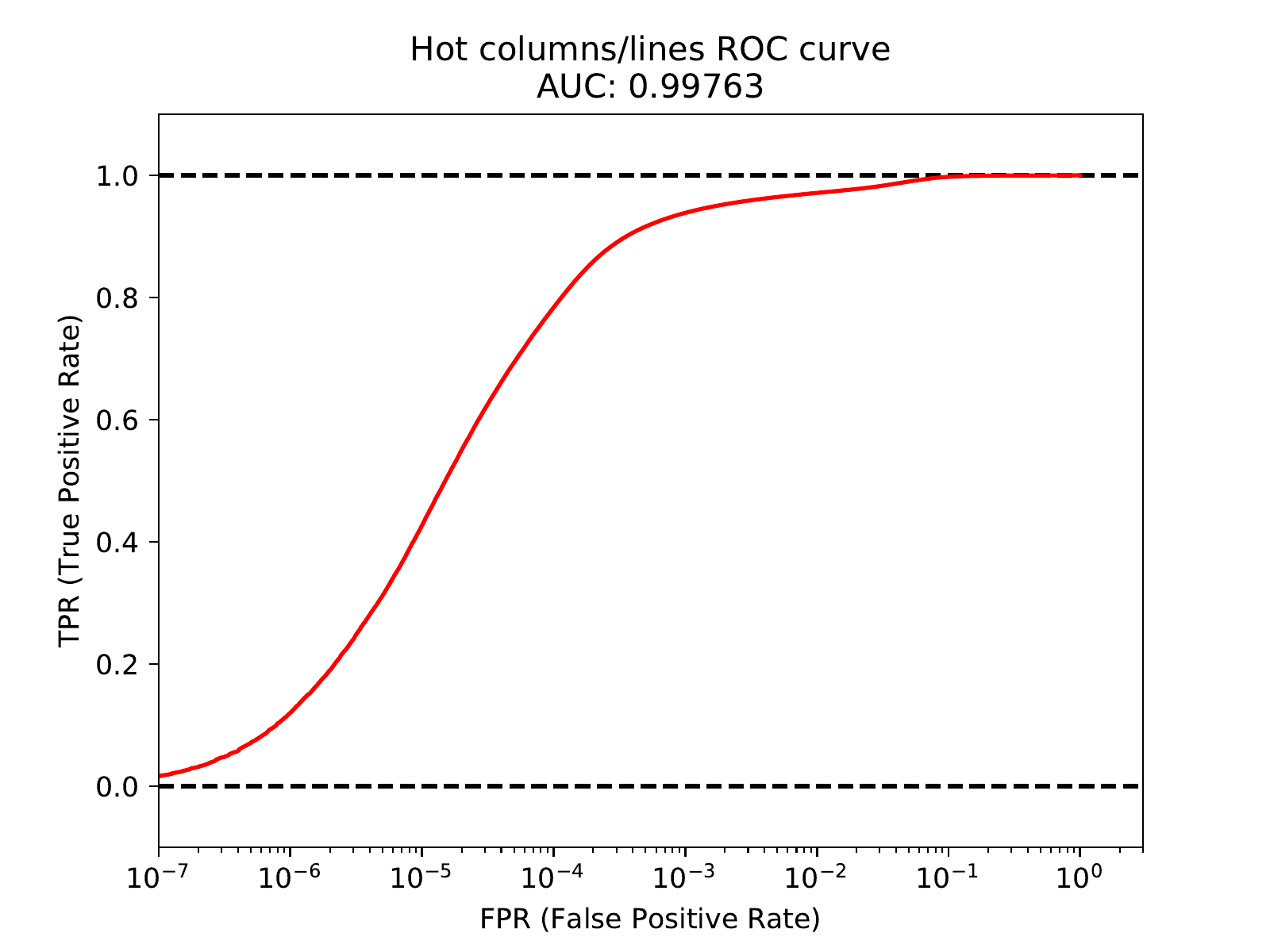}
    \end{minipage} \\
    \begin{minipage}{0.48\linewidth}
      \includegraphics[scale=0.50]{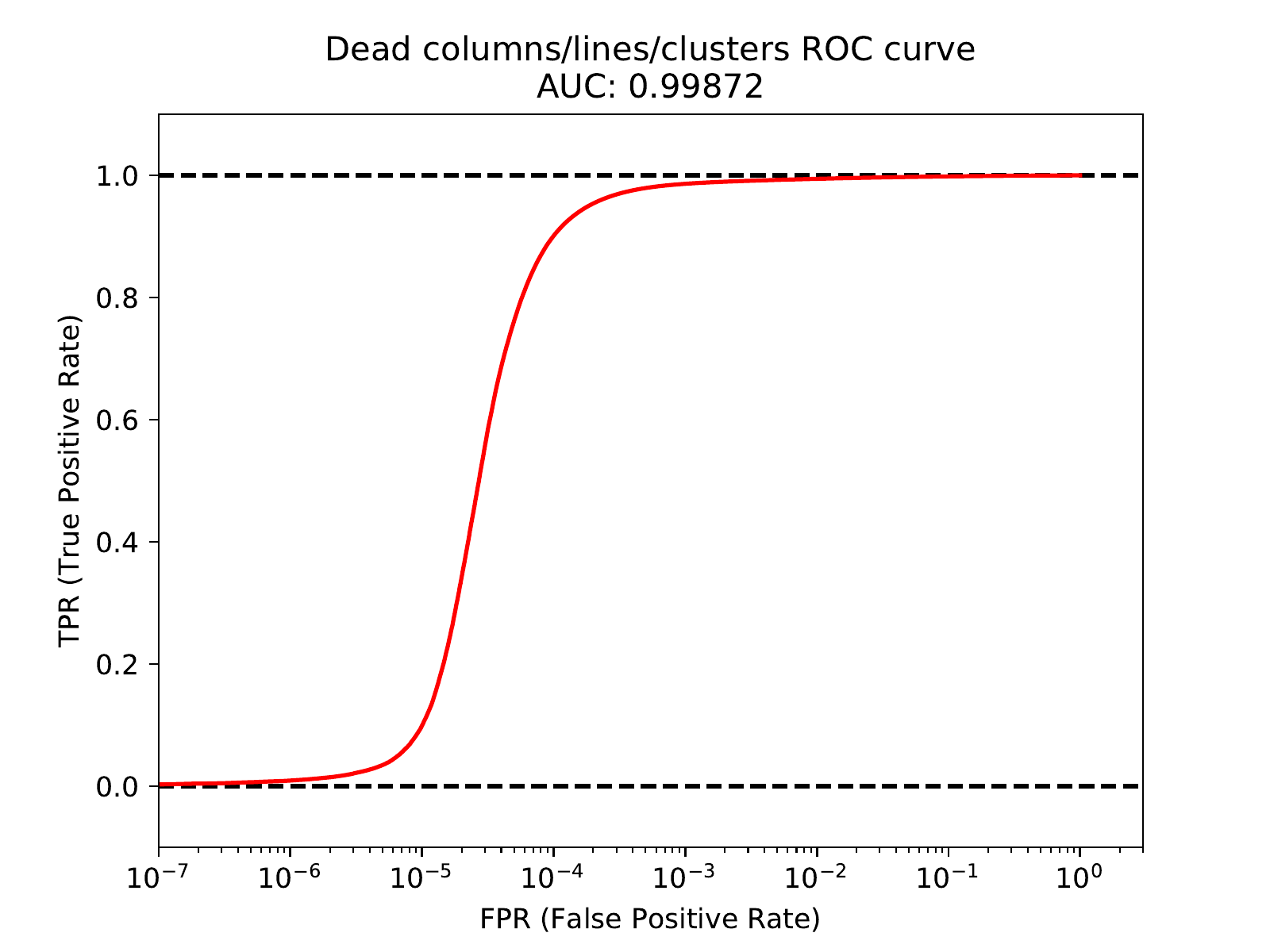}
    \end{minipage}
    \begin{minipage}{0.48\linewidth}
      \includegraphics[scale=0.50]{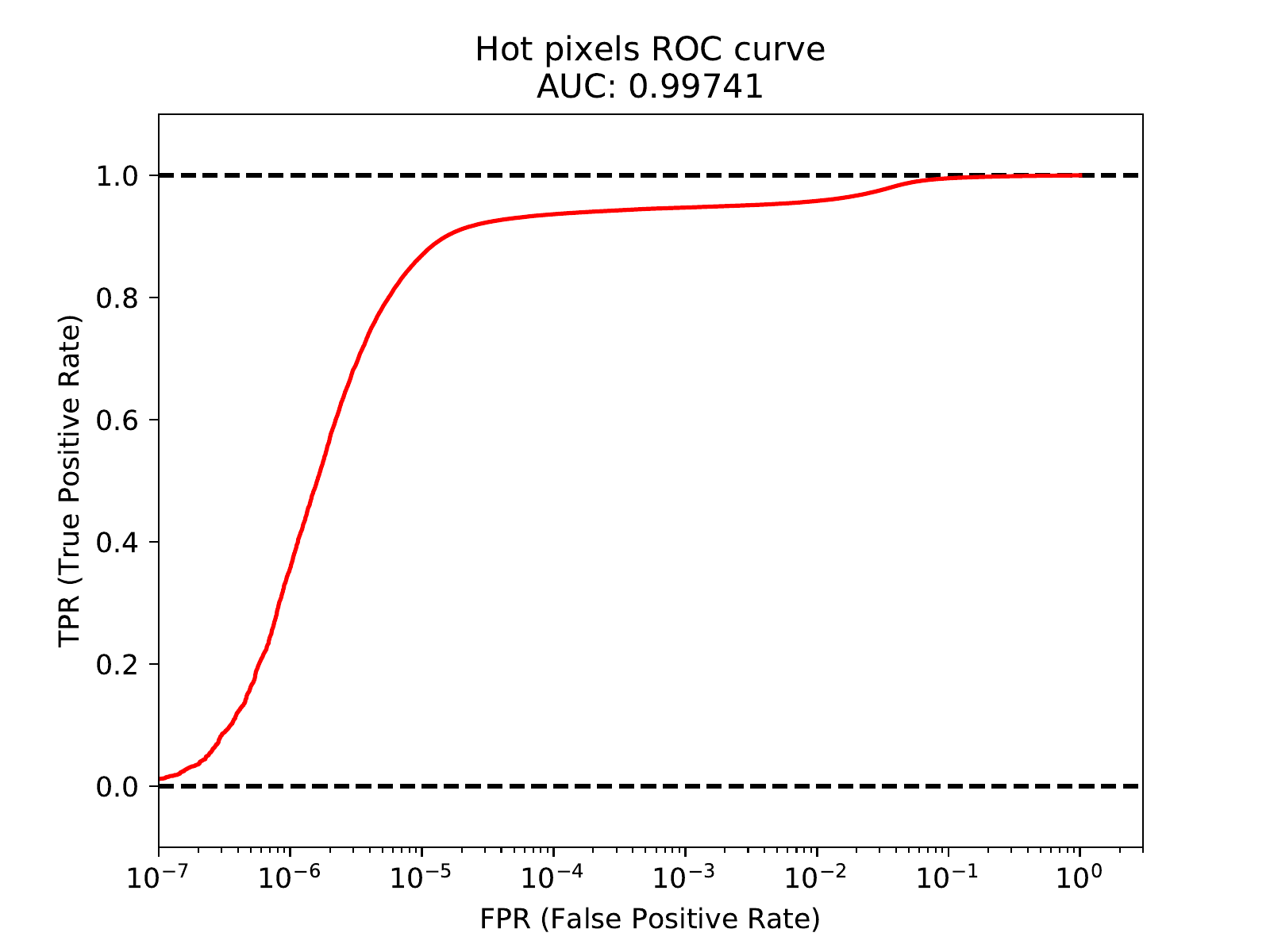}
    \end{minipage} \\
    \begin{minipage}{0.48\linewidth}
      \includegraphics[scale=0.50]{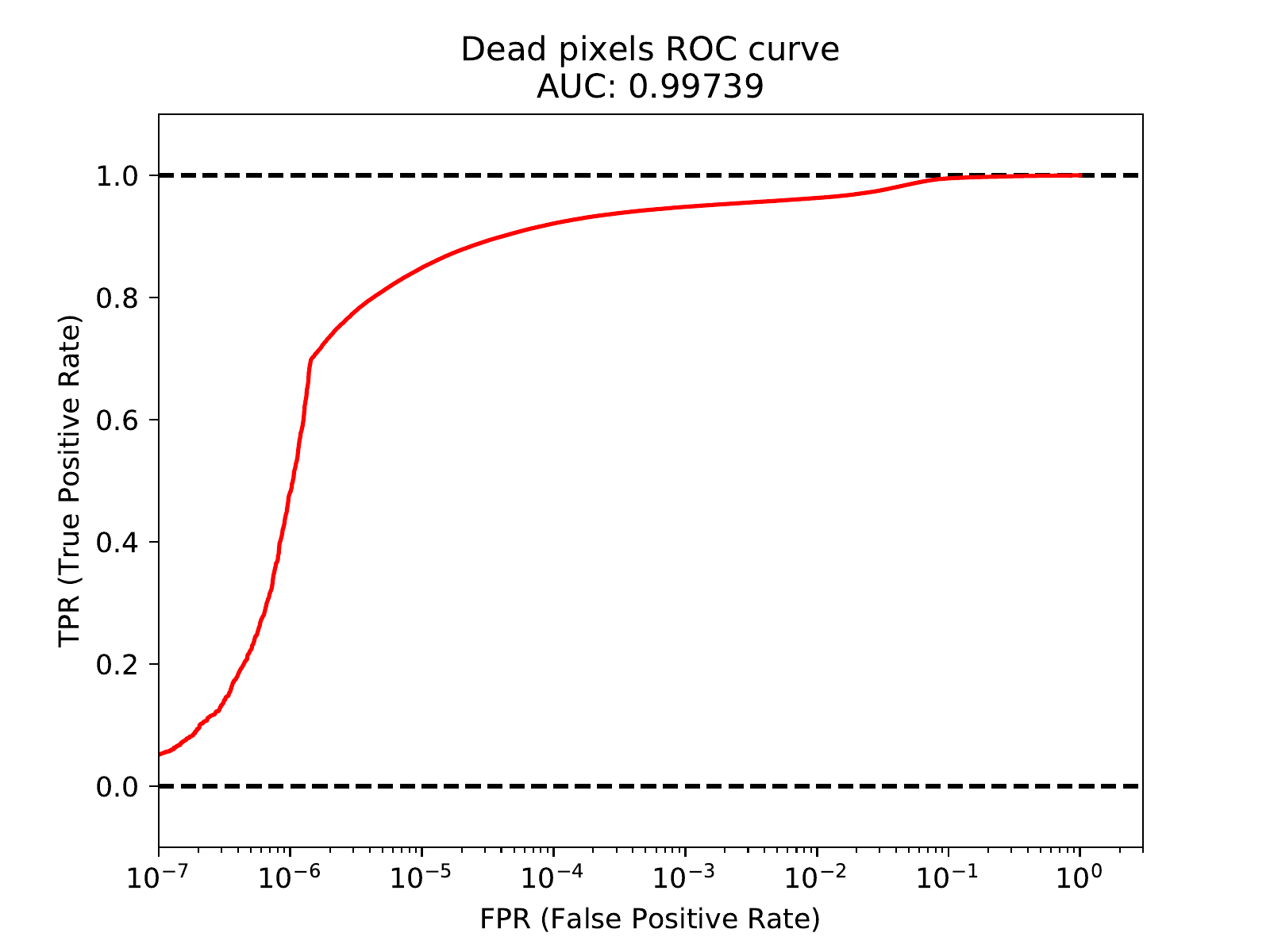}
    \end{minipage}
    \begin{minipage}{0.48\linewidth}
      \includegraphics[scale=0.50]{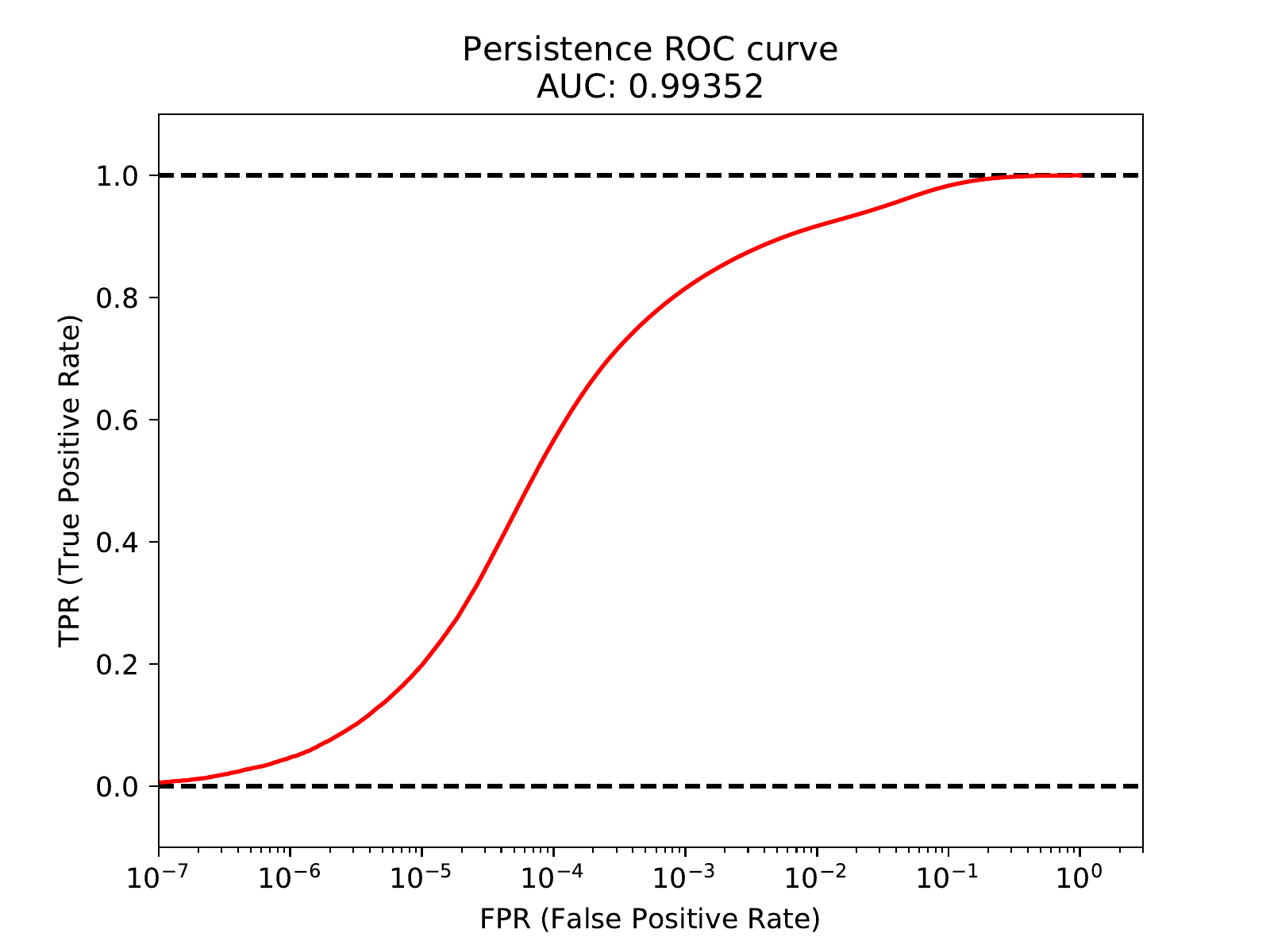}
    \end{minipage} \\
    \begin{minipage}{0.48\linewidth}
      \includegraphics[scale=0.50]{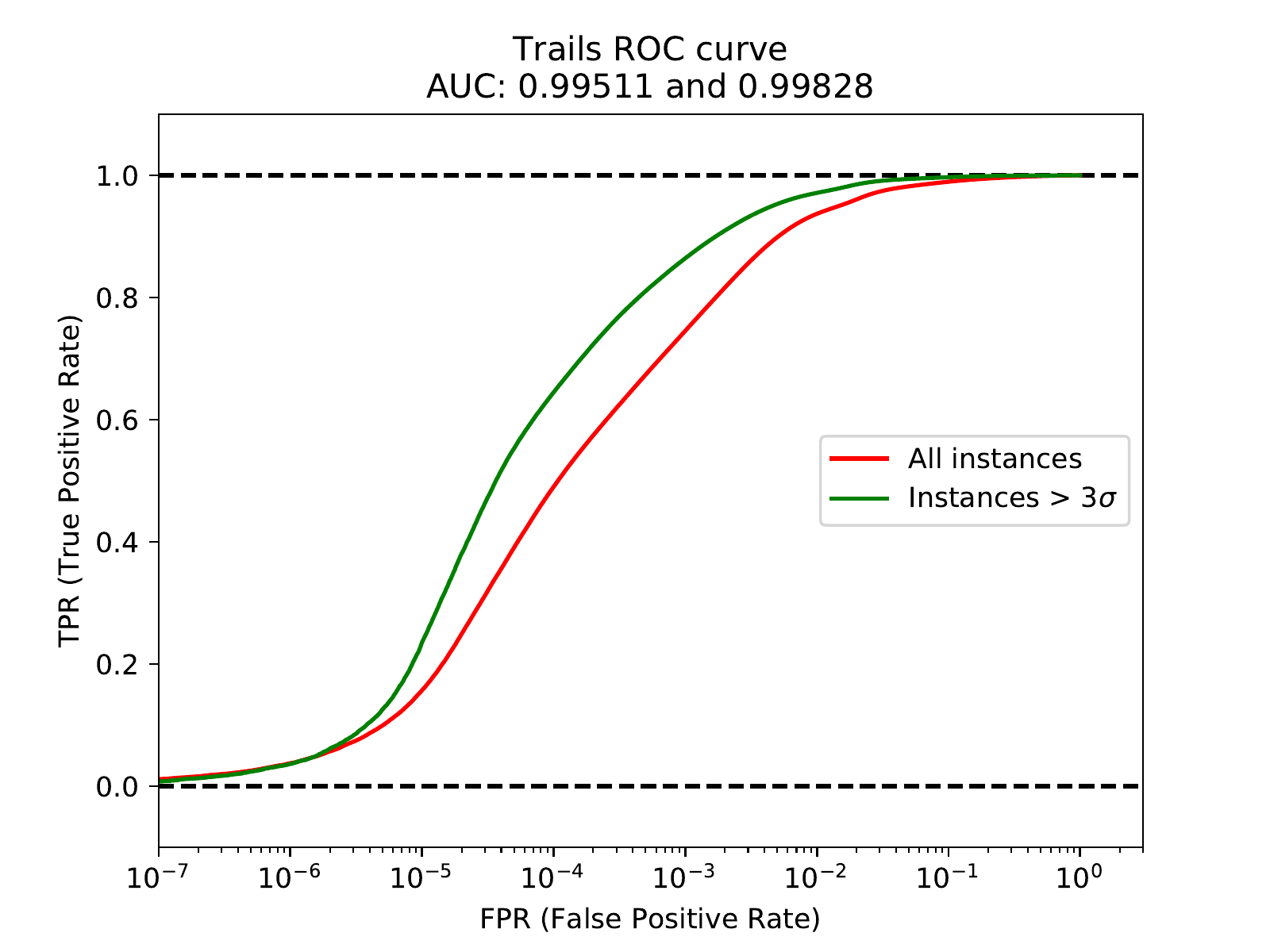}
    \end{minipage}
    \begin{minipage}{0.48\linewidth}
      \includegraphics[scale=0.50]{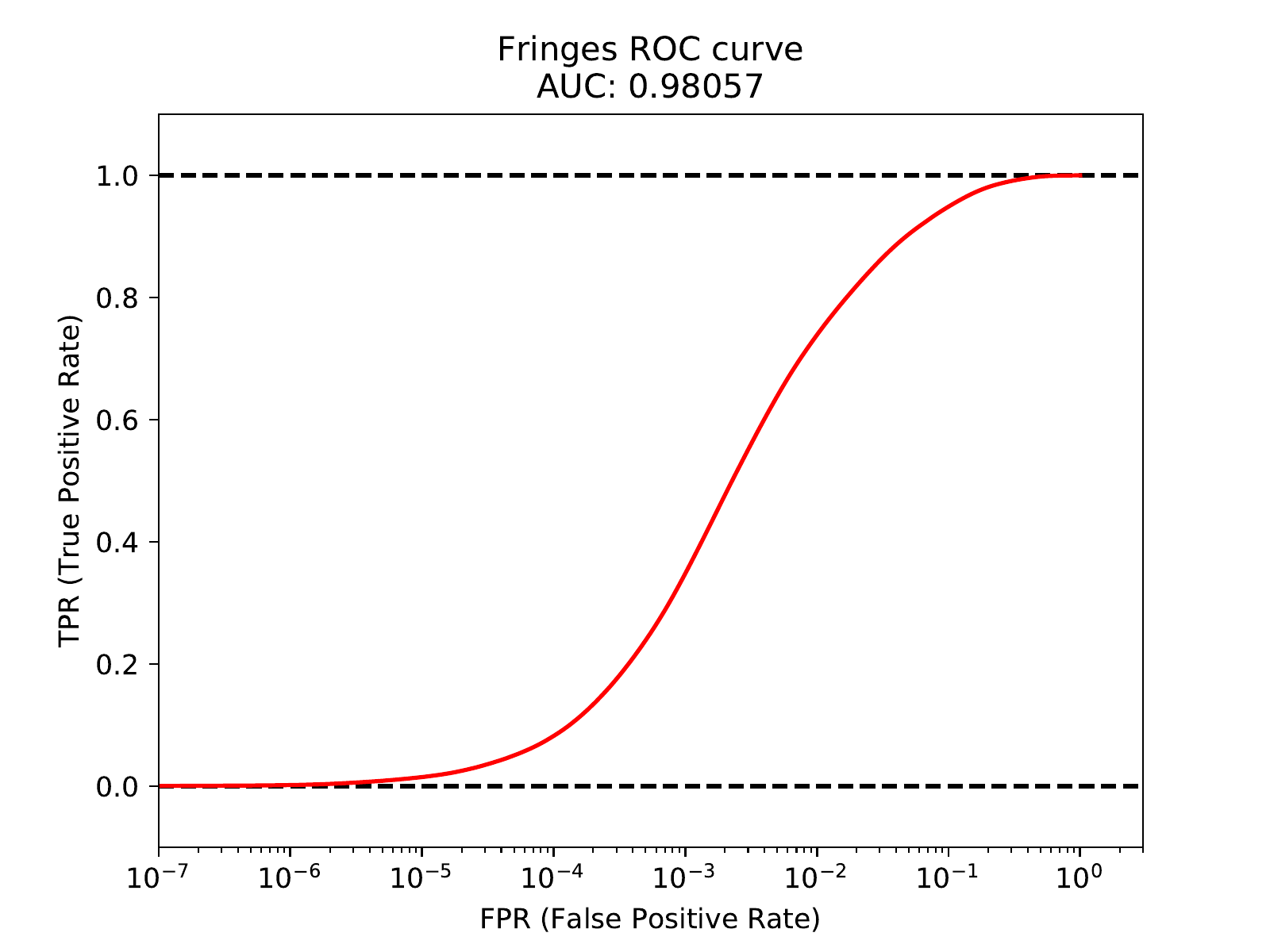}
    \end{minipage} \\
  \end{figure*}
  
  \begin{figure*}[ht]
    \begin{minipage}{0.48\linewidth}
      \includegraphics[scale=0.50]{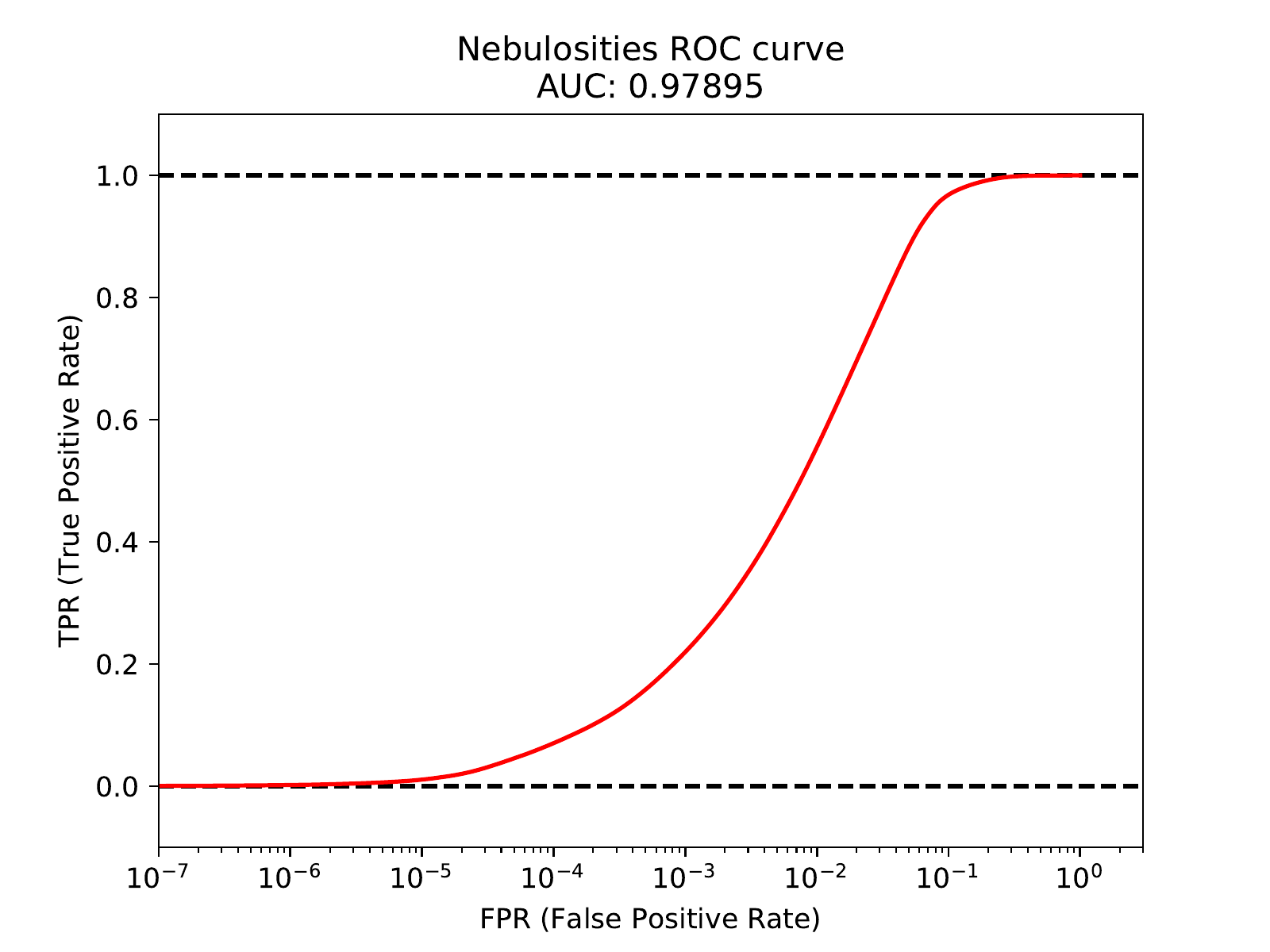}
    \end{minipage}
    \begin{minipage}{0.48\linewidth}
      \includegraphics[scale=0.50]{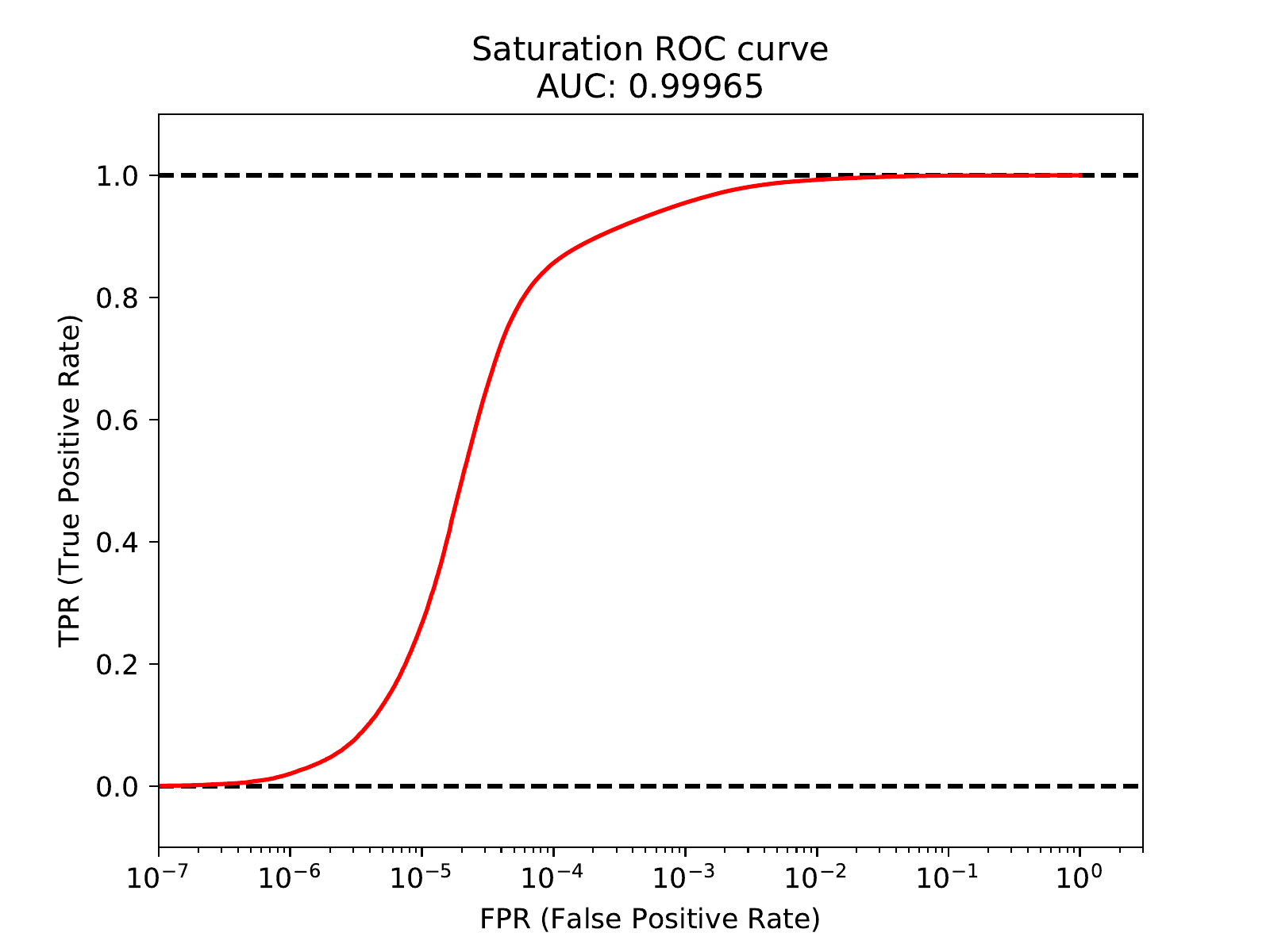}
    \end{minipage} \\
    \begin{minipage}{0.48\linewidth}
      \includegraphics[scale=0.50]{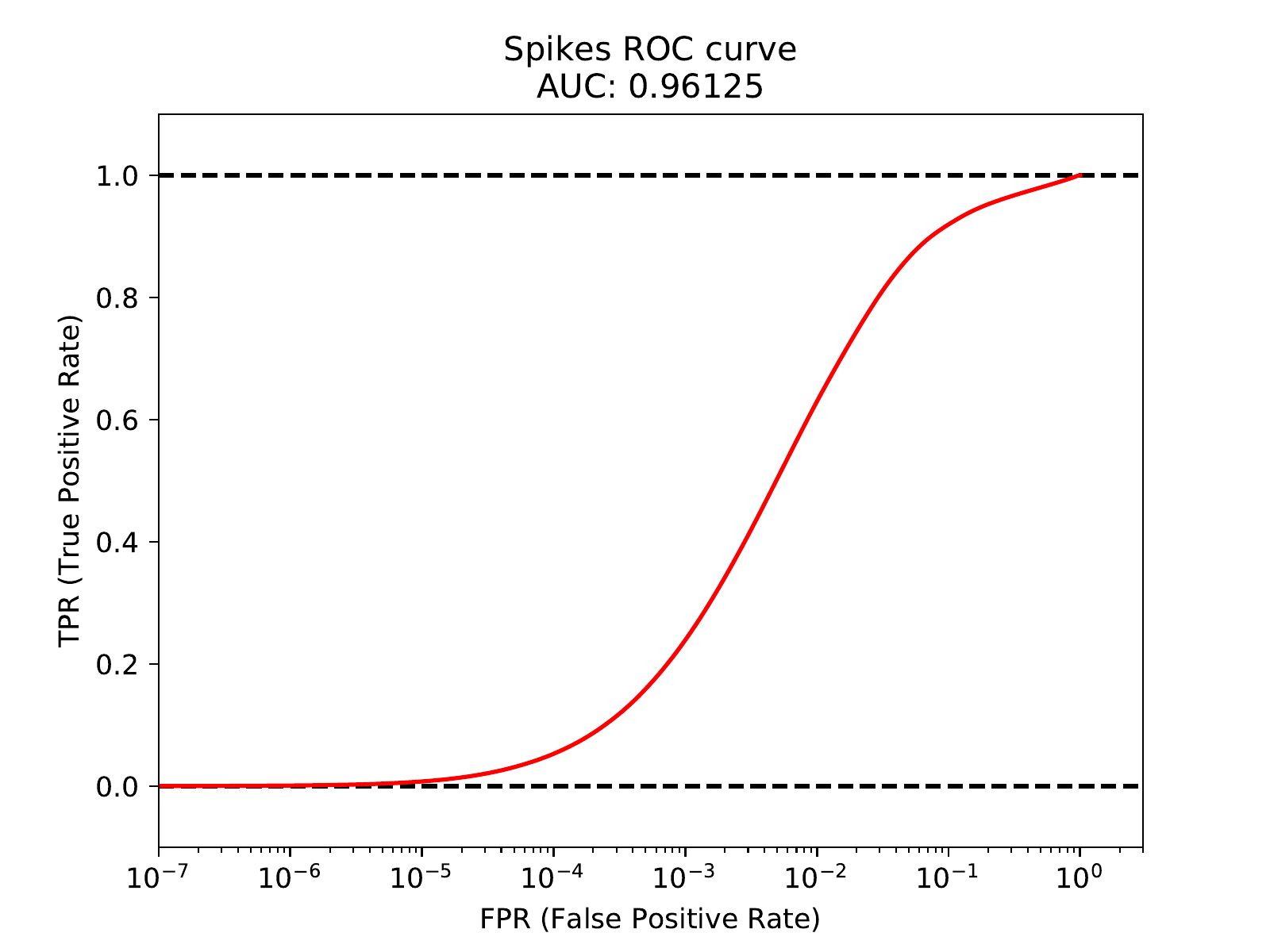}
    \end{minipage}
    \begin{minipage}{0.48\linewidth}
      \includegraphics[scale=0.50]{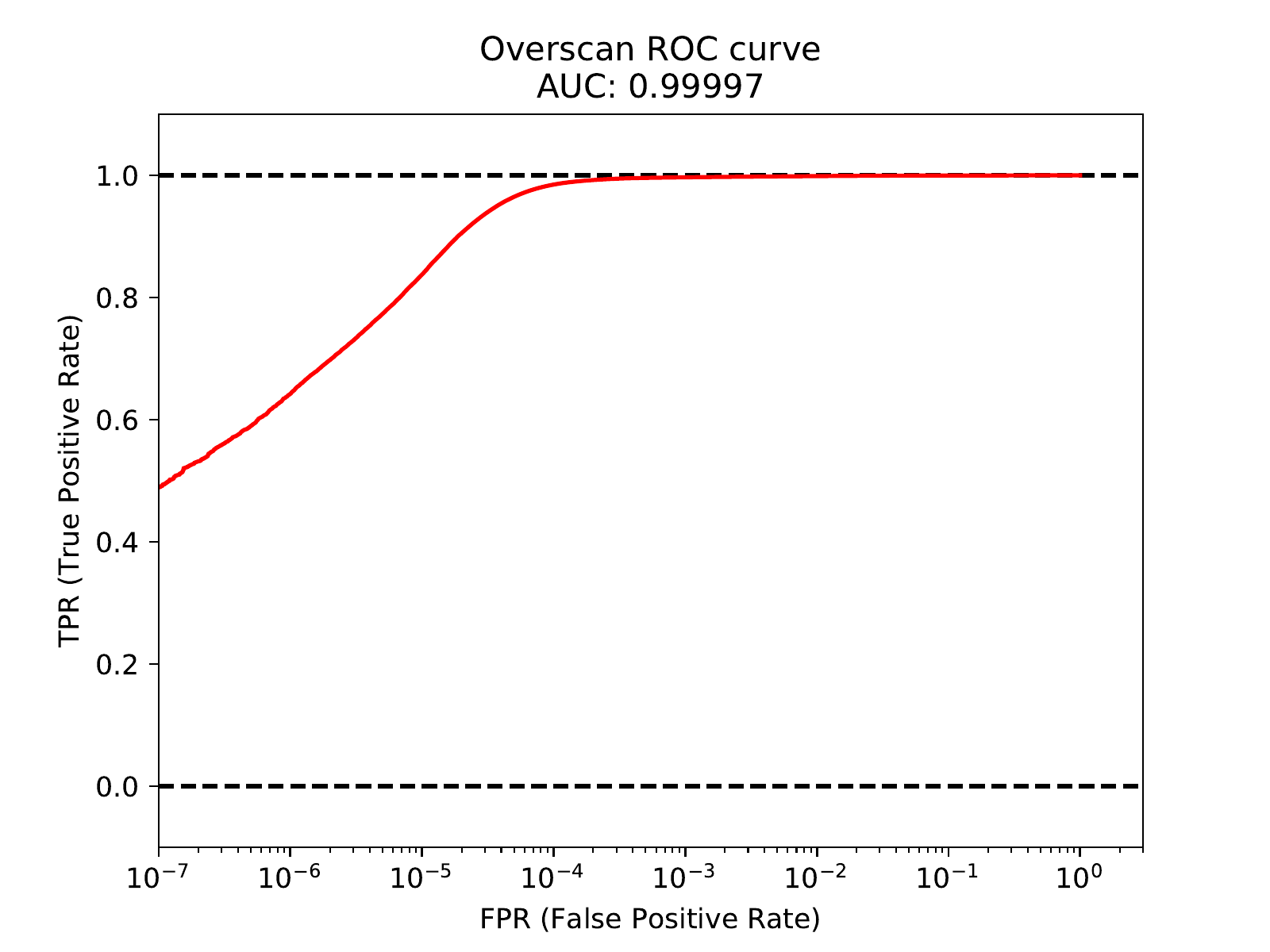}
    \end{minipage} \\
    \begin{minipage}{0.48\linewidth}
      \includegraphics[scale=0.50]{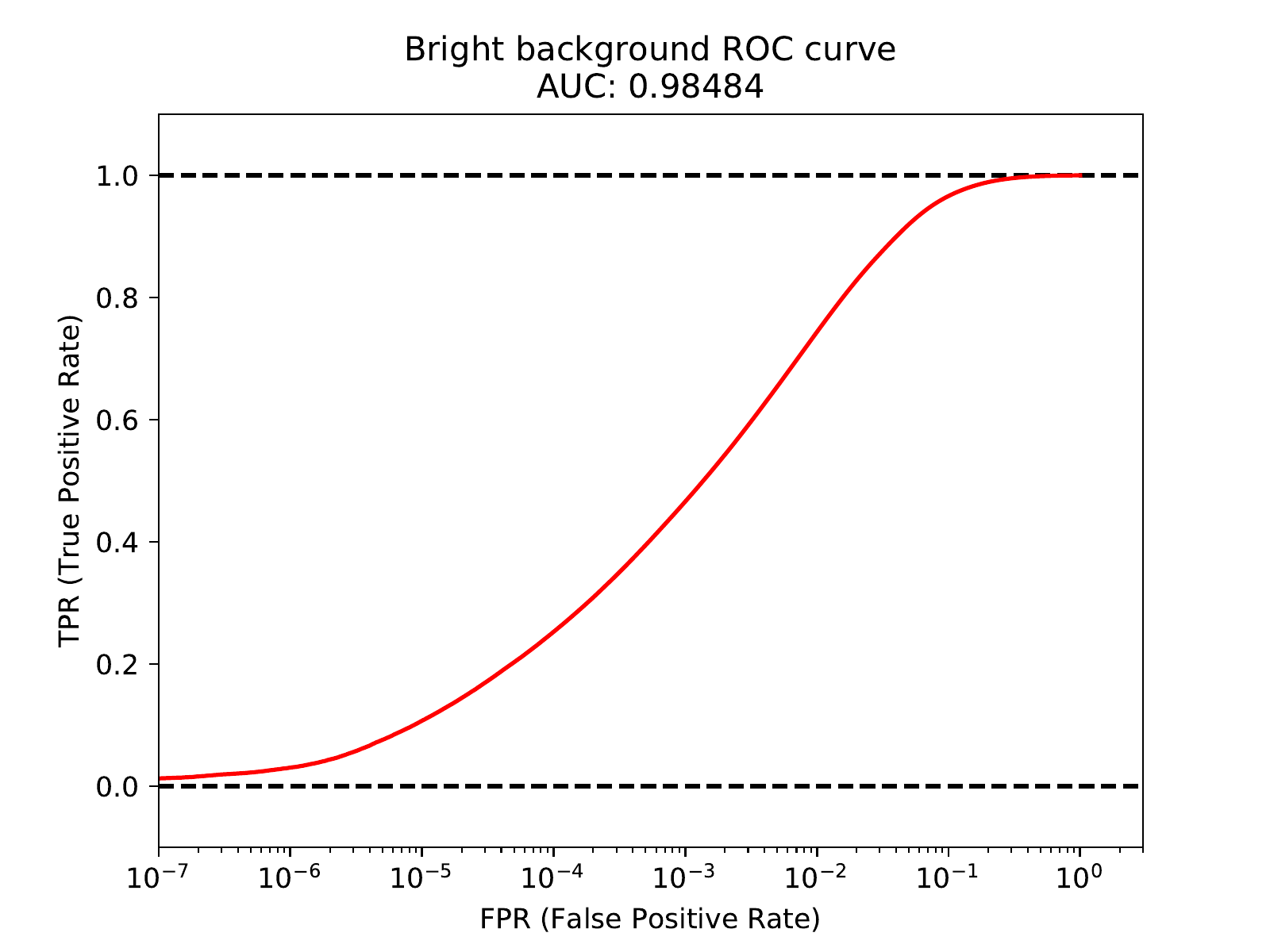}
    \end{minipage}
    \begin{minipage}{0.48\linewidth}
      \includegraphics[scale=0.50]{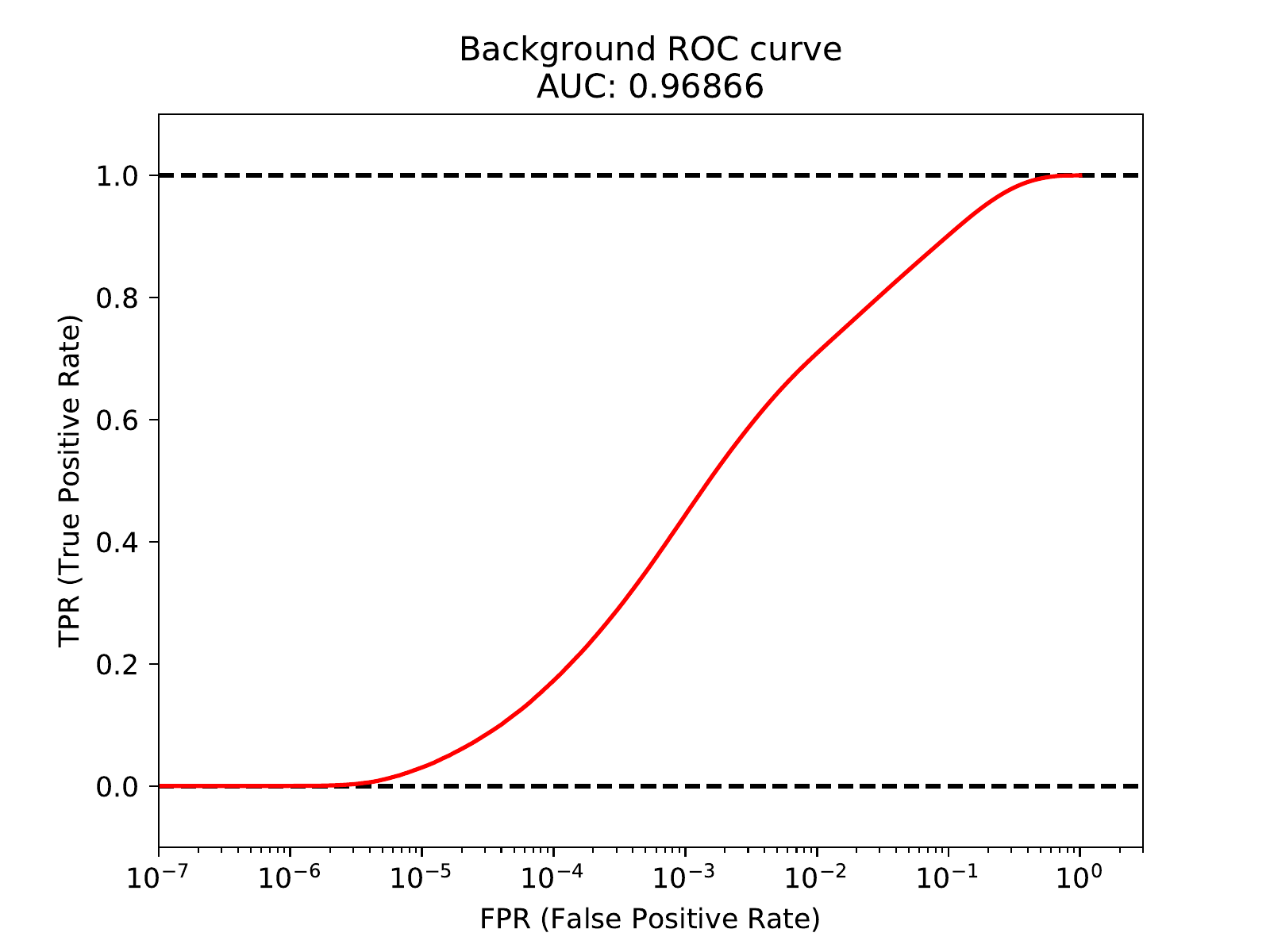}
    \end{minipage}
    \caption{ROC curves: $TPR$ vs $FPR$. The $FPR$ axis in in logarithmic scale so that very low $FPR$ are best visualized.The ROC curve and the AUC are provided for each class.}
    \label{roc}
  \end{figure*}

  \begin{figure*}[ht]
    \begin{minipage}{0.48\linewidth}
      \includegraphics[scale=0.50]{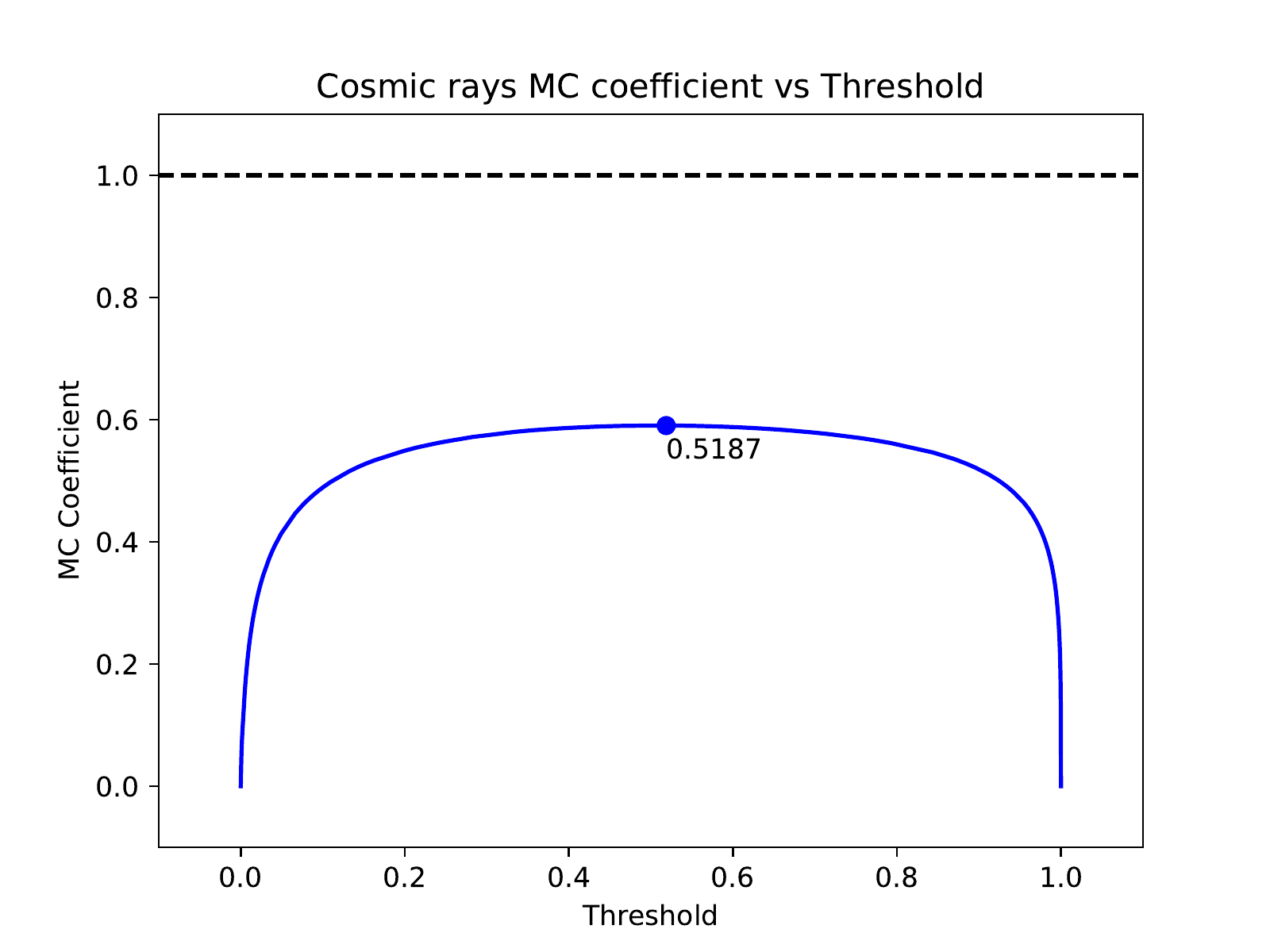}
    \end{minipage} 
    \begin{minipage}{0.48\linewidth}
      \includegraphics[scale=0.50]{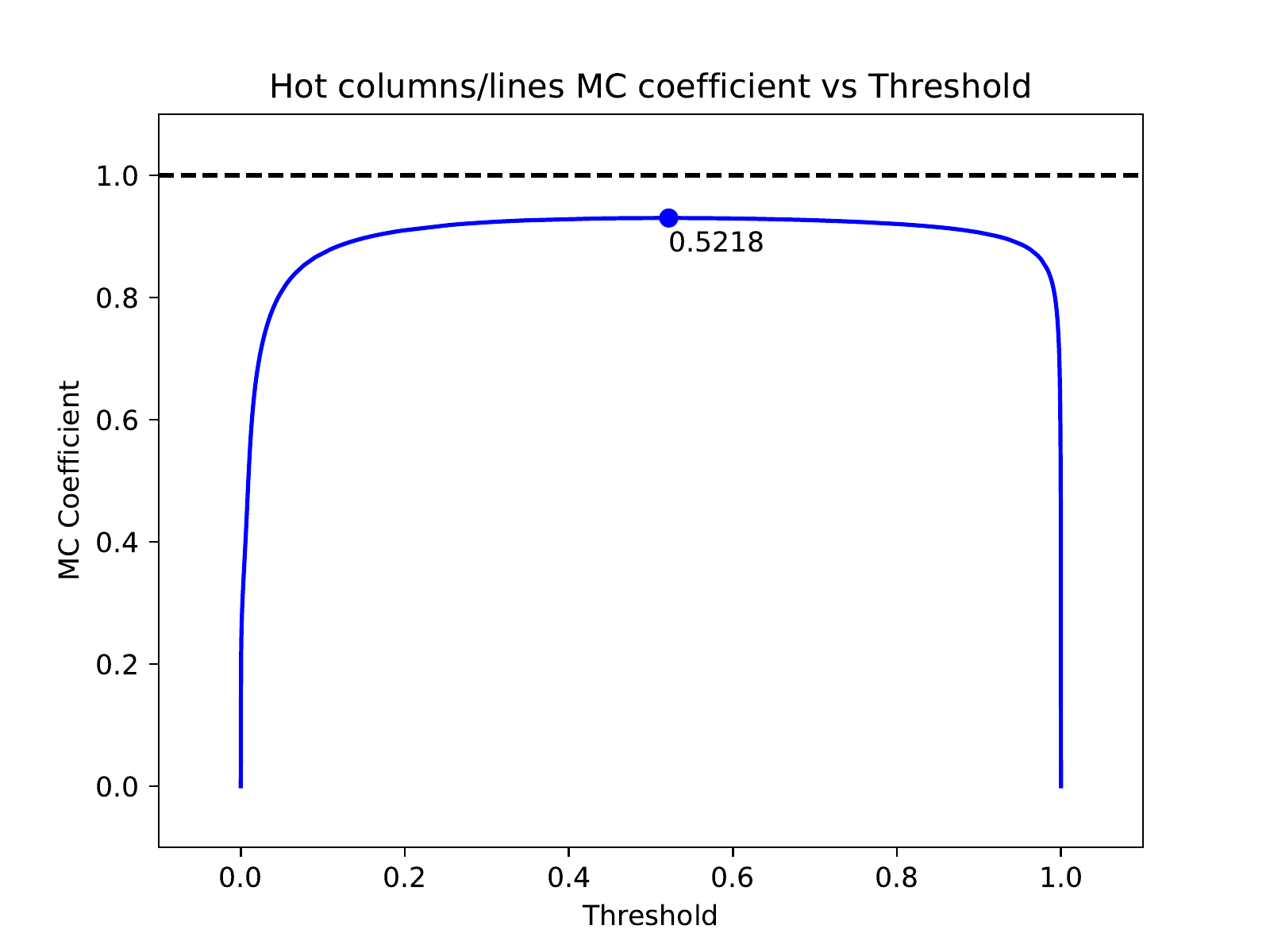}
    \end{minipage} \\
    \begin{minipage}{0.48\linewidth}
      \includegraphics[scale=0.50]{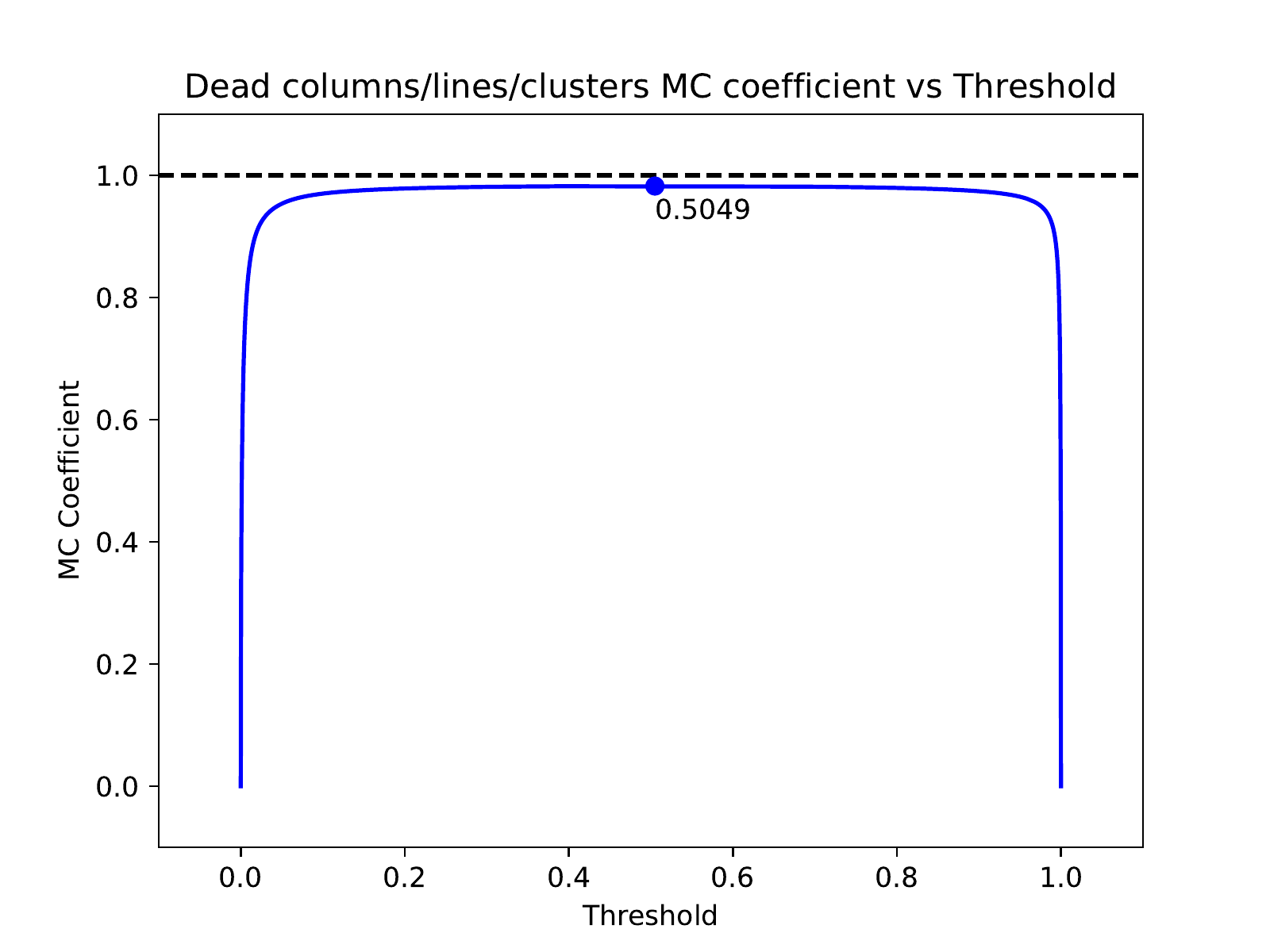}
    \end{minipage}
    \begin{minipage}{0.48\linewidth}
      \includegraphics[scale=0.50]{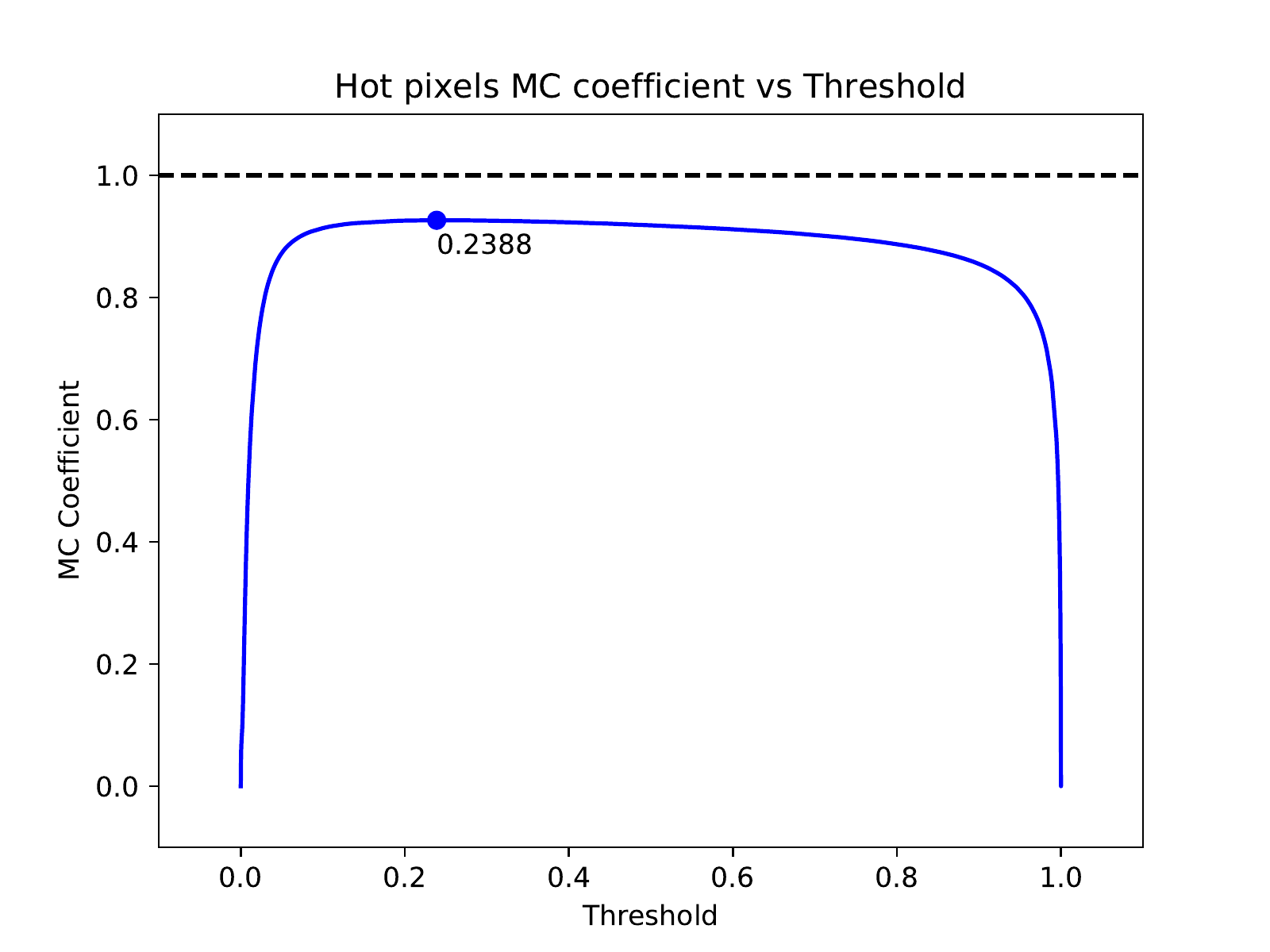}
    \end{minipage} \\
    \begin{minipage}{0.48\linewidth}
      \includegraphics[scale=0.50]{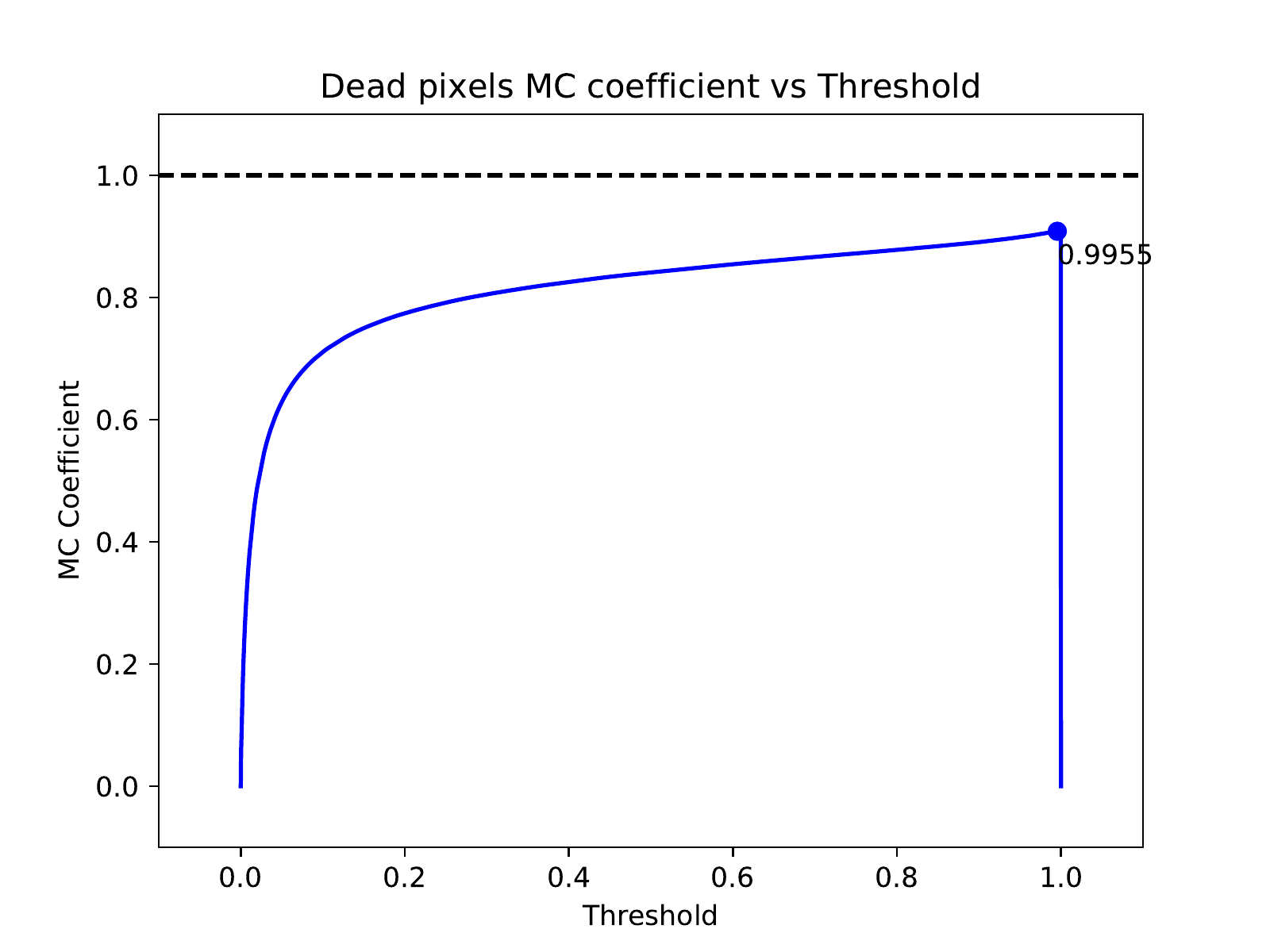}
    \end{minipage}
    \begin{minipage}{0.48\linewidth}
      \includegraphics[scale=0.50]{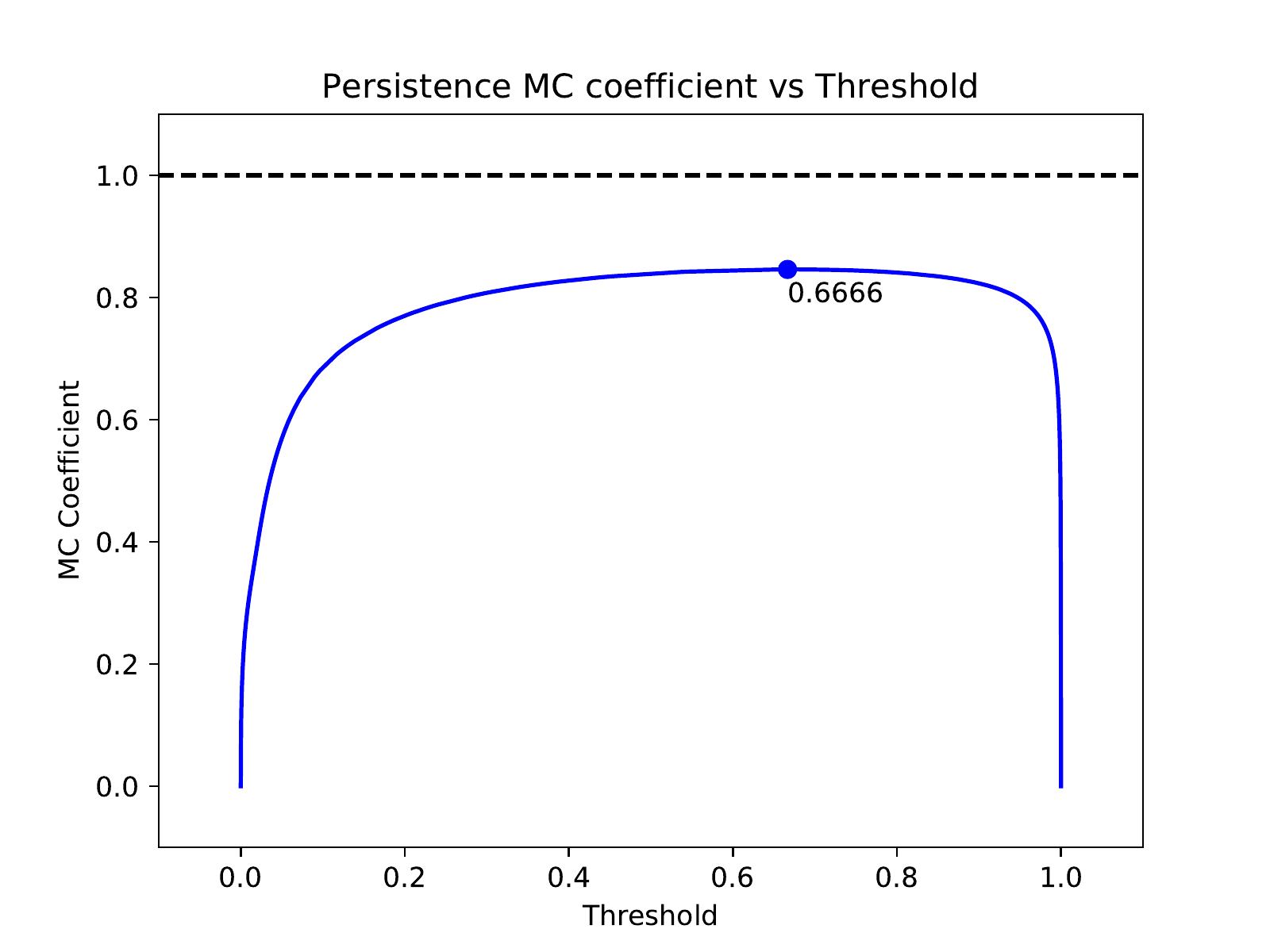}
    \end{minipage} \\
    \begin{minipage}{0.48\linewidth}
      \includegraphics[scale=0.50]{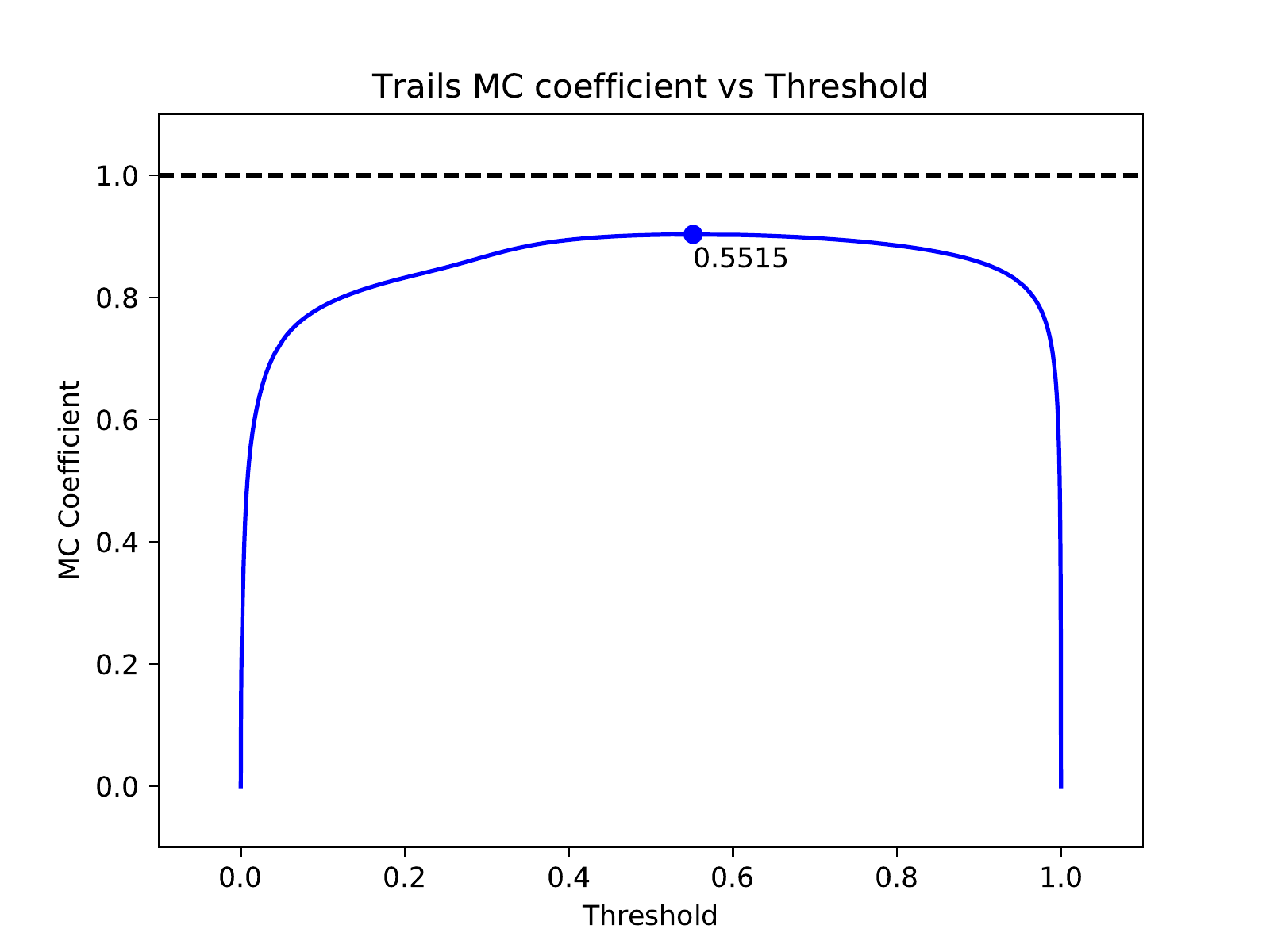}
    \end{minipage}
    \begin{minipage}{0.48\linewidth}
      \includegraphics[scale=0.50]{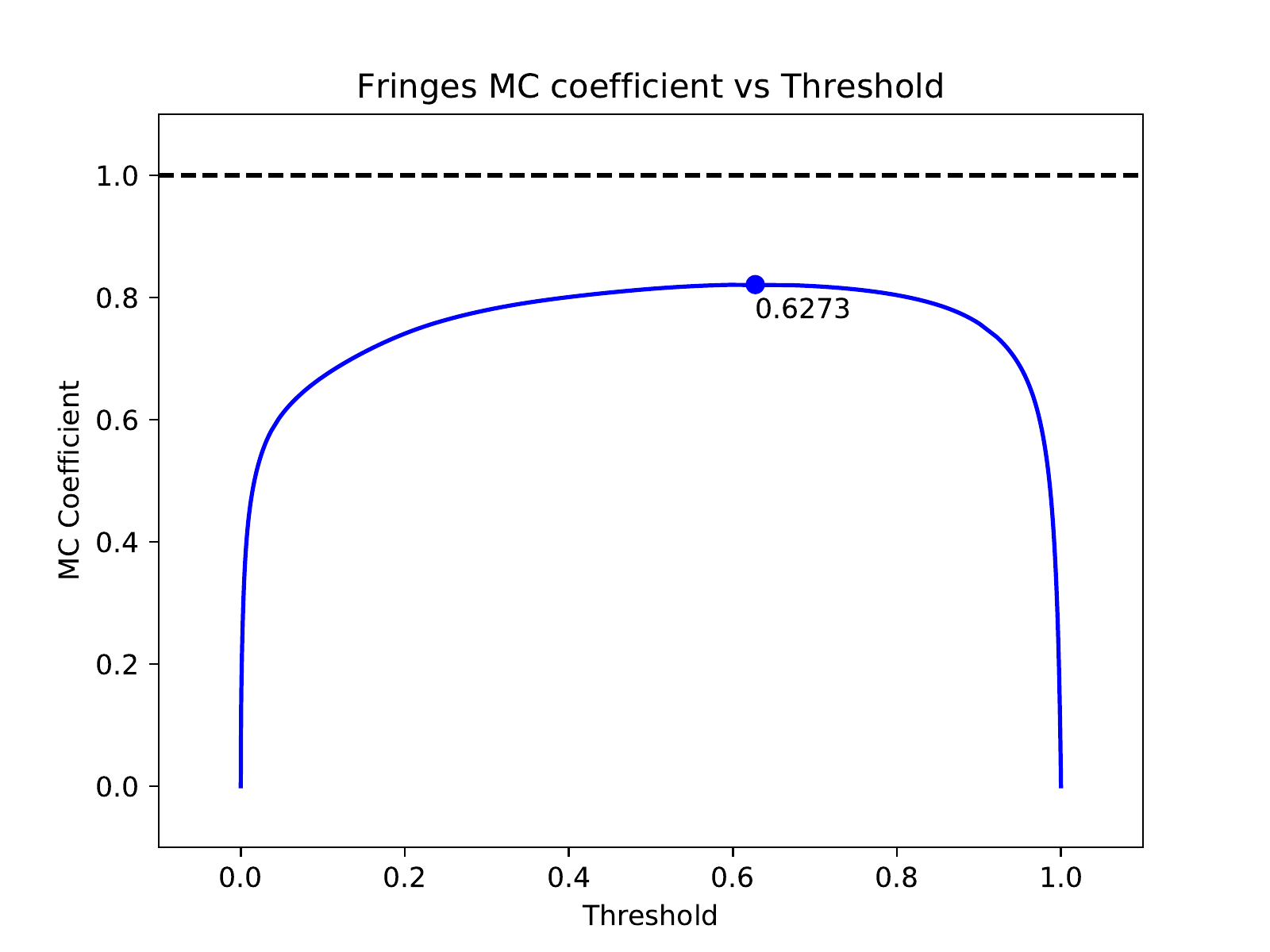}
    \end{minipage} \\
  \end{figure*}
  
  \begin{figure*}[ht]
    \begin{minipage}{0.48\linewidth}
      \includegraphics[scale=0.50]{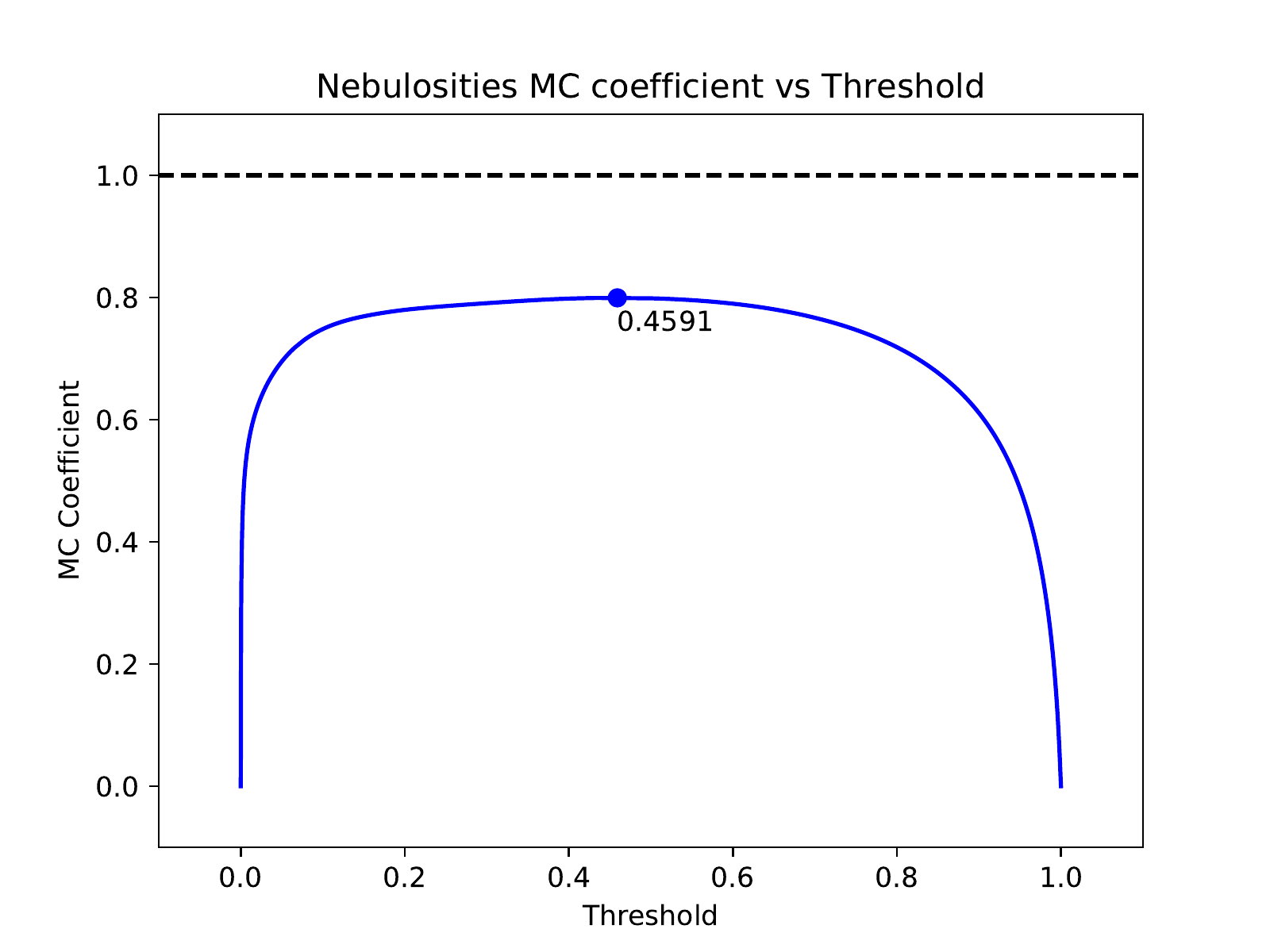}
    \end{minipage}
    \begin{minipage}{0.48\linewidth}
      \includegraphics[scale=0.50]{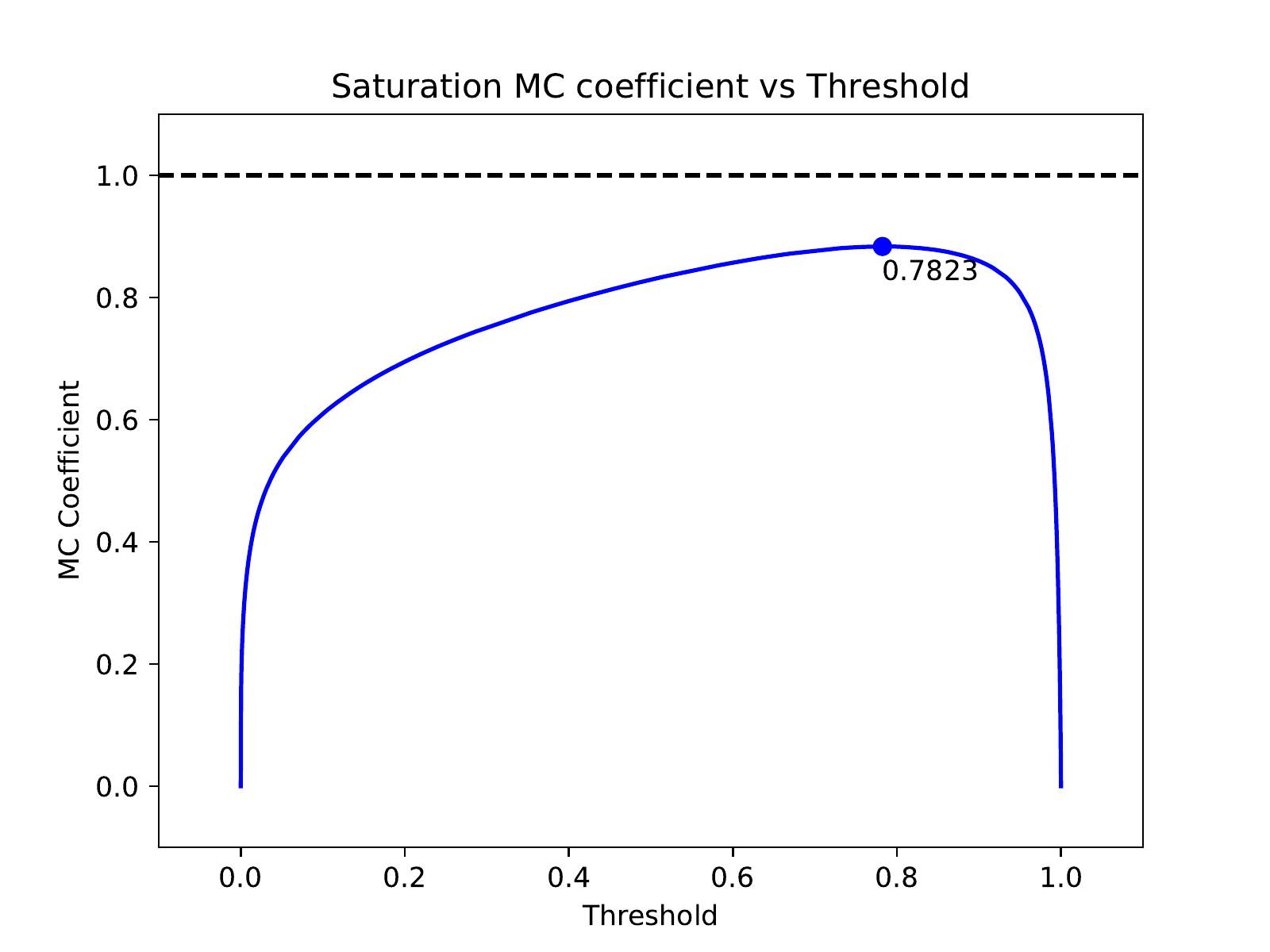}
    \end{minipage} \\
    \begin{minipage}{0.48\linewidth}
      \includegraphics[scale=0.50]{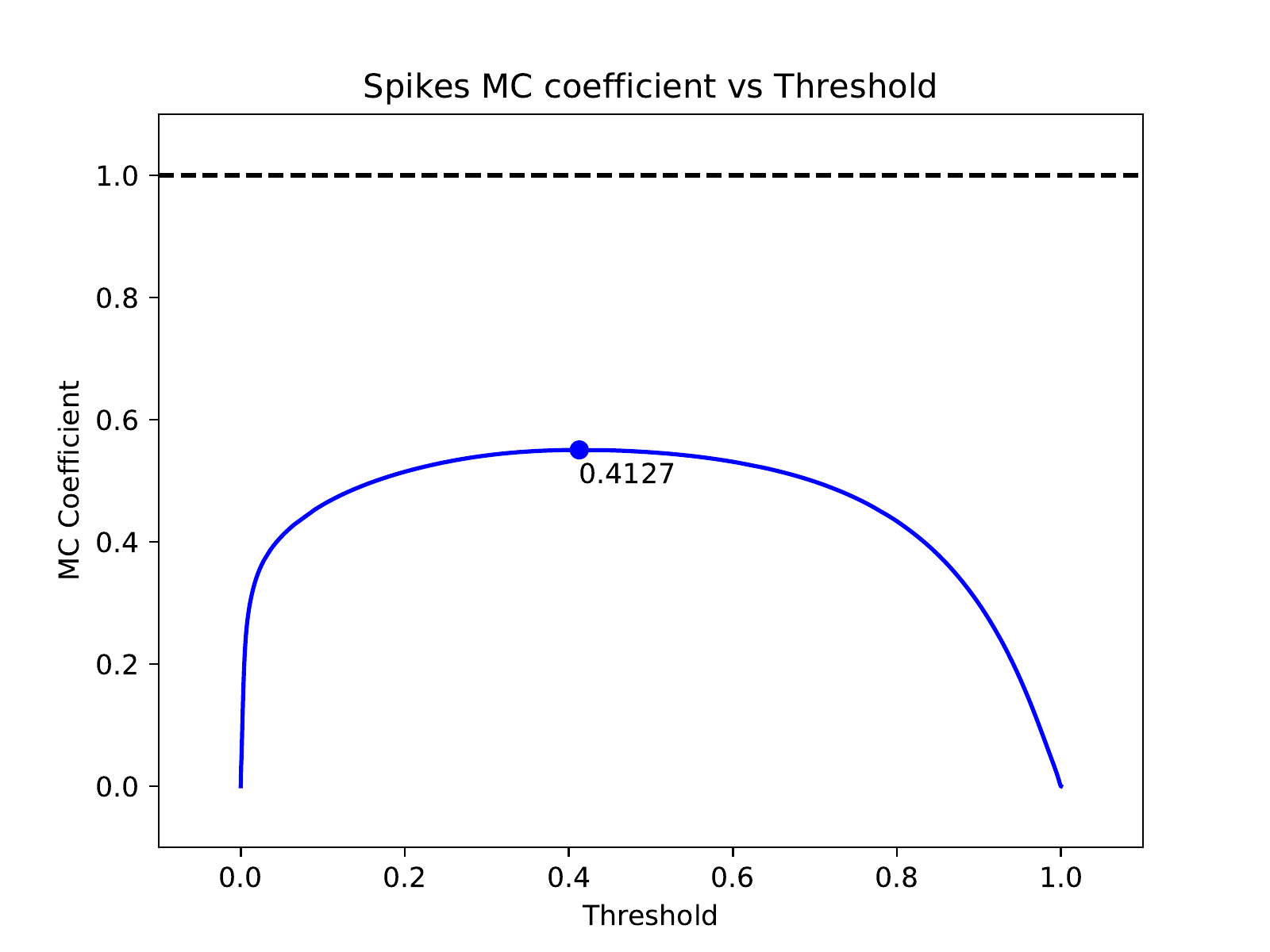}
    \end{minipage}
    \begin{minipage}{0.48\linewidth}
      \includegraphics[scale=0.50]{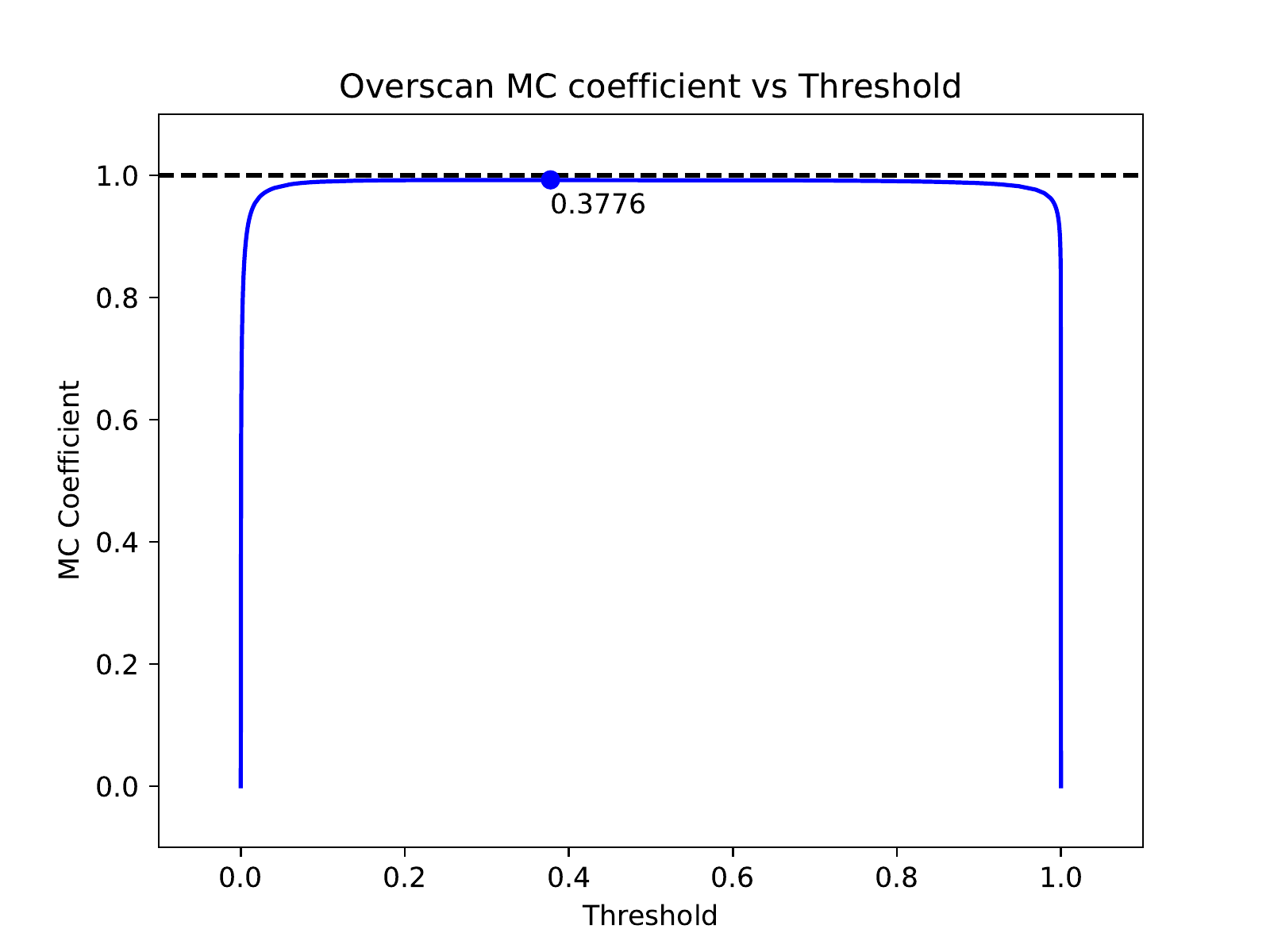}
    \end{minipage} \\
    \begin{minipage}{0.48\linewidth}
      \includegraphics[scale=0.50]{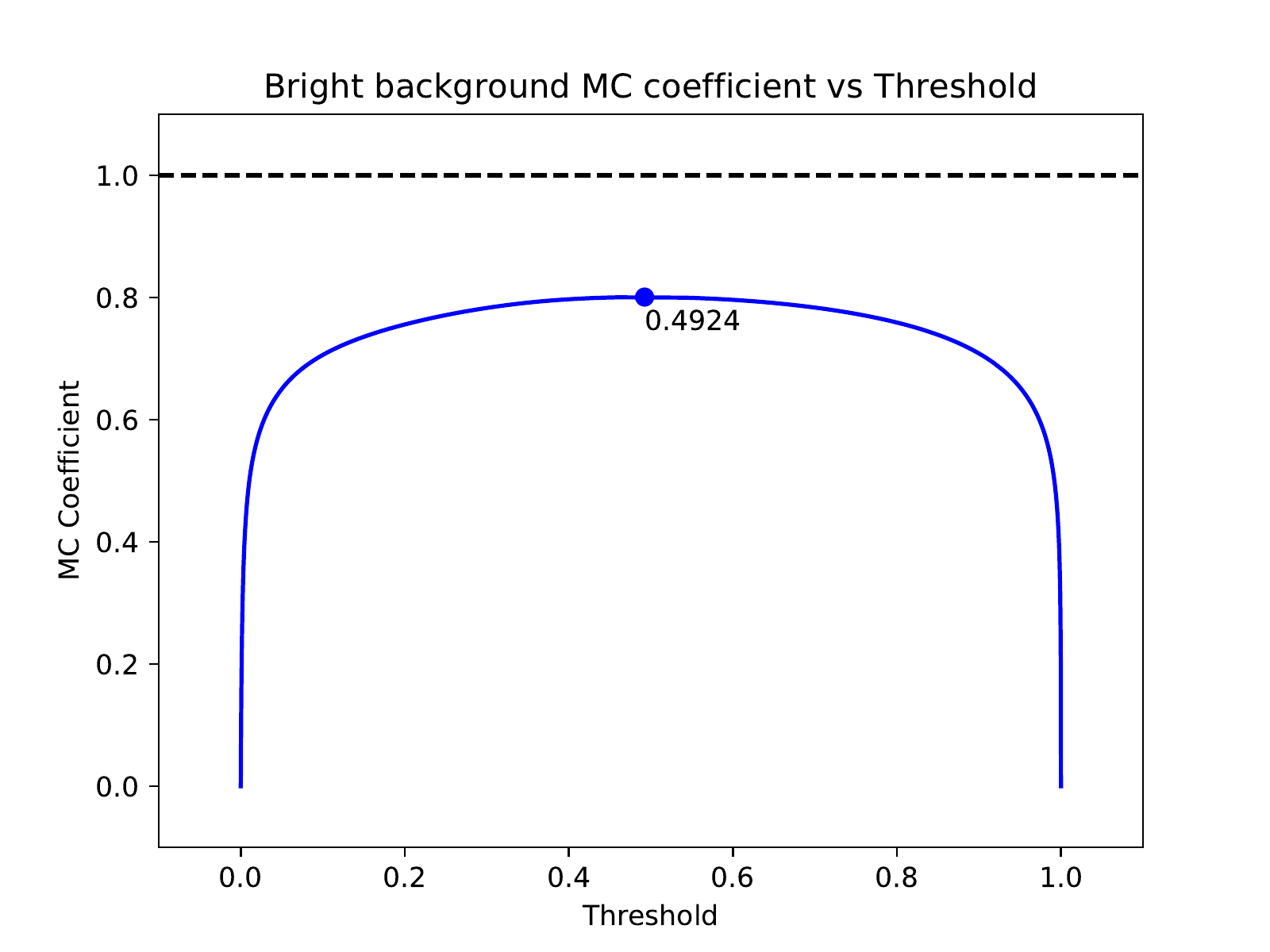}
    \end{minipage}
    \begin{minipage}{0.48\linewidth}
      \includegraphics[scale=0.50]{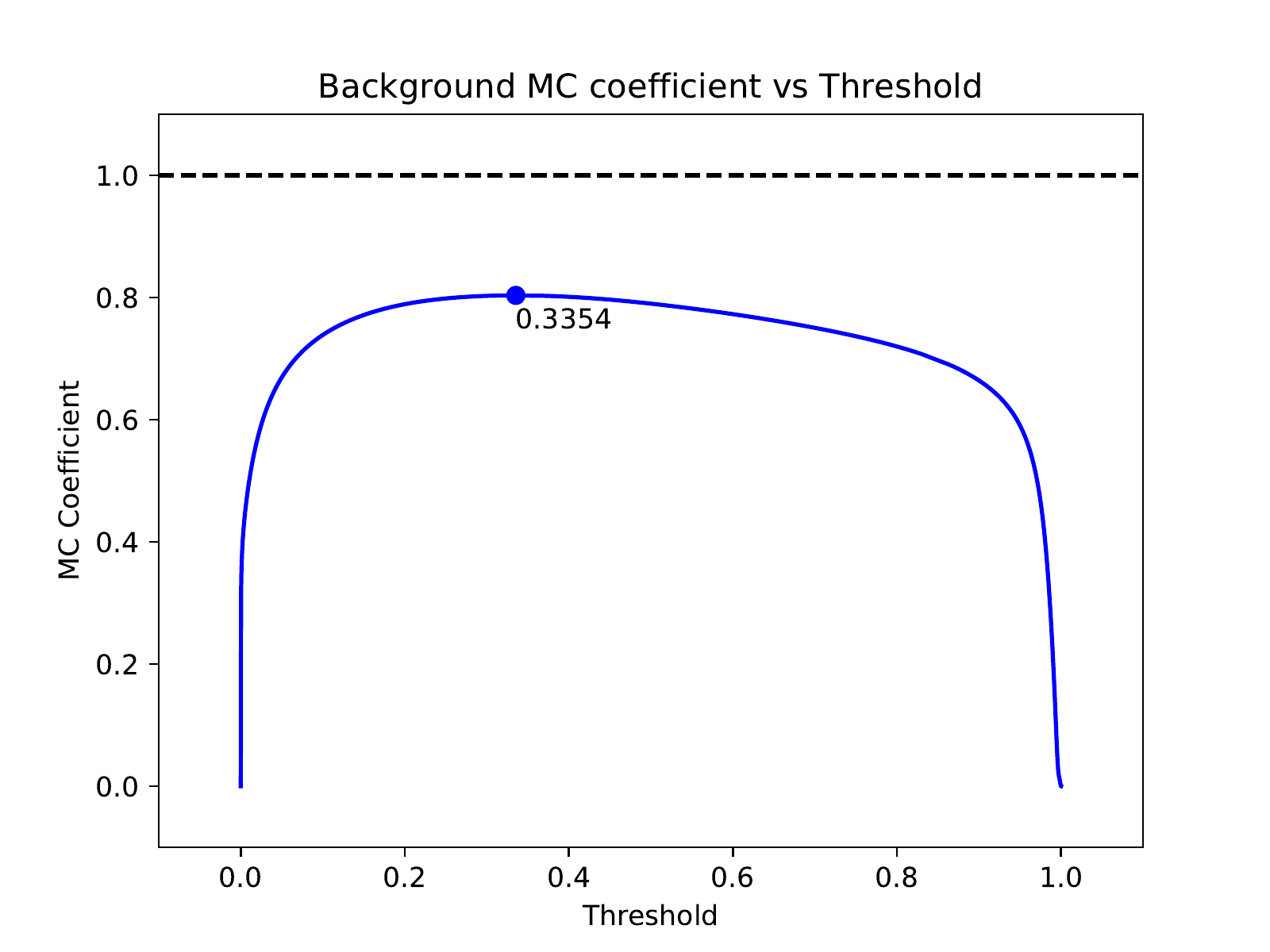}
    \end{minipage}
    \caption{MC coefficient curves: MC coefficient vs Detection threshold. On each curve is annotated the threshold for which the MC coefficient is the highest. These curves were computed using the probabilities corrected from priors using empirical training priors.}
    \label{mcc}
  \end{figure*}

  \begin{figure*}[ht]
    \begin{minipage}{0.48\linewidth}
      \includegraphics[scale=0.50]{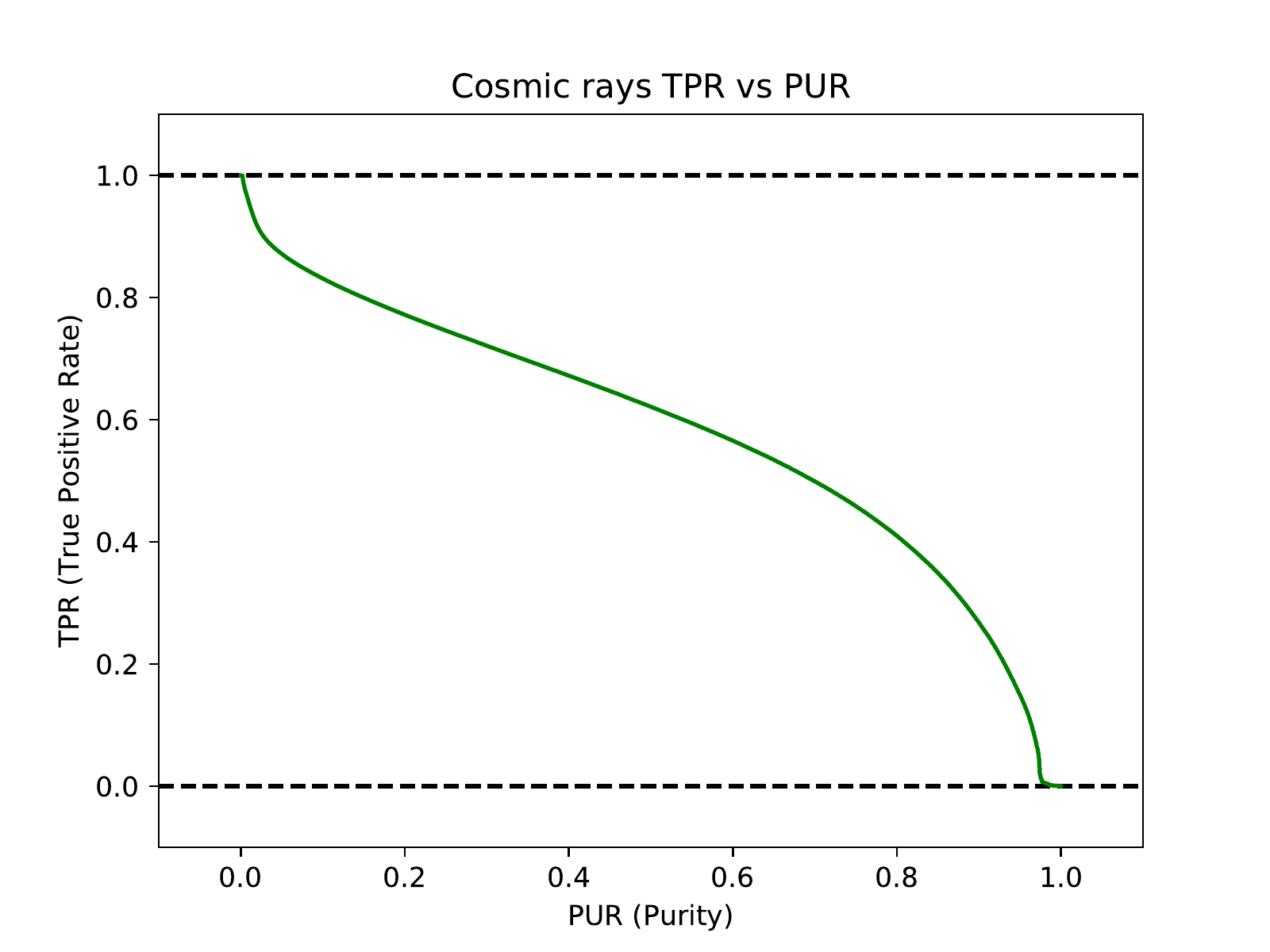}
    \end{minipage} 
    \begin{minipage}{0.48\linewidth}
      \includegraphics[scale=0.50]{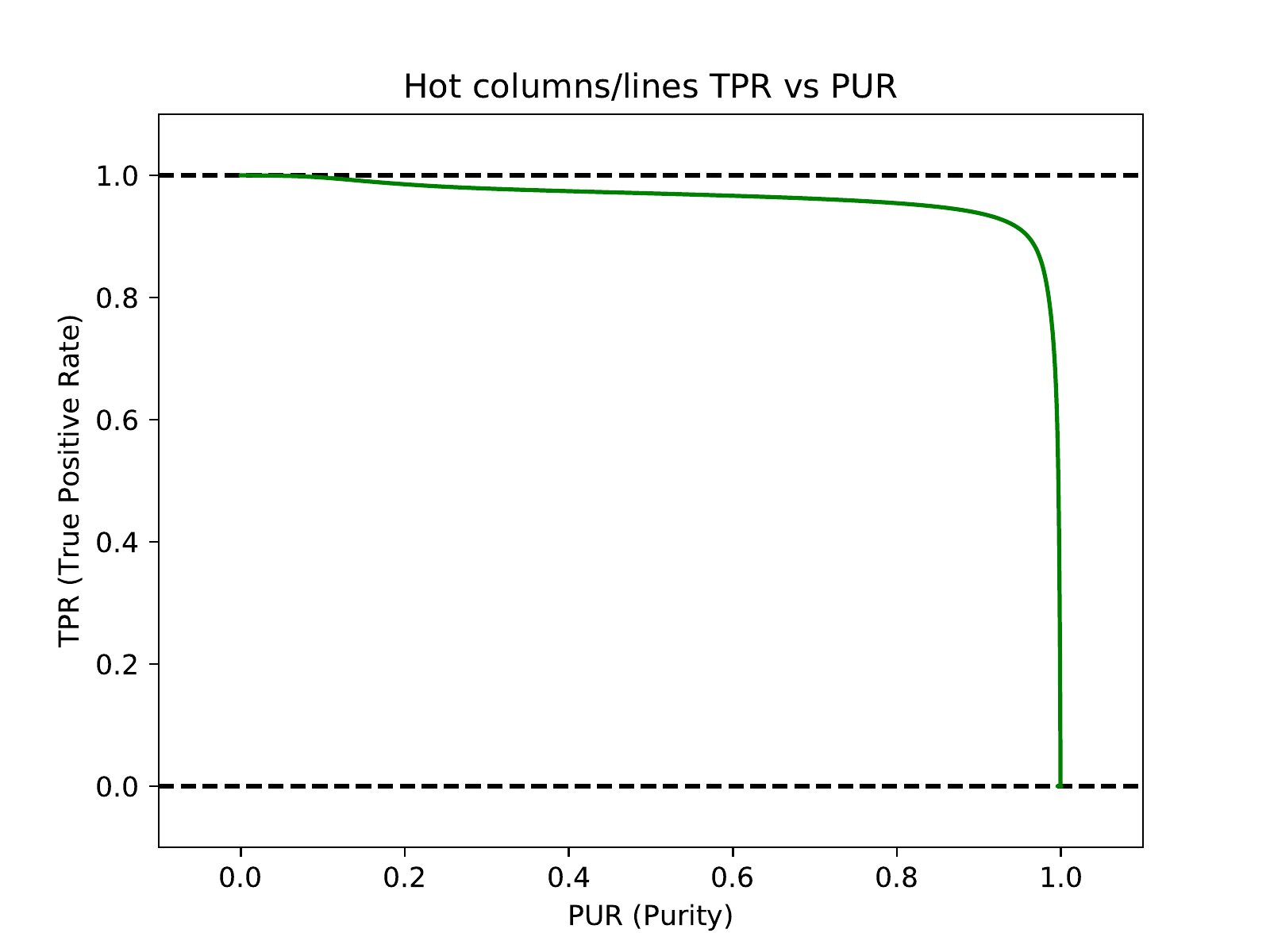}
    \end{minipage} \\
    \begin{minipage}{0.48\linewidth}
      \includegraphics[scale=0.50]{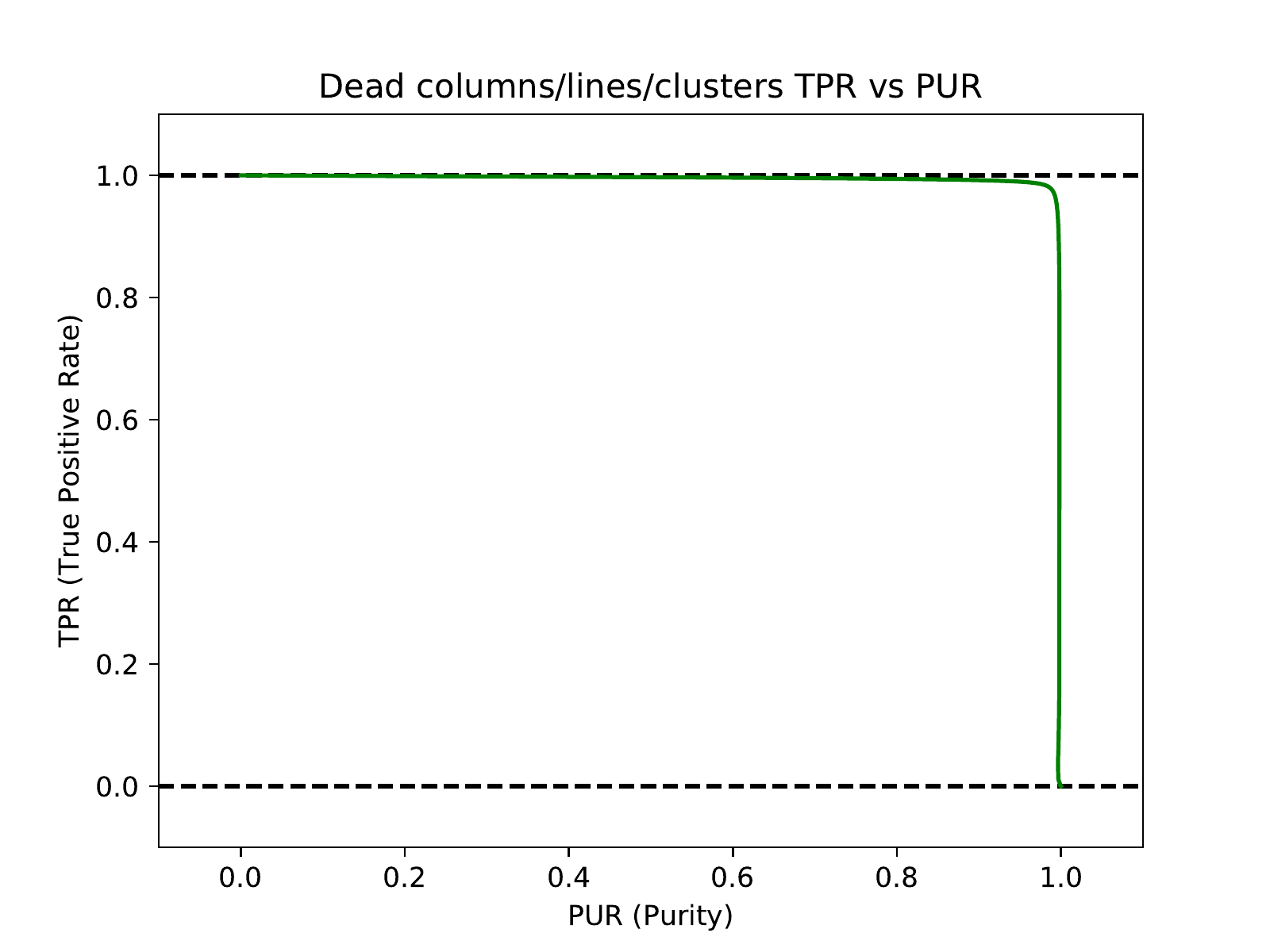}
    \end{minipage}
    \begin{minipage}{0.48\linewidth}
      \includegraphics[scale=0.50]{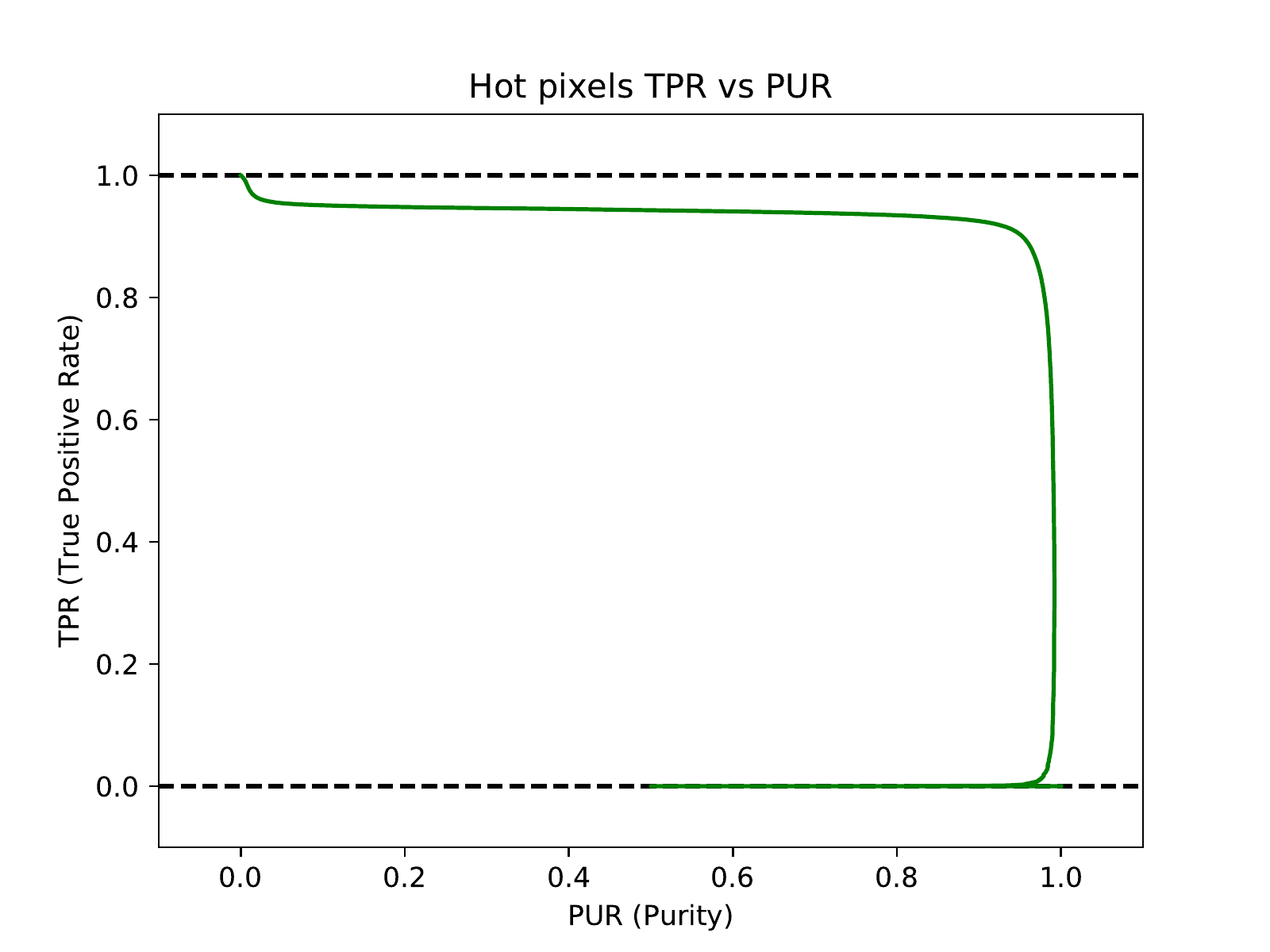}
    \end{minipage} \\
    \begin{minipage}{0.48\linewidth}
      \includegraphics[scale=0.50]{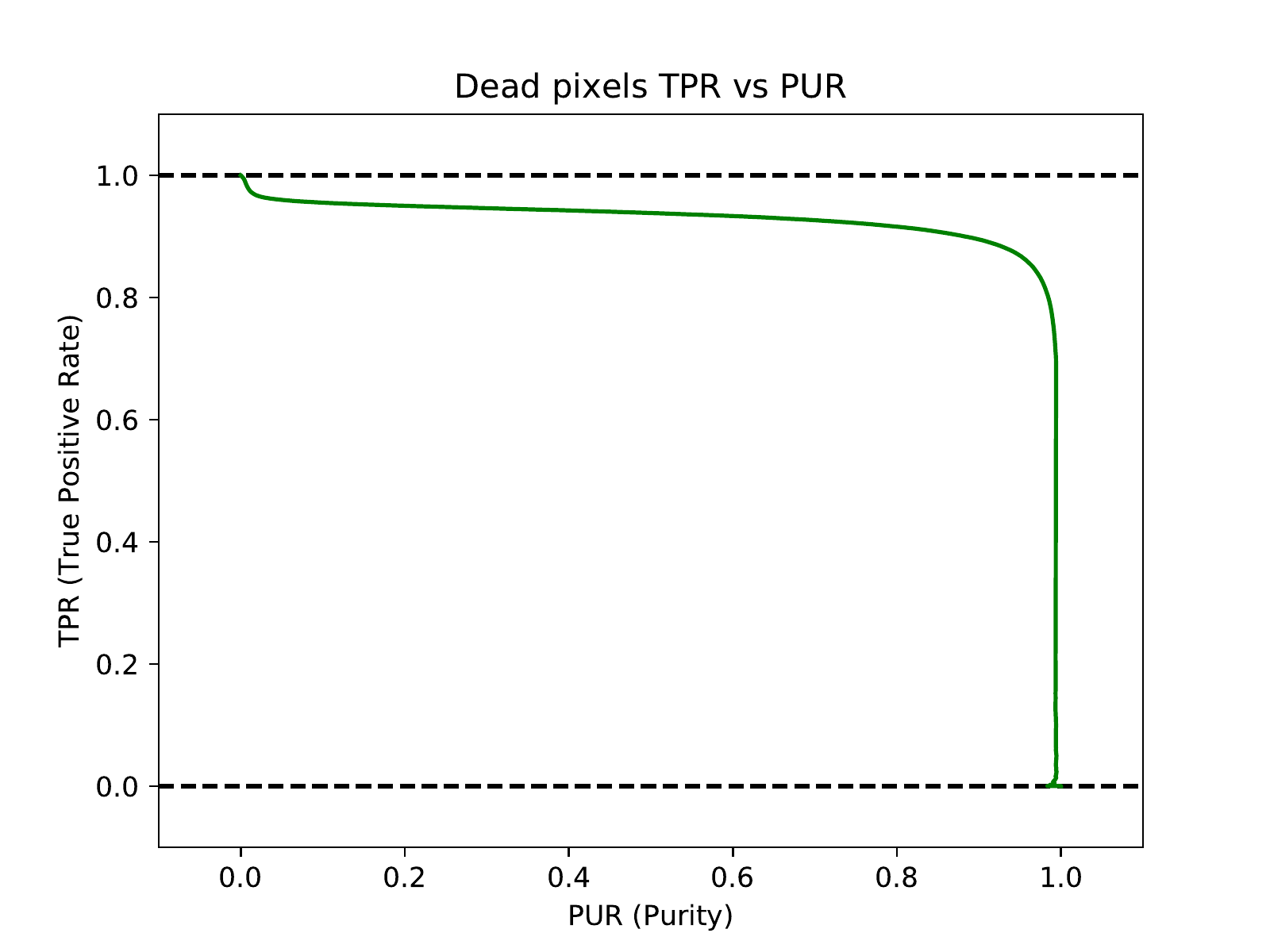}
    \end{minipage}
    \begin{minipage}{0.48\linewidth}
      \includegraphics[scale=0.50]{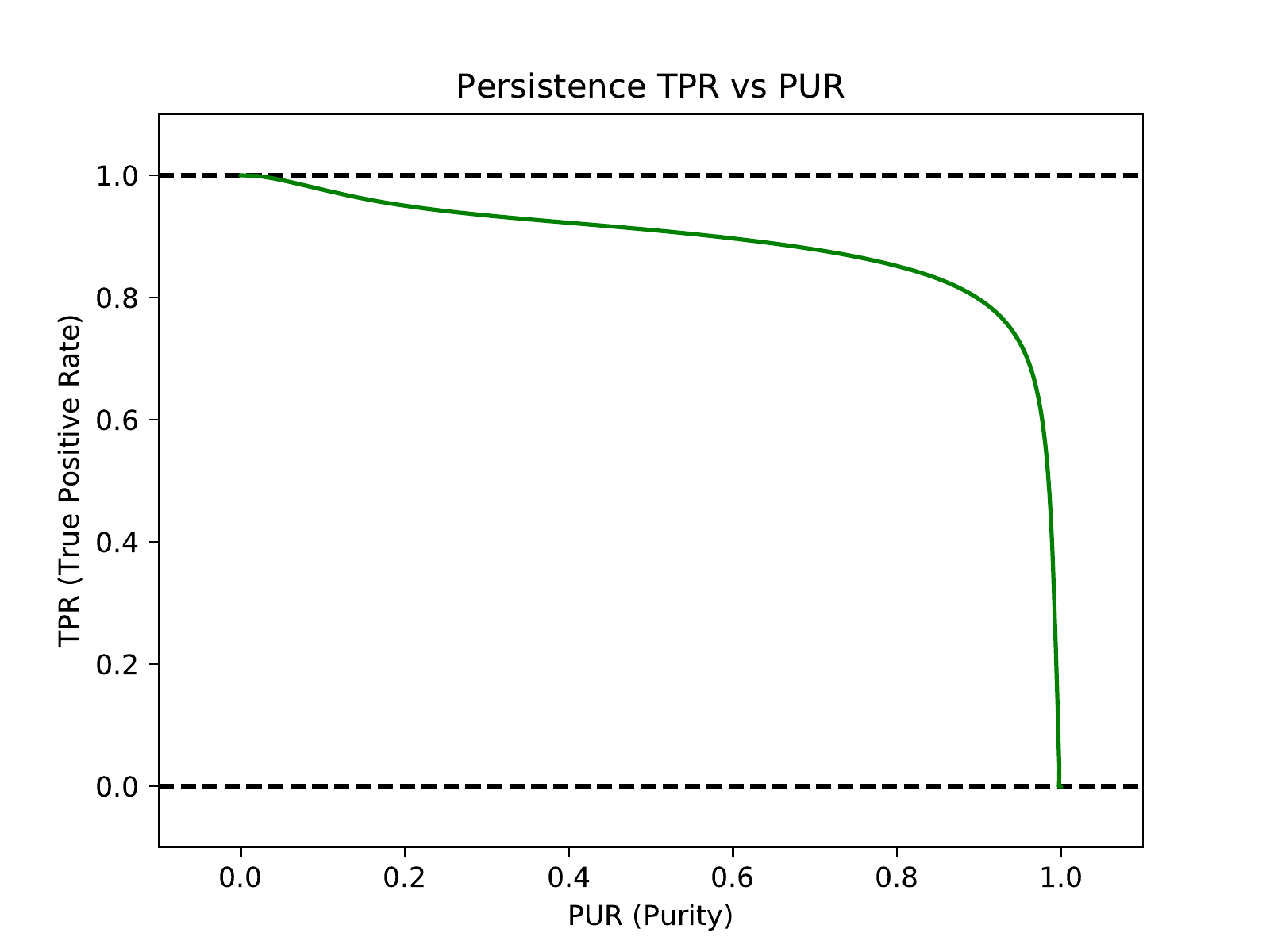}
    \end{minipage} \\
    \begin{minipage}{0.48\linewidth}
      \includegraphics[scale=0.50]{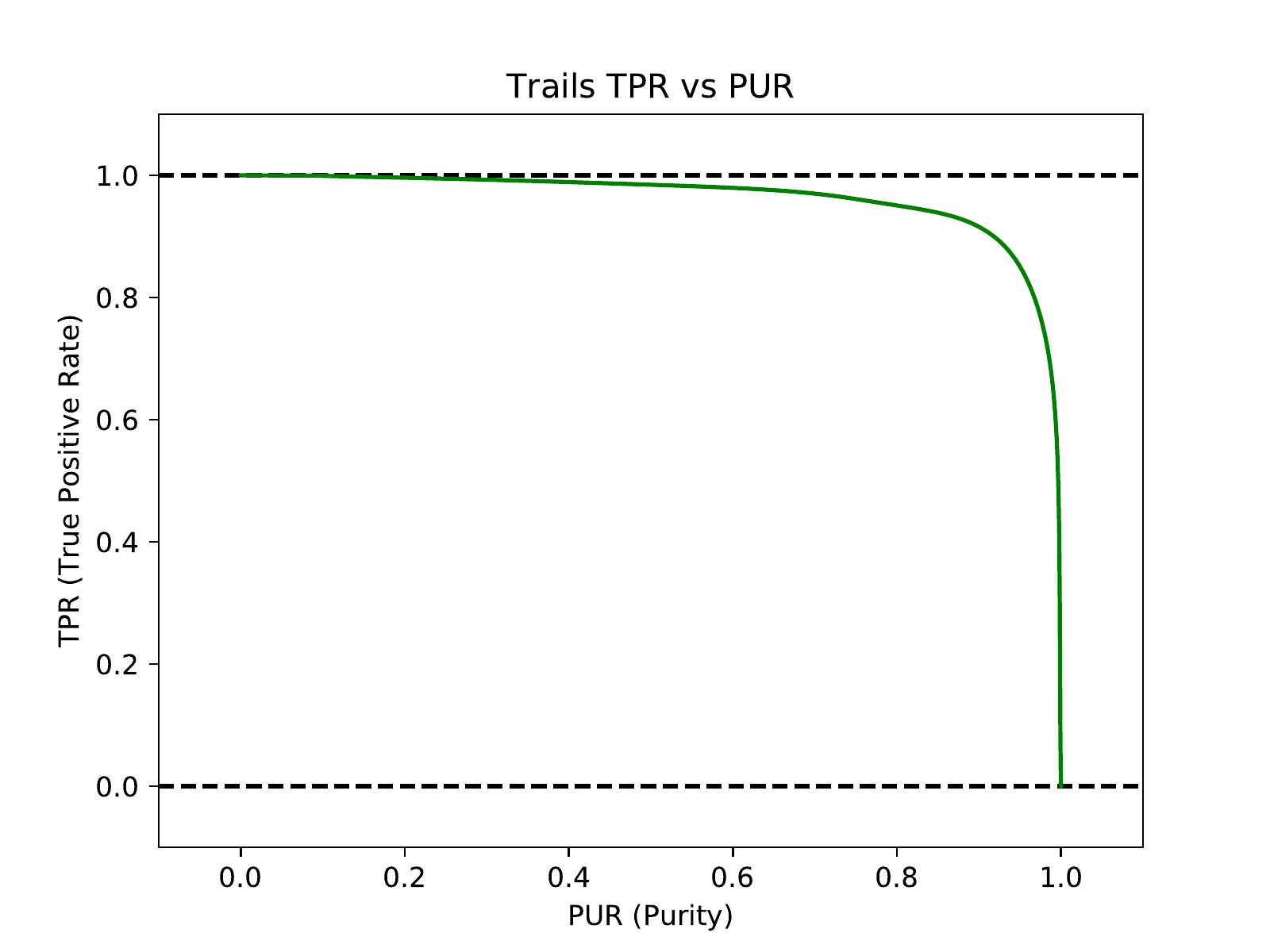}
    \end{minipage}
    \begin{minipage}{0.48\linewidth}
      \includegraphics[scale=0.50]{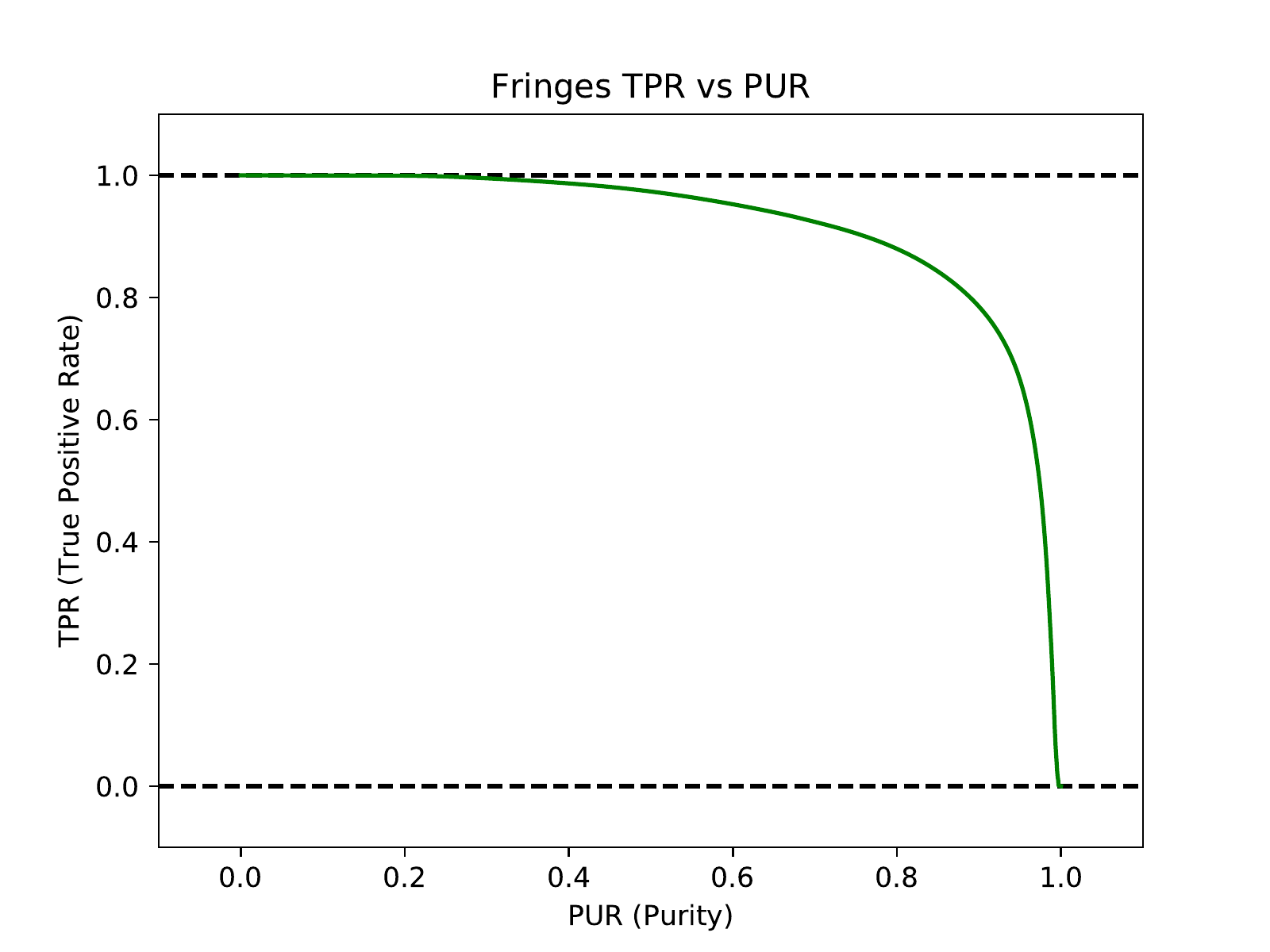}
    \end{minipage} \\
  \end{figure*}
  
  \begin{figure*}[ht]
    \begin{minipage}{0.48\linewidth}
      \includegraphics[scale=0.50]{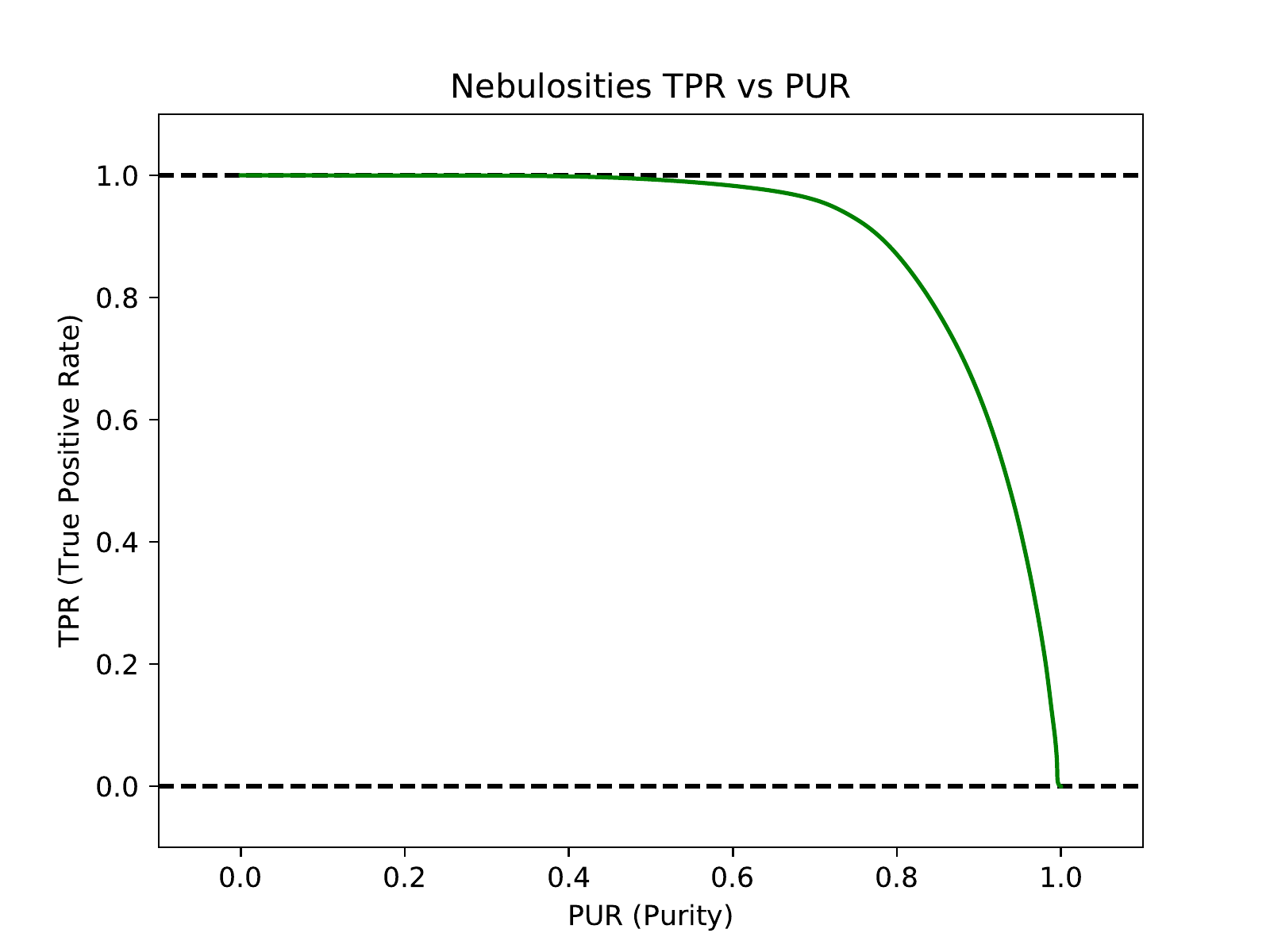}
    \end{minipage}
    \begin{minipage}{0.48\linewidth}
      \includegraphics[scale=0.50]{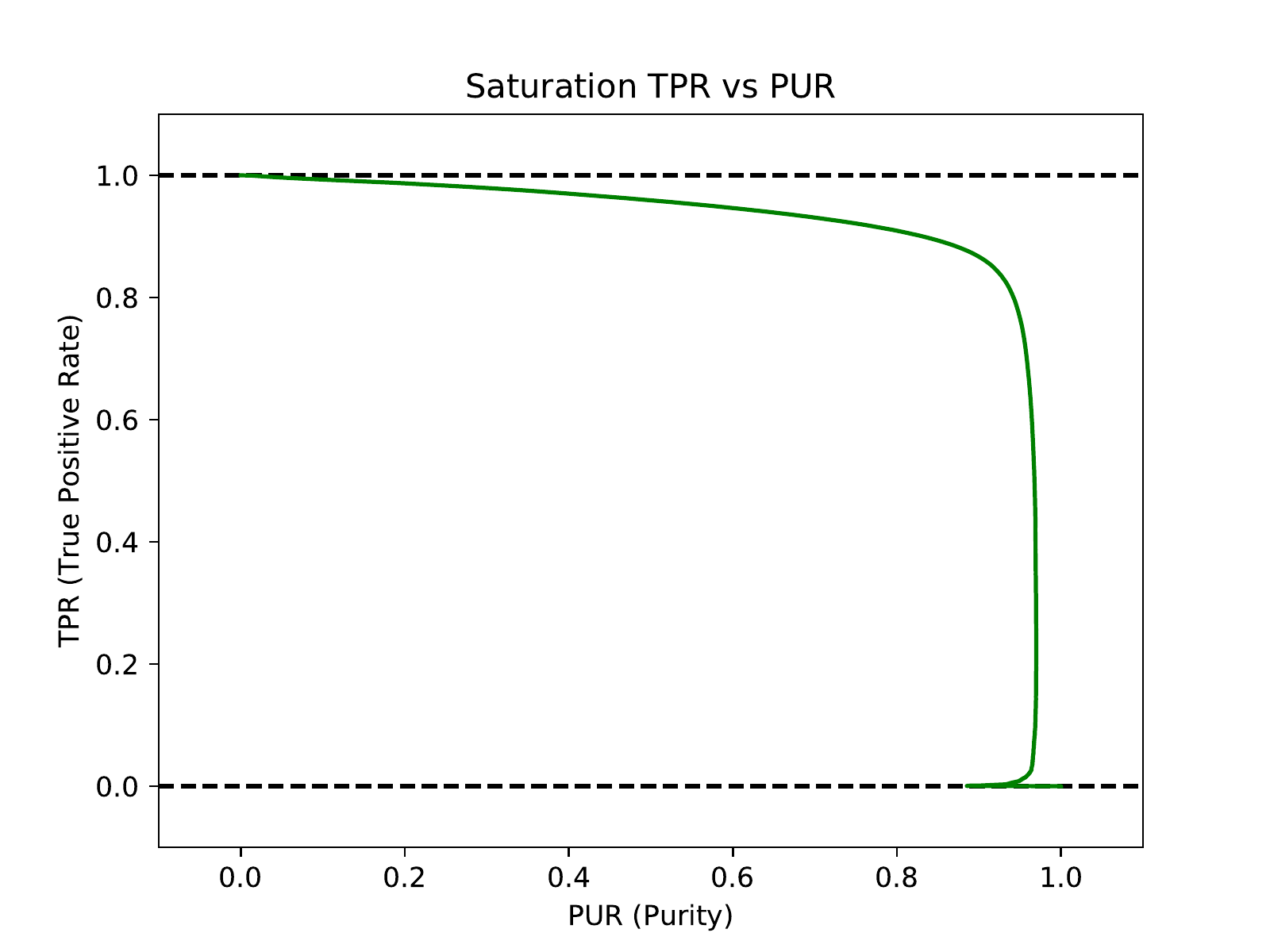}
    \end{minipage} \\
    \begin{minipage}{0.48\linewidth}
      \includegraphics[scale=0.50]{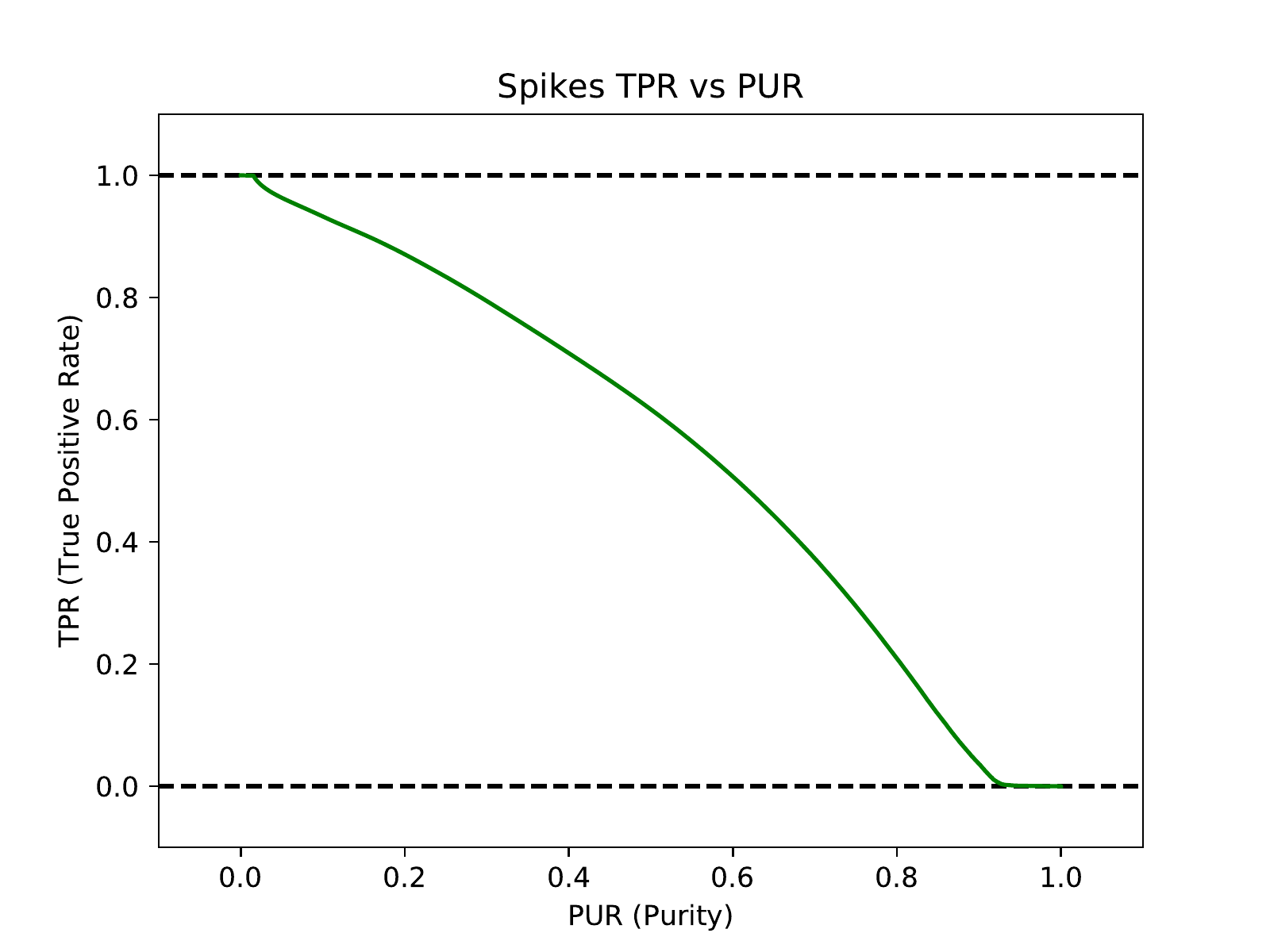}
    \end{minipage}
    \begin{minipage}{0.48\linewidth}
      \includegraphics[scale=0.50]{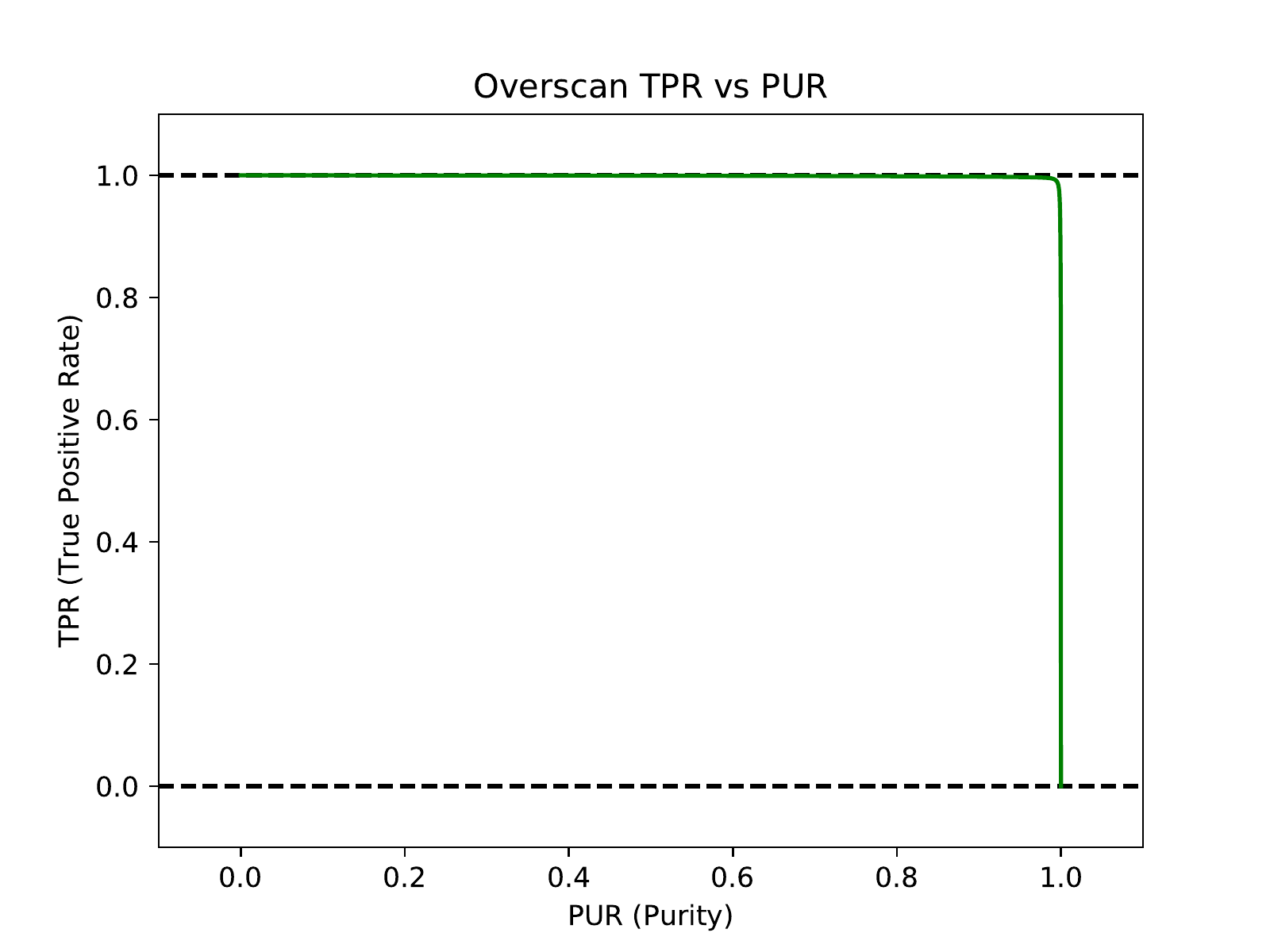}
    \end{minipage} \\
    \begin{minipage}{0.48\linewidth}
      \includegraphics[scale=0.50]{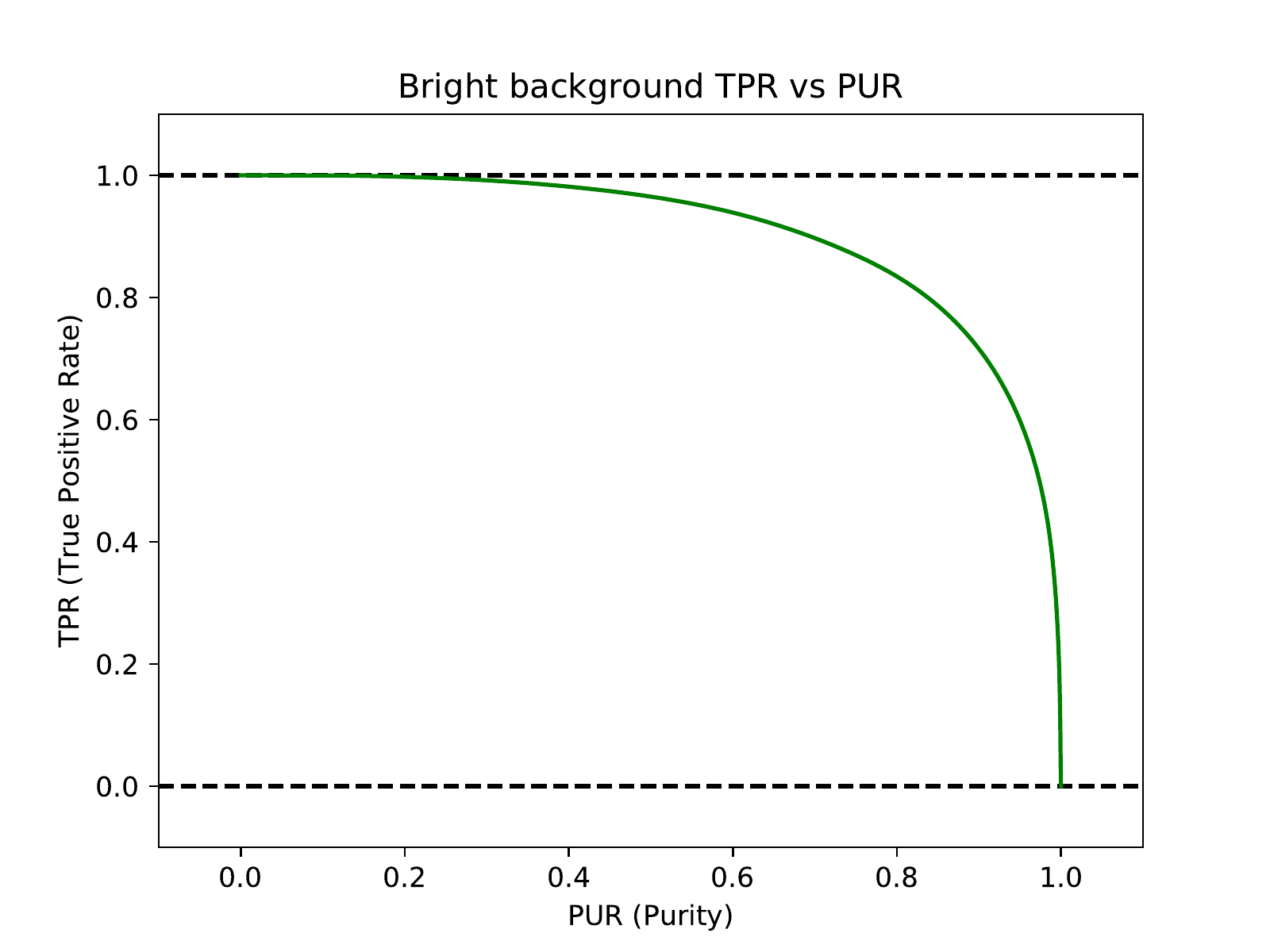}
    \end{minipage}
    \begin{minipage}{0.48\linewidth}
      \includegraphics[scale=0.50]{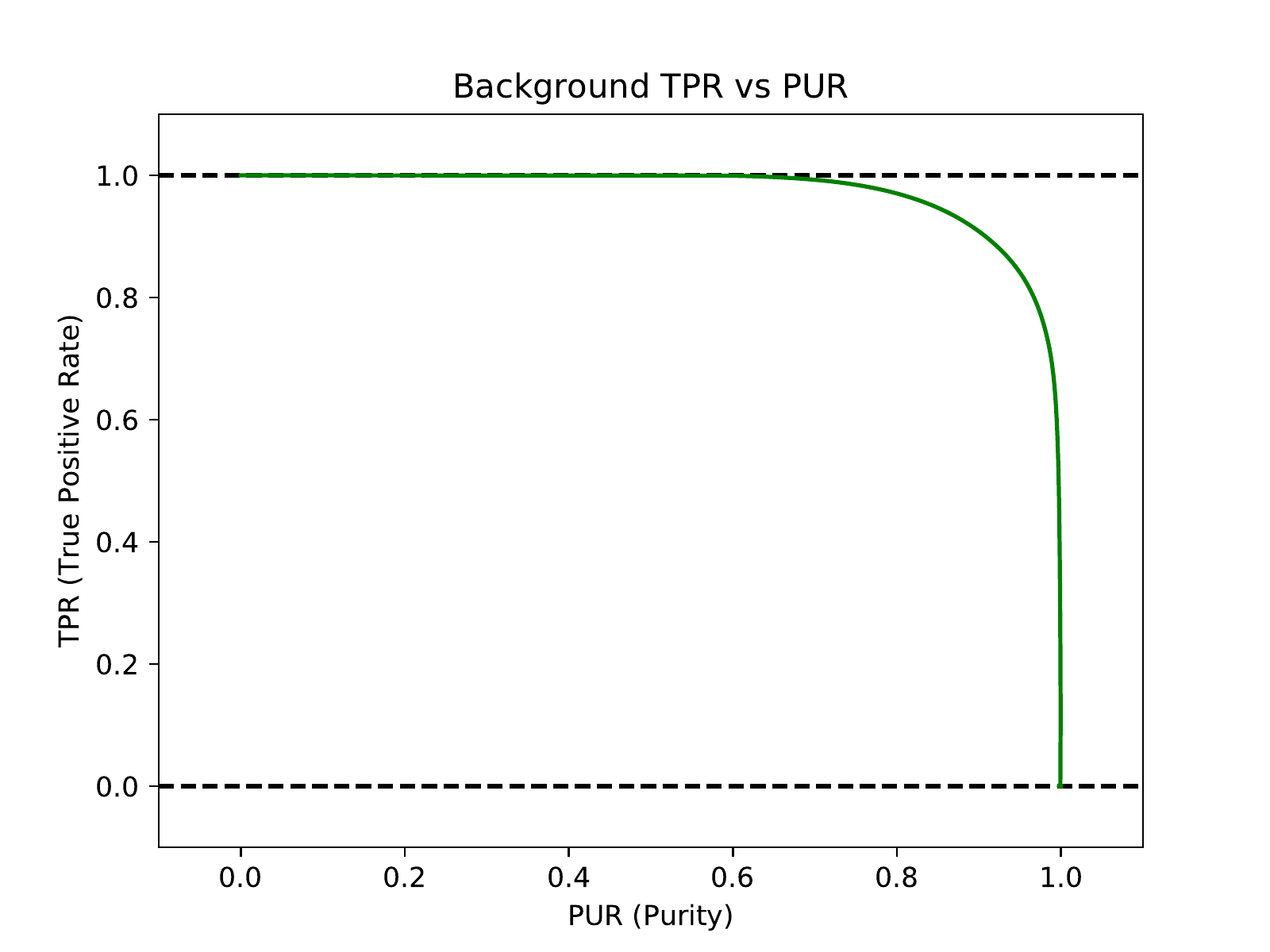}
    \end{minipage}
    \caption{Purity curves: $TPR$ vs $PUR$.}
    \label{pur}
  \end{figure*}
  
\onecolumn

  \begin{figure}[ht]
    \centering
    \begin{minipage}{0.48\linewidth}
      \includegraphics[scale=0.5]{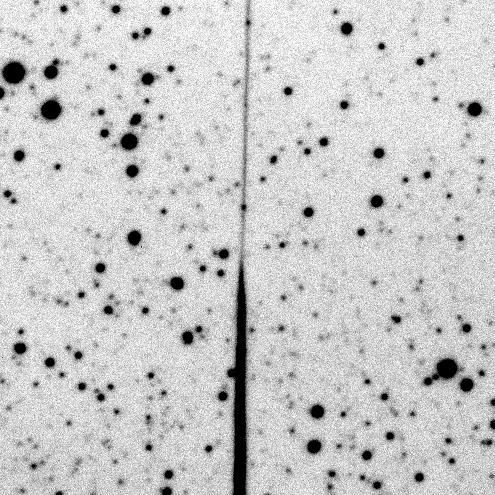}
    \end{minipage}
    \begin{minipage}{0.48\linewidth}
      \includegraphics[scale=0.5]{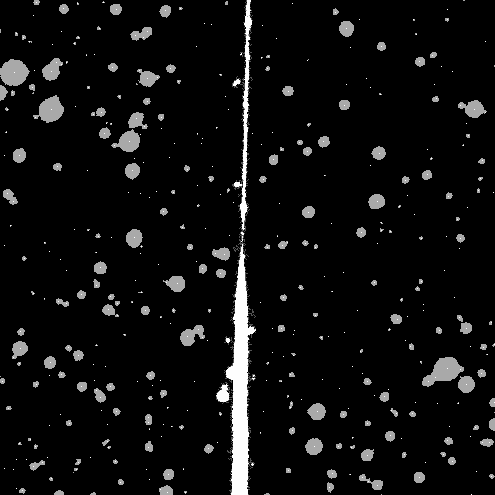} \\
      \includegraphics[scale=0.5]{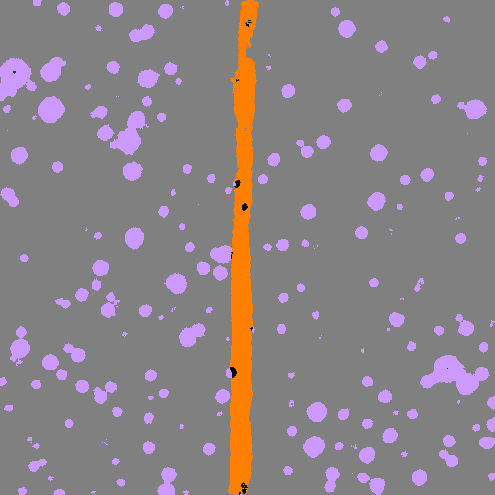}
    \end{minipage}
    \caption{Prediction example for an instrument not used in training: ZTF \citep{2019PASP..131a8002B}. Left: a science image exposure. Top right: mask from the ZTF pipeline. Bottom right: flagging by \textsc{MaxiMask}; the trail is correctly recovered. Also, \textsc{MaxiMask} CNN is able to correctly flag pixels where the trail overlaps sources whereas in the ZTF pipeline, all pixels (i.e., pixels belonging only to the trail, pixels belonging only to sources, and pixels belonging to both the trail and sources) are flagged as both trail and source.}
    \label{ztfex}
  \end{figure}  

  \begin{figure}[ht]
    \centering
    \begin{minipage}{0.48\linewidth}
      \includegraphics[scale=0.62]{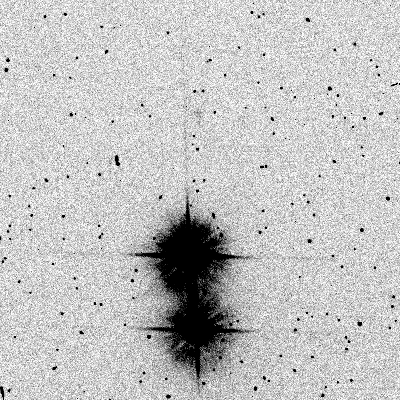}
    \end{minipage}
    \begin{minipage}{0.48\linewidth}
      \includegraphics[scale=0.62]{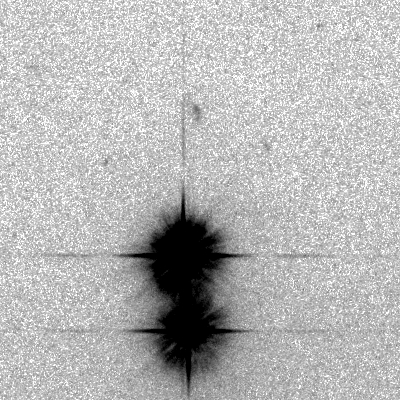} \\
      \includegraphics[scale=0.62]{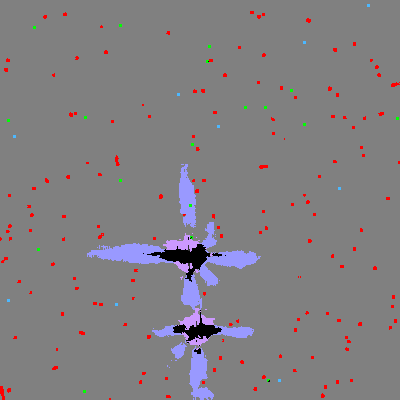}
    \end{minipage}
      
    \caption{Example of a prediction for a space instrument (HST) not used in training (ACS exposure). Left: a calibrated (flat-fielded, CTE-corrected) individual exposure of a stellar field in the Pleiades. Top right: the fully calibrated, geometrically-corrected, dither-combined image where cosmic rays and artefacts have been removed. Bottom right: \textsc{MaxiMask} contaminant identification. Each class is assigned a color so that the ground truth can be represented as a single image (red: CR, dark green: HCL, dark blue: BCL, green: HP, blue: BP, yellow: P, orange: TRL, gray: FR, light gray: NEB, purple: SAT, light purple: SP, brown: OV, pink: BBG, dark gray: BG). Pixels that belong to several classes are represented in black. For the sake of visualization, hot and dead pixel masks have been morphologically dilated so that they appear as $3\times 3$ pixel areas in this representation.}
      \label{hstex}
  \end{figure}

\begin{figure}[ht]
    \begin{minipage}{0.48\linewidth}
      \includegraphics[scale=0.62]{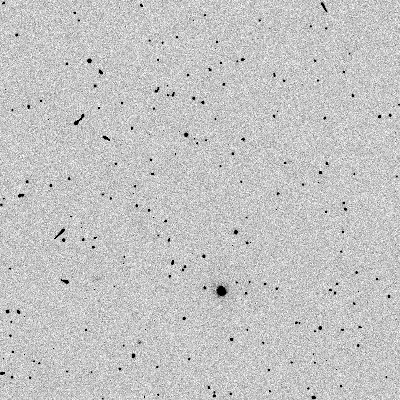}
    \end{minipage}
    \begin{minipage}{0.48\linewidth}
      \includegraphics[scale=0.62]{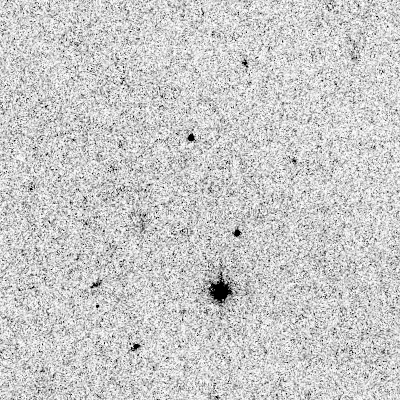} \\
      \includegraphics[scale=0.62]{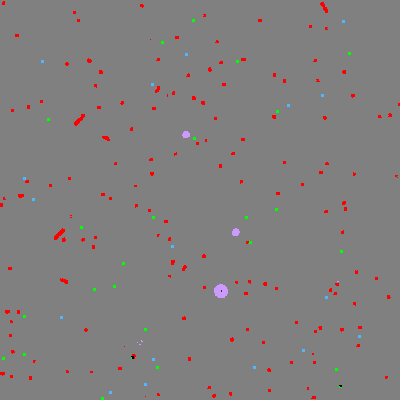}
    \end{minipage}
    \caption{Same as A~\ref{hstex} at a different location in the field to illustrate the ability of \textsc{MaxiMask} to differentiate poorly sampled stellar images from cosmic rays.}
    \label{hstex2}
\end{figure}

\end{appendix}

\end{document}